\documentclass[11pt, a4paper]{article}
\usepackage{jheppub}

\usepackage{amssymb,amsfonts,amsmath,amsthm}
\usepackage{bbm}
\usepackage{booktabs}
\usepackage{dsfont}
\usepackage{graphicx} 
\usepackage{indentfirst}
\usepackage{latexsym}
\usepackage{mathrsfs}
\usepackage{verbatim}
\usepackage{slashed, tensor}
\usepackage{hyperref}
\usepackage{ccaption}
\usepackage{subfigure}
\usepackage{rotating}
\usepackage{lscape}
\usepackage{hypernat}
\usepackage{amsopn}

\def \un{\underline}

\newcommand {\cC}{{\cal C}}
\newcommand {\cD}{{\cal D}}

\newcommand {\cF}{{\cal F}}
\newcommand {\cG}{{\cal G}}
\newcommand {\cH}{{\cal H}}

\newcommand {\cK}{{\cal K}}
\newcommand {\cL}{{\cal L}}
\newcommand {\cM}{{\cal M}}
\newcommand {\cN}{{\cal N}}
\newcommand {\cO}{{\cal O}}

\newcommand {\cR}{{\cal R}}

\newcommand {\cU}{{\cal U}}
\newcommand {\cV}{{\cal V}}
\newcommand {\cW}{{\cal W}}
\newcommand {\cX}{{\cal X}}
\newcommand {\cY}{{\cal Y}}

\def\a{\alpha}
\def\b{\beta}
\def\c{\chi}
\def\d{\delta}
\def\e{\epsilon}
\def\f{\phi}
\def\g{\gamma}
\def\G{\Gamma}

\def\l{\lambda}

\def\o{\omega}

\def\q{\theta}
\def\r{\rho}
\def\s{\sigma}
\def\t{\tau}

\def\x{\xi}

\def\D{\Delta}
\def\F{\Phi}

\def\L{\Lambda}

\def\U{\Upsilon}

\def\ri{{\rm i}}

\newcommand{\ve}{\varepsilon}

\newcommand{\pa}{\partial}                
\newcommand{\hf}{\frac12}

\newcommand{\be}{\begin{equation}}
\newcommand{\ee}{\end{equation}}
\newcommand{\bea}{\begin{eqnarray}}
\newcommand{\eea}{\end{eqnarray}}
\newcommand{\non}{\nonumber}
\newcommand{\ba}{\begin{array}}
\newcommand{\ea}{\end{array}}

\newcommand{\bm}[1]{\mbox{\boldmath$#1$}}

\def\double #1{#1{\hbox{\kern-2pt $#1$}}}

\newcommand{\ha}{{\hat{a}}}

\newcommand{\de}{{\nabla}}

\newcommand{\bbD}{{\mathbb {D}}}

\newcommand{\bsubeq}{\begin{subequations}}
\newcommand{\esubeq}{\end{subequations}}

\newcommand{\ul}{\underline}
\newcommand{\eps}{{\ve}}

\newcommand{\eol}{\notag \\}
\newcommand{\rd}{\mathrm d}
\newcommand{\gD}{{\mathbb D}}

\newcommand{\veps}{\varepsilon}

\newcommand{\RM}{R(M)}
\newcommand{\RD}{R(\mathbb D)}

\newcommand{\RJ}{R(J)}
\newcommand{\RS}{R(S)}
\newcommand{\RK}{R(K)}
\newcommand{\sRM}{R(M)}
\newcommand{\sRD}{R(\mathbb D)}

\newcommand{\sRJ}{R(J)}

\newcommand{\rep}[1]{\textbf{#1}}

\DeclareMathOperator{\Tr}{Tr}

\newcommand{\loco}{\vert}
\newcommand{\doubar}{{{\loco}\!{\loco}}}

\newcommand{\hnabla}{\hat\nabla}
\newcommand{\tgamma}{\tilde\gamma}

\makeatletter
\g@addto@macro\bfseries{\boldmath}
\makeatother

\title{The component structure of conformal supergravity invariants in six dimensions}

\author[a,b]{Daniel Butter,}
\author[c]{Joseph Novak,}
\author[d]{and Gabriele Tartaglino-Mazzucchelli}

\affiliation[a]{Nikhef Theory Group, \\
Science Park 105, 1098 XG Amsterdam, The Netherlands}

\affiliation[b]{George and Cynthia Woods Mitchell Institute for Fundamental Physics and Astronomy,\\
Texas A\&M University, College Station, TX 77843, USA}
\emailAdd{dbutter@tamu.edu}

\affiliation[c]{Max-Planck-Institut f\"ur Gravitationsphysik, Albert-Einstein-Institut, \\
Am M\"uhlenberg 1, D-14476 Golm, Germany}
\emailAdd{joseph.novak@aei.mpg.de}

\affiliation[d]{Instituut voor Theoretische Fysica, KU Leuven,\\
Celestijnenlaan 200D, B-3001 Leuven, Belgium}
\emailAdd{Gabriele.Tartaglino-Mazzucchelli@fys.kuleuven.be}

\preprint{Nikhef-2017-004}

\arxivnumber{1701.08163}

\abstract{
In the recent paper arXiv:1606.02921,
the two invariant actions for 6D $\cN=(1,0)$ conformal supergravity 
were constructed in superspace, corresponding to the supersymmetrization
of $C^3$ and $C\Box C$. In this paper, we provide the translation
from superspace to the component formulation of superconformal tensor calculus,
and we give the full component actions of these two invariants.
As a second application, we build the component form for the 
supersymmetric  $F\Box F$ action coupled to conformal supergravity.
Exploiting the fact that the $\cN=(2,0)$ Weyl multiplet has a consistent
truncation to $\cN=(1,0)$, we then verify that there is indeed only a single
$\cN=(2,0)$ conformal supergravity invariant and reconstruct
most of its bosonic terms by uplifting a certain linear combination of $\cN=(1,0)$
invariants.
}

\numberwithin{equation}{section}
\allowdisplaybreaks

\begin{document}
\maketitle


\section{Introduction}

The invariants for conformal gravity naturally arise 
in the study of conformal field theories on curved manifolds.
Deser and Schwimmer divided the
possible conformal anomalies into two families, type A and type B
\cite{DS}. Type A anomalies are topological and always involve the Euler term,
while type B anomalies are Weyl invariants built
(in the purely gravitational case) from the Riemann tensor and its derivatives.
In six dimensions, there are three independent conformal gravity invariants 
parametrizing the type B anomalies. Their Lagrangians are
\begin{align}\label{eq:ConfInvs}
\cL_1 &= C_{abcd} C^{aefd} C_e{}^{bc}{}_f~, \qquad
\cL_2 = C_{ab}{}^{cd} C_{cd}{}^{ef} C_{ef}{}^{ab}~, \eol
\cL_3 &= C_{abcd} (\d_e^a \Box - 4 \cR_e{}^a + \tfrac{6}{5} \d_e^a \cR) C^{ebcd}~,
\end{align}
where $C_{abcd}$ is the Weyl tensor and $\cR_{ab}$ is the Ricci tensor.

When superconformal field theories are under consideration, the type B anomalies
should correspond to conformal supergravity invariants
and generally the number of such invariants decreases with more supersymmetry.
In six dimensions, superconformal algebras only exist for $\cN=(n,0)$
(or $\cN=(0,n)$) \cite{Nahm}, while the requirement that conformal supergravity
does not contain higher spin fields limits $n < 3$.
Besides general interest in superconformal field theories,
$\cN=(2,0)$ models have been the focus of much interest due to their
somewhat mysterious nature.
Their existence was actually inferred by various arguments in string theory
and they are believed to provide a description of the low-energy dynamics of
multiple coincident M5-branes in M-theory.

In regards to the type B anomalies, there are two obvious questions.
First, how many supersymmetric invariants are permitted and what linear
combinations of \eqref{eq:ConfInvs} do they correspond to?
Second, what are their fully supersymmetric forms when couplings to the rest of
the Weyl multiplet of conformal supergravity are included?
While very strong evidence exists for the purely gravitational form
of these anomalies in the supersymmetric cases -- namely the existence of
two invariants in  $(1,0)$ and only one for  $(2,0)$, which we discuss below --
very little was known about their supersymmetric completions.
In principle the answer to both questions could be investigated
via indirect means by e.g. computing the conformal anomaly of various  $(1,0)$
or  $(2,0)$ matter multiplets coupled to (super)gravity, as advocated in
\cite{Beccaria:2015uta}.\footnote{Such an
approach was advocated for constructing the 4D $\cN\!=\!4$ conformal
supergravity action \cite{BuPlTs} prior to its recent direct construction
\cite{BCdWS}.}
So far only the purely gravitational part of these computations have
been performed in 6D, see e.g. \cite{BFT00, Kulaxizi:2009pz, BT15}.
Alternatively, one could construct the full supersymmetric invariants
directly.

Recently, the direct path was pursued in \cite{BKNT16}, where two  $(1,0)$ Weyl invariants were built
using superspace techniques, and a separate supercurrent analysis was
given that established that there were at most two such invariants.
One of them was observed to contain only $C^3$ terms in the particular combination
\begin{align}\label{eq:IntroC3}
- \frac{1}{8}\eps^{abcdef} \eps_{a'\!b'\!c'\!d'\!e'\!f'}
	C_{ab}{}^{a'\!b'} C_{cd}{}^{c'\!d'} C_{ef}{}^{e'\!f'}
 = 8 \cL_1 + 4 \cL_2~,
\end{align}
while the other was observed to contain $\cL_3$ at the quadratic order.
As we will see, it actually contains
additional cubic terms in the Weyl tensor in the particular combination 
\begin{align}\label{eq:IntroCBoxC}
4 \cL_1 - \cL_2 + \cL_3~.
\end{align}
We will refer to these two particular combinations as the $C^3$ and $C \Box C$
invariants from now on.

The approach of \cite{BKNT16} involved the direct construction of the two invariants
from certain conformal primary superfields. These were composites built
from the super-Weyl tensor and its derivatives within a novel superspace formulation of 
6D $\cN=(1,0)$ conformal supergravity, called conformal superspace. Inspired by earlier 
formulations in three, four, and five
dimensions \cite{Butter4DN=1, Butter4DN=2, BKNT-M1, BKNT-M5D},
it is obtained by gauging the 6D $\cN=(1,0)$ superconformal algebra in superspace.
As in those cases, the superspace torsion and curvature tensors turn out to be built solely in terms
of the super-Weyl tensor, which simplifies computations significantly.\footnote{In conventional superspace approaches, see e.\,g.\,\cite{WZ, SG, HoweTucker, Grimm, Howe1, Howe2, KLRT-M2008, HIPT,KLT-M, KT-M08, LT-M12}, the structure group contains only the Lorentz and $R$-symmetry groups, and additional torsion superfields appear.}
Conformal superspace has proven useful in the context of general 
higher-derivative supergravity actions, such as
the $\cN$-extended conformal supergravity 
actions in three dimensions for $3\leq \cN \leq 6$ \cite{BKNT-M2} and 
a new set of curvature-squared invariants in 4D $\cN=2$ supergravity \cite{BdeWKL, BdWL},
which arise from dimensional reduction of the 5D mixed gauge-gravity Chern-Simons
term. A major advantage of conformal superspace is that
at the component level, it recovers the Weyl multiplet and transformation rules of
conformal supergravity as formulated within
superconformal tensor calculus, first developed in six dimensions
by Bergshoeff, Sezgin and Van Proeyen \cite{BSVanP}, building off earlier
work in 4D \cite{KTvN2, deWvHVP1} (see also the textbook \cite{FVP}).
The 6D superconformal tensor calculus has proven useful in the construction of the
supersymmetric extension of a Riemann curvature squared term \cite{BSS1, BSS2, BR}
and, more recently, in the complete off-shell action for minimal Poincar\'e
supergravity \cite{Coomans:2011ih}. Gauged minimal  supergravity \cite{Bergshoeff:2012ax}
has also been worked out 
by coupling the minimal Poincar\'e supergravity to an off-shell vector 
multiplet. 

Our primary aims in this paper are to describe the connection between superspace and
components for 6D  $(1,0)$ conformal superspace, and to convert 
the superspace invariants given in \cite{BKNT16} to component form in the language
of superconformal tensor calculus \cite{BSVanP}.
In particular, we will focus on the full set of (bosonic) terms that supersymmetrize
\eqref{eq:IntroC3} and \eqref{eq:IntroCBoxC}. These are given respectively in
\eqref{eq:BosonicC3} and \eqref{eq:BosonicCBoxC}.
As another application, we present the component structure of an $F\Box F$ invariant
coupled to  $(1,0)$ conformal supergravity with $F$ the field strength of a Yang-Mills
multiplet. Its bosonic terms are given in \eqref{eq:FBoxF} and coincide with the
flat space construction of \cite{ISZ05}.

Although our main interest is in  $(1,0)$ supersymmetry, it turns out one can deduce
a great deal about the  $(2,0)$ invariant once the structure of the  $(1,0)$ invariants
is known. Already quite a lot of evidence (see e.g. \cite{HS98, BFT00, BT15})
suggests that there should only be one type B anomaly with  $(2,0)$ supersymmetry.
Its purely gravitational part is
\begin{align}\label{eq:Intro(2,0)}
4\cL_1 + \cL_2 + \frac{1}{3} \cL_3~,
\end{align}
and should be extendable to some $(2,0)$ conformal supergravity invariant
containing additional terms involving  other fields of the Weyl multiplet.
To our knowledge, no analysis of the off-shell supersymmetric extension of 
this term has been performed.
Another goal of this paper is to make a major step towards solving this problem.
By analyzing the structure of the two $(1,0)$ conformal supergravity invariants,
we will show that only a certain combination can be lifted to $(2,0)$
conformal supergravity; this combination has \eqref{eq:Intro(2,0)} as its purely
gravitational part.
We exploit the uplift to $(2,0)$ to give for the first time a large part of the
bosonic sector of the off-shell $(2,0)$ conformal supergravity invariant.

This paper is organized as follows. In section \ref{SCTC} we show how to recover the Weyl multiplet from 
conformal superspace and derive the associated supersymmetry transformations.
In section \ref{C3action} we describe the component structure of the $C^3$ invariant,
explicitly giving its bosonic sector.
In section \ref{CboxCaction} we consider the $C \Box C$ and $F \Box F$ invariants.
We devote  section \ref{confSUGRAInv(2,0)} to a discussion of the off-shell $\cN=(2,0)$ conformal
supergravity invariant.
Finally, we discuss our results and present some concluding remarks in section \ref{conclusion}.

We have also included a few technical appendices. In Appendix \ref{conf-superspace} we 
provide the salient details of conformal superspace and describe how the covariant derivative 
algebra is deformed under a certain redefinition of the vector covariant derivative, which 
we will use in the main body of the paper. Appendix \ref{MatchConventions} provides a 
brief prescription for how to relate our notation and conventions to those already appearing
in the literature. Along with the arXiv submission, we have included a separate file containing
the building blocks (including all fermionic terms) for the $(1,0)$
invariants constructed in sections \ref{C3action} and \ref{CboxCaction}.


\section{The 6D $\cN=(1,0)$ Weyl multiplet from superspace}
\label{SCTC}

In this section, we will show how to reduce the superspace formulation
of conformal supergravity in \cite{BKNT16} to components. Let us first
elaborate on the component structure of  
the Weyl multiplet of 6D $\cN=(1,0)$ conformal supergravity,
first developed in \cite{BSVanP}.
Within the superconformal tensor calculus framework \cite{BSVanP}, one
gauges the superconformal algebra in spacetime. Associated respectively
with local translations, $Q$-supersymmetry, $\rm SU(2)$ $R$-symmetry, and
dilatations are the vielbein $e_m{}^a$, the gravitino $\psi_m{}^\alpha_i$,
the SU(2) gauge field $\cV_m{}^{ij}$, and a dilatation gauge field $b_m$.
The remaining gauge symmetries are associated with composite connections:
these are the spin connection $\omega_m{}^{cd}$,
the $S$-supersymmetry connection $\phi_m{}_\a^i$,
and the special conformal connection $\frak{f}_m{}_a$.
To ensure that the last three connections are composite,
one imposes conventional constraints (which are in general not unique) on the vielbein
curvature $R(P)_{mn}{}^a$, the gravitino curvature
$R(Q)_{mn}{}^\a_i$,
and the conformal Lorentz curvature $R(M)_{mn}{}^{ab}$.
However, the independent one-forms cannot furnish an off-shell representation
of a conformal supersymmetry algebra as the bosonic and fermionic
degrees of freedom do not match. 
An off-shell representation
is achieved by  introducing three covariant fields:
a real anti-self-dual tensor $T^-_{abc}$,
a chiral fermion $\chi^{\a i}$, and a real scalar field $D$
which deform the supersymmetry algebra, the curvatures and the
constraints imposed on the curvatures in a 
consistent 
way \cite{BSVanP}.
This procedure can be considered as a bottom-up approach where
step-by-step one builds up a consistent
off-shell multiplet for conformal supergravity.

In conformal superspace the superconformal algebra is manifestly gauged off-shell from
the very beginning. Rather than
construct a multiplet of gauge fields and covariant matter fields which must possess the same
supersymmetry algebra (usually with modifications due to the curvature tensors and
covariant fields), one must completely determine the supersymmetry algebra
directly by solving superspace Bianchi identities. Typically these identities
are solved by a single superfield, which encodes
all component curvature tensors along with the covariant fields necessary
for off-shell closure. Once the solution is found,
the component fields and their supersymmetry transformations can be obtained directly
by projecting to spacetime; the resulting component structure turns out to match
that constructed in components directly.
This method therefore can be viewed as a top-down approach.


\subsection{Component fields and curvatures from superspace}

We begin by identifying the various component fields of the
6D $\cN=(1,0)$ Weyl multiplet \cite{BSVanP} 
within the geometry of conformal superspace \cite{BKNT16}.
For the one-forms, this identification is particularly easy as
each component one-form is in direct correspondence with some superspace one-form
and these can be connected in a straightforward way.

Let us start with the vielbein ($e_m{}^a$) and gravitino ($\psi_m{}^\a_i$).
These appear as the $\q=0$ projections of the coefficients of $\rd x^m$ in the supervielbein
$E^A = (E^a, E^\a_i) = \rd z^M\, E_M{}^A$,\footnote{Recall that
$z^M=(x^m,\q^\mu_\imath)$ are coordinates for a local parametrization of 
6D $\cN=(1,0)$ conformal superspace, see \cite{BKNT16} and Appendix 
\ref{conf-superspace}.}
\begin{align}
e_m{}^a(x) := E_m{}^a(z) \loco~, \qquad
\psi_m{}^\alpha_i(x) := 2 \, E_m{}^\alpha_i(z) \loco~,
\end{align}
where a single vertical line next to a superfield denotes setting $\q=0$.
This operation can be written in a coordinate-independent way using the so-called
double-bar projection \cite{Baulieu:1986dp, BGG01}
\begin{align}
e{}^a = \rd x^m e_m{}^a = E^a\doubar~,~~~~~~
\psi^\a_i = \rd x^m \psi_m{}^\a_i =  2 E^\a_i \doubar ~,
\end{align}
where the double bar denotes setting $\q = \rd \q = 0$.\footnote{In more mathematical
language, the double-bar projection is the pullback of the inclusion map embedding
spacetime into superspace.}
In like fashion, the remaining fundamental and composite one-forms correspond to
double-bar projections of superspace one-forms,
\begin{align}
	\cV^{kl} := \Phi{}^{kl} \doubar~, \quad
	b := B\doubar ~, \quad
\omega^{cd} := \Omega{}^{cd} \doubar ~, \quad  
\phi_\g^k := 2 \,\frak F{}_\g^k\doubar~, \quad
\frak{f}{}_c := \frak{F}{}_c\doubar~. 
\end{align}

The covariant matter fields are contained within the super-Weyl tensor $W_{abc}$
and its independent descendants. 
We define the three covariant component fields as\footnote{We have chosen the coefficients such that
 $T^-_{abc}$, $\chi^{\a i}$ and  $D$ exactly correspond to the covariant matter fields 
 of the Weyl multiplet introduced in \cite{BSVanP}.
We always denote the anti-self-dual covariant field as $T^-_{abc}$ 
to avoid confusion with the superspace torsion tensor $T_{ab}{}^c$.}
\bsubeq
\bea
T^-_{abc} &:=& -2 W_{abc}\loco \ ,
\\
\chi^{\a i} &:=& \frac{15}{2}X^{\a i} \loco=-\frac{3\ri}{4}\de^i_\b W^{\a\b}\loco~,
\\
D&:=&
\frac{15}{2}Y \loco=
-\frac{3\ri}{16}\de_\a^k\de_{\b k}W^{\a\b}\loco
~.
\eea
\esubeq
There are three additional independent descendant fields:
the dimension-3/2 fermionic field $\cX_{\a}^i{}^{\b\g}:=X_{\a}^i{}^{\b\g}|$, and
the dimension-2 bosonic fields $\cY_\a{}^\b{}^{kl}:=Y_\a{}^\b{}^{kl}|$ and 
$\cY_{\a\b}{}^{\g\d}:=Y_{\a\b}{}^{\g\d}|$.\footnote{The descendant fields of $W_{abc}$ are defined 
in Appendix \ref{conf-superspace}.}
These will turn out to be composite and expressible directly in terms of the component
curvatures. The differential constraints on the superfield $W_{abc}$ forbid
any independent component fields at higher dimension \cite{BKNT16}.

It should be mentioned that one can impose a Wess-Zumino gauge to fix
the $\q$ expansions of the superspace gauge one-forms, 
so that they are completely determined by the above component fields.
This ensures that the entire content of the superspace
geometry is
encoded in the
independent physical fields. In practice,
using the above definitions eliminates the need to do this explicitly.

In conformal superspace the covariant exterior derivative is defined as
\be\label{eq:covD}
\nabla = E^A \nabla_A =
\rd - \frac{1}{2} \Omega{}^{cd} M_{cd}
	- B \mathbb D
	- \Phi{}^{kl} J_{kl}
	- \mathfrak F{}_{\a}^{ i} S^{\a}_{ i}
	- {\frak F}_{a} K^a~,
\ee
with $\de_A$ the covariant derivatives.
By taking the double bar projection of $\de$,
we can define the component $\de_a$ to coincide with
the projection of the superspace derivative $\nabla_a \loco$,
\begin{align}
e_m{}^a \nabla_a = \pa_m
	- \frac{1}{2} \psi_m{}^\a_i \nabla_{\a}^i \loco
	- \frac{1}{2} \omega_m{}^{cd} M_{cd}
	- b_m \mathbb D
	- \cV_m{}^{kl} J_{kl}
	- \frac{1}{2} \phi_m{}_{\a}^{ i} S^{\a}_{ i}
	- {\frak f}_m{}_{a} K^a
	~.
\end{align}
The projected spinor covariant derivative $\nabla_{\a}^i\loco$
corresponds to the generator of $Q$-supersymmetry, and is defined so that
if $\cU = U\vert$, then $\nabla_\a^i| \cU := (\nabla_\a^i U) \loco$.
Note that there is no ambiguity for the other generators as e.g.
$M_{c d}\cU =(M_{c d} U)\loco$, and so local diffeomorphisms,
$Q$-supersymmetry transformations, and so forth descend naturally
from their corresponding rule in superspace \eqref{SUGRAtransformations-tensor},
which can be written
\begin{align}
\delta \cU =
\xi^a \nabla_a \cU
+\xi^\alpha_i \nabla_\alpha^i \loco \cU
+ \hf \l^{cd} M_{cd} \cU
+ \l^{kl} J_{kl} \cU
+ \s \mathbb D\cU
+ \eta_\g^k S^\g_k \cU
+ \l_{ c} K^c \cU
~.
\label{SUGRAtransformations-tensor-components}
\end{align}
Note that in spacetime we can choose to parametrize local superconformal
transformations either with a covariant diffeomorphism, generated by $\xi^a \nabla_a$,
or as a normal diffeomorphism, generated by $\xi^m \pa_m$. For $Q$-supersymmetry
on the other hand, the natural choice in spacetime involves the covariant spinorial
derivative $\nabla_{\alpha}^i \loco$ rather than the $\q$-derivative.

The algebraic structure of these operators descends straightforwardly from superspace.
For example, the component supercovariant curvature tensors arise by projecting
\eqref{gauging-SCA-2_c},
\begin{align}
[\nabla_a, \nabla_b]
	&= 
	- R(P)_{ab}{}^c \nabla_{c}
	- R(Q)_{a b}{}^{\g}_k \nabla_\g^k|
	- \frac{1}{2} \RM_{ab}{}^{cd} M_{cd}
	- \RJ_{ab}{}^{kl} J_{kl}
	\eol & \quad
	- \RD_{ab} \gD 
	- \RS_{ab}{}_{\g}^{k} S^{\g}_{ k}
	- \RK_{ab}{}_c K^c
	~,
\end{align}
where we have introduced the expressions
$R(P)_{a b}{}^{c} = T_{a b}{}^c \loco$ and
$R(Q)_{a b}{}^\g_k = T_{ab}{}^\g_k \loco$
for the lowest components of the superspace torsion tensors
to match the usual component nomenclature, while
$\RM_{ab}{}^{cd}$, $\RJ_{ab}{}^{kl}$, $\RD_{ab}$, $\RS_{ab}{}_{\g}^{k}$ and $\RK_{ab}{}_c$
are the lowest components of the corresponding superspace curvatures.

The constraints on the superspace curvatures determine
how the covariant fields of the 
Weyl multiplet should appear within these curvatures; in other words,
the superspace geometry dictates how to supercovariantize
a given component curvature. Let us illustrate this by deriving the
explicit form of $R(P)_{ab}{}^c:=T_{ab}{}^c|$.
Consider the double bar projection of the torsion two-form $T^c$, eq. \eqref{torCurExp_a}.
This can be evaluated either in terms of its explicit definition,
\bea
T^c\doubar
:= \cD E^c \doubar = 
\cD e^c
= \rd x^n \wedge\rd x^m\, \cD_{[m} e_{n]}{}^c
~,
\label{T_111}
\eea
where we have introduced  the spin, dilatation, and $\rm SU(2)$ covariant derivative
\begin{align}
\cD_m &:= \pa_m
	- \frac{1}{2} \omega_m{}^{cd} M_{cd}
	- b_m \mathbb D
	- \cV_m{}^{kl} J_{kl}~,  \qquad
\cD_a := 
e_a{}^m \cD_m ~,
\end{align}
or in terms of its tangent space decomposition
\begin{align}
T^c\doubar &= \frac{1}{2} E^A \wedge E^B T_{BA}{}^c \doubar \eol
&= \hf \rd x^n\wedge\rd x^m
\Big(
e_m{}^ae_n{}^b
T_{ab}{}^c|
+e_m{}^a\psi_n{}^\b_jT_{a}{}_\b^j{}^c|
-\frac{1}{4}\psi_m{}^\a_i\psi_n{}^\b_jT_\a^i{}_\b^j{}^c|
\Big)
 \ .
 \label{T_222}
\end{align}
Equating \eqref{T_111} and \eqref{T_222} and solving for $R(P)_{ab}{}^c$ leads to
\bsubeq\label{sup-conp_curvs_1}
\bea
R(P)_{ab}{}^c 
&=&
2e_a{}^me_b{}^n\cD_{[m} e_{n]}{}^c
+\psi_{[a}{}^\b_jT_{b]}{}_\b^j{}^c|
+\frac{1}{4}\psi_{[a}{}^\a_i\psi_{b]}{}^\b_jT_\a^i{}_\b^j{}^c|
~.
\label{sup-conp_curvs_1-a}
\eea
Proceeding in the same way for the other curvature two-forms gives
\bea
R(Q)_{ab}{}^\g_k
&=&
\hf\Psi_{ab}{}^\g_k
+\ri \,(\tilde{\g}_{[a})^{\g\d} \phi_{b]}{}_{\d k} 
+\psi_{[a}{}^\b_jT_{b]}{}_\b^j{}^\g_k|
+\frac{1}{4}\psi_a{}^\a_i\psi_b{}^\b_jT_\a^i{}_\b^j{}^\g_k|
~,
\label{sup-conp_curvs_1-b}
\\
R(\mathbb D)_{ab} 
&=& 
2e_a{}^me_b{}^n\pa_{[m}b_{n]}
 +4 \mathfrak{f}_{[a}{}_{b]}
-\psi_{[a}{}^\a_i \phi_{b]}{}_\a^i
+\psi_{[a}{}^\b_j\sRD_{b]}{}_\b^j |
\non\\
&&
+\frac{1}{4}\psi_{[a}{}^\a_i\psi_{b]}{}^\b_j\sRD_\a^i{}_\b^j |
~,
\label{sup-conp_curvs_1-c}
\\
R(M)_{ab}{}^{cd} 
&=&
\cR_{ab}{}^{cd}
+8 \d_{[a}^{[c} \mathfrak{f}_{b]}{}^{d]} 
-\psi_{[a}{}^\a_j  \f_{b]}{}_\b^j (\g^{cd})_\a{}^\b
+\psi_{[a}{}^\b_j\sRM_{b]}{}_\b^j{}^{cd} |
\non\\
&&
+\frac{1}{4}\psi_a{}^\a_i\psi_b{}^\b_j\sRM_\a^i{}_\b^j{}^{cd} |
~,
\label{sup-conp_curvs_1-d}
\\
R(J)_{ab}{}^{kl} 
&=&
\cR_{ab}{}^{kl}
+4 \psi_{[a}{}^{\g (k} \f_{b]}{}_{\g}^{l)} 
+\psi_{[a}{}^\b_j\sRJ_{b]}{}_\b^j{}^{kl} |
+\frac{1}{4}\psi_a{}^\a_i\psi_b{}^\b_j\sRJ_\a^i{}_\b^j{}^{kl} |
~,
\label{sup-conp_curvs_1-e}
\eea
\esubeq
where
\bsubeq
\bea
\Psi_{ab}{}^\g_k
&:=&
2e_a{}^me_b{}^n\cD_{[m}\psi_{n]}{}^\g_k
~,
\label{Psi}
\\
\cR_{ab}{}^{cd}
&:=&
\cR_{ab}{}^{cd}(\o)
=
e_a{}^me_b{}^n\Big(
2\pa_{[m}\o_{n]}{}^{cd}
-2\o_{[m}{}^{ce} \o_{n]}{}_e{}^d
\Big)
~,
\label{Romega}
\\
\cR_{ab}{}^{kl}
&:=&
\cR_{ab}{}^{kl}(\cV)
=
e_a{}^me_b{}^n\Big(
2\pa_{[m}\cV_{n]}{}^{kl}
+2\cV_{[m}{}^{p(k} \cV_{n]}{}_p{}^{l)}
\Big)
~.
\eea
\esubeq
The last two terms in
each of the curvatures \eqref{sup-conp_curvs_1}
involve the covariant fields of the Weyl multiplet
and from a component perspective are necessary for off-shell superconformal covariance.
From a superspace perspective, their structure is instead dictated by
the superspace geometry. Using the torsion and curvatures of \cite{BKNT16},
their explicit forms are
\bsubeq\label{comp-curv_2-0}
\bea
R(P)_{ab}{}^c 
&=&
2e_a{}^me_b{}^n\cD_{[m} e_{n]}{}^c
+\frac{\ri}{2}\psi_{[a}{}_i\g^c\psi_{b]}{}^i
~,
\label{comp-curv_2-0-a}
\\
R(Q)_{ab}{}_k
&=&
\hf\Psi_{ab}{}_k
+\ri \,\tilde{\g}_{[a}\phi_{b]}{}_{ k} 
+\frac{1}{12}T^-_{cde}\tilde{\g}^{cde}\g_{[a}\psi_{b]}{}_k
~,
\label{comp-curv_2-0-b}
\\
R(\mathbb D)_{ab} 
&=& 
2e_a{}^me_b{}^n\pa_{[m}b_{n]}
 +4 \mathfrak{f}_{[a}{}_{b]}
-\psi_{[a}{}_i \phi_{b]}{}^i
- \frac{\ri}{6}\psi_{[a}{}_j \g_{b]} \chi^{ j} 
~,
\label{comp-curv_2-0-c}
\\
R(M)_{ab}{}^{cd} 
&=&
 \cR_{ab}{}^{cd}
+8 \d_{[a}^{[c} \mathfrak{f}_{b]}{}^{d]} 
-\psi_{[a}{}_j\g^{cd} \f_{b]}{}^j 
+2\ri\psi_{[a}{}_j\g_{b]} R(Q)^{cd}{}^{ j}
-\frac{\ri}{6}\psi_{[a}{}_j\g_{b]}\tilde{\g}^{cd}\chi^{j} 
~,~~~~~~~
\label{comp-curv_2-0-d}
\\
R(J)_{ab}{}^{kl} 
&=&
\cR_{ab}{}^{kl}
+4\psi_{[a}{}^{(k} \f_{b]}{}^{l)} 
+\frac{2\ri}{3}\psi_{[a}{}^{(k}\g_{b]} \chi^{ l)}
~.
\label{comp-curv_2-0-e}
\eea
\esubeq
Here we have suppressed spinor indices for legibility.
Note that in  \eqref{comp-curv_2-0-d} we have used the fact that the component field
$\cX_{\a }^k{}^{\b\g} $ turns out to be composite
\bea
(\g_{ab})_\b{}^\a \cX_{\a }^k{}^{\b\g} 
=
 R(Q)_{ab}{}^{\g k}
-\frac{1}{10} (\tilde{\g}_{ab})^\g{}_\b \chi^{\b k}
~. 
\label{g-traceless-T}
\eea
For the sake of brevity, we
do not present here the expressions for $R(S)_{ab}{}^k$ and $R(K)_{ab}{}_c$.


\subsection{Analysis of the curvature constraints}

It has already been mentioned that the component spin, $S$-supersymmetry,
and special conformal connections turn out to be composite. This property
arises here just as in the purely component framework \cite{BSVanP} because
of constraints on some of the curvature tensors. In our framework, these
constraints are already imposed at the superfield level and lead to\bsubeq\label{compConstraintsTR}
\bea
R(P)_{ab}{}^c &=&2T^-_{ab}{}^c
~,
\label{compConstraintsTR_1}
\\
\g^bR(Q)_{ab}{}_k
&=&
 \frac{1}{2} \g_a\chi_k 
~,
\label{compConstraintsTR_2}
\\
\RM_{ab}{}^{cb} 
&=&
-\frac{1}{3}\d_{a}^c  \,D
+ \nabla^{b} T^-_{ba}{}^{c}
~,
\label{compConstraintsTR_3}
\eea
\esubeq
where
\bea
\nabla_d T^-_{abc}
=
\cD_d T^-_{abc}
+\frac{\ri}{15}(\g_{abc})_{\a\b}\psi_d{}^{\a}_k\chi^{\b k}
+\frac{\ri}{2}(\g_{abc})_{\a\b} \psi_d{}^{\g}_k\cX_\g^k{}^{\a\b} 
~.
\label{deW}
\eea

The first constraint \eqref{compConstraintsTR_1} determines the spin connection to be
\begin{align}
\omega_{a}{}_{bc}
=
\omega(e)_{a}{}_{bc}
-2\eta_{a[b}b_{c]}
-\frac{\ri}{4}\psi_b{}^k\g_a\psi_c{}_k
-\frac{\ri}{2}\psi_a{}^k\g_{[b}\psi_{c]}{}_k
+T^-_{abc}
~,
\label{omega}
\end{align}
where $\omega(e)_{a}{}_{bc} = -\frac{1}{2} (\cC_{abc} + \cC_{cab} - \cC_{bca})$
is the usual torsion-free spin connection given
in terms of the anholonomy coefficient
$\cC_{mn}{}^a := 2 \,\partial_{[m} e_{n]}{}^a$.
It is important to note that the spin connection $\omega_{abc}$
possesses not only the usual fermionic torsion, due to the gravitino terms,
but also bosonic torsion from the covariant field $T^-_{abc}$. In particular,
this means that there is 
non-trivial dependence on $T^-_{abc}$ nested in every covariant derivative $\de_a$ and $\cD_a$.

The second constraint \eqref{compConstraintsTR_2} is solved by
\bea
\phi_m{}^{k}
&=&
\frac{\ri}{16}\Big(
\g^{bc}\g_m-\frac{3}{5}\g_m\tilde{\g}^{bc}
\Big)
\Big(
\Psi_{bc}{}^{k}
+\frac{1}{6}T^-_{def}\tilde{\g}^{def}\g_{[b}\psi_{c]}{}^{k}
\Big)
-\frac{\ri}{10}\g_m\chi^{k} 
~.
\label{eq:SConn} 
\eea
Reinserting this into the original expression for $R(Q)$ gives
\bea
R(Q)_{ab}{}_k
=
\hf\Pi_{ab}{}^{cd}
\Big(
\Psi_{cd}{}_k
+\frac{1}{6}T^-_{efg}\tilde{\g}^{efg}\g_{[c}\psi_{d]}{}_k
\Big)
+\frac{1}{10}\tilde{\g}_{ab}\chi_{k} 
~,
\label{eq:RQ}
\eea
where $\Pi_{ab}{}^{cd}$ is the projection operator onto gamma-traceless
spinor-valued two-forms,\footnote{For the projection operator $\Pi_{ab}{}^{cd}$ in other dimensions, 
see also e.\,g.\,\cite{BKNT-M5D,BrandtCMN}.}
\begin{align}
\Pi_{ab}{}^{cd} &:=
\frac{3}{5}\d_a^{[c}\d_b^{d]}
+\frac{3}{10}\d_{[a}^{[c}\tilde{\g}_{b]}{}^{d]}
+\frac{1}{40}\ve_{ab}{}^{cdef}\tilde{\g}_{ef}
~, \eol
\g^a\Pi_{ab}{}^{cd}&= \Pi_{ab}{}^{cd}\tilde{\g}_c=0
~,~~~
\Pi_{ab}{}^{ef}\Pi_{ef}{}^{cd}=\Pi_{ab}{}^{cd}
~.
\end{align}
Note that eq. \eqref{g-traceless-T} can then be expressed as
\bea
\cX_{ab}{}^{\g k}
:= \frac{1}{2} (\g_{ab})_\b{}^\a \cX_{\a}^k{}^{\b\g}
=
\hf(\Pi_{ab}{}^{cd})^\g{}_\d{R}(Q)_{cd}{}^{\d k}
~.
\label{XRQ0}
\eea
This relates  the field $\cX_{\a }^k{}^{\b\g}$  to the $\g$-traceless part 
of the gravitino field strength.

The third constraint \eqref{compConstraintsTR_3} is solved by
\bea
\mathfrak{f}_{a}{}^{b}
&=&
-\frac{1}{8} \cR_{a}{}^{b}(\o)
+\frac{1}{80}\d_{a}^{b}\cR(\o)
-\frac{1}{60}\d_{a}^bD
+\frac{1}{8} \nabla^{c}T^-_{ca}{}^{b}
\non\\
&&
+\frac{\ri}{8}\psi_{c}{}_j\g_{a}R(Q)^{bc}{}^j
+\frac{1}{8}\psi_{[a}{}_j\g^{bc}\f_{c]}{}^j 
-\frac{1}{80}\d_{a}^{b}\psi_{c}{}_j\g^{cd}\f_{d}{}^j 
\non\\
&&
-\frac{\ri}{96}\psi_{c}{}_j\g_{a}{}^{bc}\chi^{j} 
-\frac{\ri}{48}\psi_{a}{}_j\g^b\chi^{j} 
+\frac{\ri}{80}\d_{a}^{b}\psi_{c}{}_j{\g}^{c}\chi^{j} 
~,
\label{eq:KConn}
\eea
where we have defined $\cR_{a}{}^{b}(\o):= \cR_{ac}{}^{b c}(\o)$ and $\cR(\o) := \cR_a{}^a(\o)$.
Inserting this back into $R(M)_{ab}{}^{cd}$ leads to a quite involved expression;
its bosonic part is
\bea
R(M)_{ab}{}^{cd} 
&=&
C(\omega)_{ab}{}^{cd}
-\frac{2}{15}\d_{a}^{[c}\d_{b}^{d]}D
+\cD^{e}T^-_{e[a}{}^{[c}\d_{b]}^{d]}
+ (\text{explicit gravitino terms})
~.~~~~~~
\label{eq:RM_0}
\eea
Here
$C(\omega)_{ab}{}^{cd}:=
 \cR(\o)_{ab}{}^{cd}
-\d_{[a}^{[c} \cR(\o)_{b]}{}^{d]}
+\frac{1}{10}\d_{[a}^{[c}\d_{b]}^{d]}\cR(\o)$
is the traceless part of the tensor $\cR(\omega)_{ab}{}^{cd}$.
It is important to observe that $C(\omega)_{ab}{}^{cd}$ is not quite the usual component Weyl tensor
due to the presence of the bosonic torsion in the spin connection.

We have already mentioned that 
the super-Weyl tensor superfield includes, besides
the covariant matter fields,
three other independent descendants, which turn out to be composite. 
The dimension-3/2 fermionic field $\cX_{\a}^i{}^{\b\g}$ was already analyzed, see eq. \eqref{XRQ0},
and is related to the gamma-traceless part of $R(Q)_{a b}{}^{\alpha i}$.
The dimension-2 bosonic fields
$\cY_{ab}{}^{cd} := \frac{1}{4}(\g_{ab})_\g{}^\a(\g^{cd})_\d{}^\b\cY_{\a\b}{}^{\g\d}$
and
$\cY_{ab}{}^{kl}:=\hf(\g_{ab})_\b{}^\a\cY_{\a}{}^\b{}^{kl}$
are given respectively by the traceless part of $\RM_{a b}{}^{cd}$ and the SU(2) curvature
$\RJ_{ab}{}^{i j}$,
\begin{subequations}
\begin{align}
\cY_{ab}{}^{cd} 
&=
\RM_{ab}{}^{cd} 
+ \frac{2}{15} \d_{a}^{[c} \d_{b}^{d]} D
-2\nabla_{[a} T^-_{b]}{}^{cd}
-2\nabla^e T^-_{e [a}{}^{[c} \d^{d]}_{b]}
~, \\*
\cY_{ab}{}^{kl}
 &=  \RJ_{ab}{}^{kl}~.
\end{align}
\end{subequations}

It should be mentioned that the constraint
$\RD_{ab}=\nabla^c T^-_{abc}$, derived from superspace,
actually holds identically after substituting the expression for ${\frak f}_m{}_c$ into $R(\mathbb D)$.
Although we do not provide the analysis here, the same is true for the 
$\RS$ and $\RK$ curvatures, which are determined in terms of the other curvatures 
due to Bianchi identities. We will return to this point at the end of the next subsection.

Now let us note
an interesting feature of the expressions \eqref{compConstraintsTR}.
In contrast to the constraints employed in \cite{BSVanP},
these are $S$-invariant. 
The reason is that the superspace constraints of \cite{BKNT16} were chosen so that
the superspace derivatives $\nabla_{\a}^i$ and $\nabla_a$ have the same algebra with 
$S^\a_{i}$ and $K^a$ as in the 6D $\cN=(1,0)$ superconformal algebra.
This simplicity comes with the
price that the composite connections have nontrivial dependence on the fields
$T^-_{abc}$, $\chi^i$ and $D$. This renders component expressions more involved than one might 
desire. Therefore, instead of completing the analysis here and deriving the supersymmetry transformations  {\rm etc.},
we will
make a different choice of conventional constraints
to remove the dependence on the covariant fields from the connections.


\subsection{Different choices of conventional constraints}
\label{new-frames}

Let us consider the following redefinitions of the composite connections,
\bsubeq\label{new-frame-components}
\bea
\hat{\o}_m{}^{bc}
&=&
\o_m{}^{bc}
-\hf \lambda_1 e_m{}^aT^-_{a}{}^{b c} 
~,
\label{new-frame-components-a}
\\
\hat{\phi}_m{}_\b^j 
&=&
\phi_m{}_\b^j 
- \frac{4\ri}{15}\lambda_2 (\g_m)_{\b\g} \chi^{\g j}
~,
\label{new-frame-components-b}
\\
\hat{\frak{f}}_m{}_b
&=&
\frak{f}_m{}_b
+e_m{}^a\Big(
\frac{2}{15}\lambda_3  \eta_{ab}D
-\hf \lambda_4 \de^c T^-_{a c b}
+\frac{1}{4}\lambda_5 T^-_{a}{}^{cd} T^-_{cdb} 
\Big)
~,
\label{new-frame-components-c}
\eea
\esubeq
where $\l_1$, $\l_2$, $\l_3$, $\l_4$ and $\l_5$ are real constant parameters.
A specific choice can eliminate the covariant matter fields from the connections, but let us
remain more general for the moment.
These redefinitions can be interpreted as a change of frame in superspace
by redefining the vector covariant derivative
\bea
\hat{\de}_a
&=& \nabla_a
- \hf\l_1 W_{a}{}^{b c} M_{bc}
- \ri\l_2 (\gamma_a)_{\alpha \beta} X^{\alpha j} S^{\beta}_j
- \l_3 YK_a
\non\\
&&
-\l_4 \de^b W_{a b c} K^c
-\l_5 W_{a}{}^{ e f} W_{e f c} K^c
~,
\label{new-vector-derivatives}
\eea
while keeping the spinor covariant derivative the same $ \hat{\de}_\a^i=\de_\a^i$.
A detailed analysis of the modifications to the conformal superspace geometry 
associated with $\hat{\de}_A$ 
is relegated to Appendix \ref{conf-superspace}.
Here we focus on the implications for the component structures.

First of all, 
by comparing \eqref{omega} and \eqref{new-frame-components-a}, it is clear that
the dependence on $T^-_{abc}$ in the spin connection can be eliminated by choosing
\bea
\l_1=2 ~~~~~~
\Longrightarrow
~~~~~~
\hat{R}(P)_{ab}{}^c=0
~,
\eea
eliminating the bosonic torsion.
In terms of this new spin connection, 
\bea
\mathfrak{f}_{a}{}^{b}
&=&
-\frac{1}{8} \cR_{a}{}^{b}(\hat{\o})
+\frac{1}{80}\d_{a}^{b}\cR(\hat{\o})
+\frac{1}{4} \nabla^{c}T^-_{ca}{}^{b}
-\frac{1}{8}T^-_{acd}T^-{}^{cdb}
-\frac{1}{60}\d_{a}^bD
+\cdots
~,
\eea
where $\cR_a{}^b(\hat{\o})$ and $\cR(\hat{\o})$ are defined in terms of the tensor $\cR_{ab}{}^{cd}(\hat{\o})$
as in eq. \eqref{Romega}.
Now shifting
$\frak{f}_{m c}\to\hat{\frak{f}}_{m c}$ as in 
\eqref{new-frame-components-c}
with $\l_4=-\hf$,
we can get rid of the term $\nabla^{c}T^-_{ca}{}^{b}$.
In what follows,
we will keep the choices $\l_1=2$ and $\l_4=-\frac{1}{2}$ fixed.

It is now straightforward to reapply the same component reduction procedure of the previous subsection
but in the ``hat'' frame.
Using \eqref{sup-conp_curvs_1-a}--\eqref{sup-conp_curvs_1-e} together with 
\eqref{new-frame_spinor-spinor}--\eqref{new-frame_vector-vector}
leads to\footnote{As discussed in Appendix \ref{conf-superspace},
the structure functions $\hat{f}_{\ul a b}{}^{\ul c}$ may induce nontrivial corrections
to the new curvatures. In our case, the $[S^\a_i,\hat{\de}_a]$ commutator is deformed,
but induces 
modifications only in the expressions for the $\hat{R}(S)$ and $\hat{R}(K)$ curvatures.
Up to hats, eqs. \eqref{sup-conp_curvs_1-a}--\eqref{sup-conp_curvs_1-e} are formally unchanged
and apply also to the general frame.}
\bsubeq\label{comp-curv_2-1}
\bea
\hat{R}(P)_{ab}{}^c 
&=&
0
~,
\label{comp-curv_2-1-a}
\\
\hat{R}(Q)_{ab}{}_k
&=&
\frac{1}{2}\hat{\Psi}_{ab}{}_k
+\ri \tilde{\g}_{[a}\hat{\phi}_{b]}{}_{k} 
+\frac{1}{24}T^-_{cde}\tilde{\g}^{cde}\g_{[a}\psi_{b]}{}_k
~,
\label{comp-curv_2-1-b}
\\
\hat{R}(\mathbb D)_{ab}
&=&
2e_a{}^me_b{}^n\pa_{[m}b_{n]}
 +4 \hat{\mathfrak{f}}_{[a}{}_{b]}
+\psi_{[a}{}^{ i} \hat{\phi}_{b]}{}_{ i}
+\frac{4\ri}{15} \Big(\frac{5}{8}+\l_2\Big)\psi_{[a}{}^j\g_{b]} \chi_{ j} 
~,
\label{comp-curv_2-1-c}
\\
\hat{R}(M)_{ab}{}^{cd} 
&=&
\cR_{ab}{}^{cd}(\hat{\o})
+8 \d_{[a}^{[c} \hat{\mathfrak{f}}_{b]}{}^{d]} 
+\ri\psi_{[a}{}_j\g_{b]}\hat{R}(Q)^{cd}{}^j
+2\ri\psi_{[a}{}_j\g^{[c}\hat{R}(Q)_{b]}{}^{d]}{}^{ j}
\non\\
&&
-\psi_{[a}{}_j  \g^{cd}\hat{\f}_{b]}{}^j
-\frac{8\ri}{15}\Big(\frac{5}{8}+\l_2\Big)  \d_{[a}^{[c}\psi_{b]}{}_j \g^{d]}\chi^{j} 
+\frac{\ri}{2} \psi_{[a}{}^{ j}\g^e\psi_{b]}{}_j\,T^-_{e}{}^{cd} 
~,
\label{comp-curv_2-1-d}
\\
\hat{R}(J)_{ab}{}^{kl} 
&=&
\cR_{ab}{}^{kl}(\cV)
+4\psi_{[a}{}^{ (k} \hat{\f}_{b]}{}^{l)} 
+\frac{16\ri}{15}\Big(\frac{5}{8}+\l_2\Big)\psi_{[a}{}^{ (k}\g_{b]} \chi^{ l)}
~,
\label{comp-curv_2-1-e}
\eea
\esubeq
where we have introduced the derivatives
\be \label{hatDder}
\hat{\cD}_m = \partial_m 
- \hf \hat\omega_m{}^{bc} M_{bc}
- b_m \bbD 
- \cV_m{}^{ij} J_{ij} 
\ , \quad 
\hat{\cD}_a = e_a{}^m \hat{\cD}_m \ ,
\ee
together with the gravitini field strength,
 $\hat{\Psi}_{ab}{}^\g_k=e_a{}^me_b{}^n\hat{\cD}_{[m}\psi_{n]}{}^\g_k$
as in \eqref{Psi}.
The new superspace curvature constraints now lead to
\bsubeq \label{componentConstTRv2}
\bea
\hat R(P)_{ab}{}^c &=& 0~, 
\label{componentConstTRv2-a}
\\
\g^b\hat{R}(Q)_{ab}{}_k
&=&
\frac{4}{3}\Big(\l_2+\frac{3}{8}\Big)\g_a\chi_k 
~,
\label{componentConstTRv2-b}
\\
{\hat R}(M)_{ac}{}^{bc} &=&
\frac{8}{3}\Big(\lambda_3 - \frac{1}{8}\Big) \d_{a}^b D
+ 2\Big(\lambda_5 - \frac{1}{2}\Big) T^-_{a c d} T^-{}^{bcd} 
~.
\label{componentConstTRv2-c}
\eea
\esubeq
These constraints are no longer $S$-invariant.
This is a consequence of the redefinition of the composite 
connections, which deforms their $S$-supersymmetry
transformations.
It is interesting to observe that for all values of $\l_2,\l_3$ and $\l_5$,
the new dilatation curvature is zero while the SU(2) curvature is unchanged,
\begin{align}
\hat R(\mathbb D)_{a b} = 0~, \qquad
\hat R(J)_{ab}{}^{kl} = R(J)_{ab}{}^{kl} =\frac{1}{2}(\g_{ab})_\b{}^\a\cY_{\a}{}^\b{}^{kl}
~.
\end{align}
The constraints \eqref{componentConstTRv2} are solved by\bsubeq
\bea
\hat{\o}_{a}{}_{bc}
&=&
\omega(e)_{a}{}_{bc}
-2\eta_{a[b}b_{c]}
-\frac{\ri}{4}\psi_b{}^k\g_a\psi_c{}_k
-\frac{\ri}{2}\psi_a{}^k\g_{[b}\psi_{c]}{}_k
~,
\\
\hat{\phi}_m{}^{k}
&=&
\frac{\ri}{16}\Big(
\g^{bc}\g_m-\frac{3}{5}\g_m\tilde{\g}^{bc}
\Big)
\Big(
\hat{\Psi}_{bc}{}^{k}
+\frac{1}{12}T^-_{def}\tilde{\g}^{def}\g_{[b}\psi_{c]}{}^{k}
\Big)
-\frac{4\ri}{15}\Big(\frac{3}{8}+\l_2\Big)\g_m\chi^{k} 
~,~~~~~~
\label{eq:SConn-new} 
\\
\hat{\mathfrak{f}}_{a}{}^{b}
&=&
-\frac{1}{8}\cR_{a}{}^{b}(\hat{\o})
+\frac{1}{80}\d_{a}^{b}\cR(\hat{\o})
+\frac{2}{15}\Big(\lambda_3 - \frac{1}{8}\Big)\d_{a}^b D
+\frac{1}{4} \Big(\lambda_5 - \frac{1}{2}\Big) T^-_{a e f} T^{-}{}^{bef}
+\frac{1}{8}\psi_{[a}{}_j  \g^{bc}\hat{\f}_{c]}{}^j
\non\\
&&
-\frac{1}{80}\d_{a}^{b}\psi_{c}{}_j  \g^{cd}\hat{\f}_{d}{}^j
+\frac{\ri}{16}\psi_{c}{}_j\g_{a}\hat{R}(Q)^{bc}{}^j
+\frac{\ri}{8}\psi_{c}{}_j\g^{[b}\hat{R}(Q)_{a}{}^{c]}{}^{ j}
-\frac{\ri}{10}\Big(\frac{5}{24}+\l_2\Big)\psi_{a}{}_j\g^b\chi^j 
\non\\*
&&
+\frac{\ri}{30}\Big(\frac{3}{8}+\l_2\Big)  \d_{a}^{b}\psi_{c}{}_j \g^{c}\chi^{j} 
+\frac{\ri}{16}\psi_{a}{}^{ j}\g_c\psi_{d}{}_j\,T^-{}^{bcd} 
-\frac{\ri}{160}\d_{a}^{b} \psi_{c}{}^{ j}\g_d\psi_{e}{}_j\,T^-{}^{cde} 
~.
\label{eq:KConn-new}
\eea
\esubeq
One may confirm that these are equivalent to \eqref{new-frame-components}
with  $\l_1=2$ and $\l_4=-\tfrac{1}{2}$.

Reinserting the composite $S$-supersymmetry connection 
\eqref{eq:SConn-new}
into  
 eq. \eqref{comp-curv_2-1-b}
gives
 \bea
\hat{R}(Q)_{ab}{}_k
=
\hf{\Pi}_{ab}{}^{cd}
\Big(
\hat{\Psi}_{cd}{}_k
+\frac{1}{12}T^-_{efg}\tilde{\g}^{efg}\g_{[c}\psi_{d]}{}_k
\Big)
+\frac{4}{15}\Big(\frac{3}{8}+\l_2\Big)\tilde{\g}_{ab}\chi_{k} 
~.
\label{eq:RQ-new}
\eea
In the new frame the component field
$\cX_{ab}{}^{\g k}:= \frac{1}{2} (\g_{ab})_\b{}^\a \cX_{\a}^k{}^{\b\g}$
becomes
\begin{align}
\cX_{ab}{}^k
&=\hf \hat{R}(Q)_{ab}{}^k
-\frac{2}{15}\Big(\frac{3}{8} +\l_2\Big)\tilde{\g}_{ab}\chi^k \eol
&= \hf\Pi_{ab}{}^{cd} \hat{R}(Q)_{c d}{}^k
=
\frac{1}{4}\Pi_{ab}{}^{cd}
\Big(
\hat{\Psi}_{cd}{}^k
+\frac{1}{12}T^-_{efg}\tilde{\g}^{efg}\g_{[c}\psi_{d]}{}^k
\Big)
~.\label{XRQ}
\end{align}
It is important to note that the component field $\cX_{\a }^k{}^{\b\g}$
is unchanged in going to the new frame; that is, the equations
\eqref{XRQ0} and \eqref{XRQ} are completely equivalent -- only the definition
of $\hat R(Q)$ has changed.
Other useful relations, which follow from the constraint \eqref{componentConstTRv2-b}, are\bsubeq
\bea
\tilde\g^{ab}\hat{R}(Q)_{ab}{}_k
&=&
-8\Big(\l_2+\frac{3}{8}\Big)\chi_k 
~,
\\
\g^{cde}\hat{R}(Q)_{be}{}^{k}
&=&
2\g^{[c}\hat{R}(Q)_{b}{}^{d]}{}^{k}
+\frac{4}{3}\Big(\l_2+\frac{3}{8}\Big)\g^{cd}\g_b\chi^{ k} 
~,
\\
\g^{ef}\g_{abc}\hat{R}(Q)_{ef}{}^{ k}
&=&
-24\g_{[a}\hat{R}(Q)_{bc]}{}^{ k}
+8\Big(\l_2+\frac{3}{8}\Big)\g_{abc}\chi^k
~,
\\
\g_{[a}\hat{R}(Q)_{bc]}{}^{ k}
&=&
-\frac{1}{6}\ve_{abcdef}\g^{[d}\hat{R}(Q)^{ef]}{}^{ k}~.
\eea
\esubeq
It is clear that a particularly simple choice of frame is
\bea
\l_2=-\frac{3}{8}
~~~~~
\Longrightarrow
~~~~~
\hat{R}(Q)_{ab}{}_k
=
\hf{\Pi}_{ab}{}^{cd}
\Big(
\hat{\Psi}_{cd}{}_k
+\frac{1}{12}T^-_{efg}\tilde{\g}^{efg}\g_{[c}\psi_{d]}{}_k
\Big)
~.
\eea 
In this case the $\hat{R}(Q)_{ab}{}_k$ curvature is $\gamma$-traceless.
On the other hand, it can be proven that the choice $\l_2=-\frac{5}{16}$ corresponds
to the conventional 
constraint for $\hat{R}(Q)$ that was employed in \cite{BSVanP}; we refer the reader to 
Appendix \ref{MatchConventions} for more details.

In principle, one can also reinsert the expression \eqref{eq:KConn-new} for
$\hat{\frak{f}}_{m}{}_c$ into $\hat{R}(M)_{ab}{}^{cd}$; in practice, we
are mainly interested in the bosonic terms. These are
\bea
\hat{R}(M)_{ab}{}^{cd} 
&=&
C(\hat{\omega})_{ab}{}^{cd}
+\frac{16}{15}\Big(\lambda_3 - \frac{1}{8}\Big) \d_{a}^{[c}\d_{b}^{d]} D
+ 2 \Big(\lambda_5 - \frac{1}{2}\Big)\d_{[a}^{[c} T^-_{b] e f} T^{-}{}^{d]ef}
\non\\
&&
+ (\text{explicit gravitino terms})
~,
\label{new=hatRM}
\eea
where
$C(\hat{\omega})_{ab}{}^{cd}:=
 \cR(\hat{\o})_{ab}{}^{cd}
-\d_{[a}^{[c} \cR(\hat{\o})_{b]}{}^{d]}
+\frac{1}{10}\d_{[a}^{[c}\d_{b]}^{d]}\cR(\hat{\o})$
coincides with the usual Weyl tensor when $\psi_m{}^i$ and $b_m$ vanish.
This implies that
\begin{align}
{\cY}_{a b}{}^{cd}
&= {\hat R}(M)_{ab}{}^{cd}
-\frac{16}{15}\Big(\lambda_3 - \frac{1}{8}\Big)\d_{a}^{[c}\d_{b}^{d]} D
-2\Big(\lambda_5 - \frac{1}{2}\Big) T^-_{a b e} T^-{}^{ecd} \eol*
&=  C(\hat \omega)_{a b}{}^{c d} + (\text{explicit gravitino terms}) ~.
\label{YRM}
\end{align}
A special choice of $\l_3$ and $\l_5$ makes the Lorentz curvature traceless,
\bea
\lambda_3=\frac{1}{8}
~,~~~
\lambda_5 = \frac{1}{2}
~~~~~
\Longrightarrow
~~~~~
{\hat R}(M)_{ab}{}^{cd}=\cY_{a b}{}^{cd}
=C(\hat{\omega})_{ab}{}^{cd}
+\cdots
~.~~~~~~
\eea

The particular choices of $\lambda_2$, $\lambda_3$ and $\lambda_5$ we have discussed
maximally simplify the component curvatures and connections.
We will refer to this choice as the ``traceless'' frame. It is associated with the
following conventional constraints
\bea
\hat R(P)_{ab}{}^c =0
~,~~~~~~
\g^b\hat{R}(Q)_{ab}{}_k
=0
~,~~~~~~
{\hat R}(M)_{ac}{}^{bc} 
=0
~.
\eea
Note that an alternative choice recovers the conventional constraints employed 
by Bergshoeff et al. \cite{BSVanP}, up to changes in notation described
in Appendix \ref{MatchConventions}.
We summarize these two particular choices
in Table \ref{table-coefficients}.

\begin{table*}[t]
\centering
\renewcommand{\arraystretch}{1.4}
\begin{tabular}{@{}cccc@{}} \toprule
& Traceless && Bergshoeff et al.\\ \midrule
$\lambda_1$ & $2$ && $2$ \\
$\lambda_2$ & $-\frac{3}{8}$ && $-\frac{5}{16}$ \\
$\lambda_3$ & $\frac{1}{8}$ && $\frac{5}{32}$ \\
$\lambda_4$ & $-\frac{1}{2}$ && $-\frac{1}{2}$ \\
$\lambda_5$ & $\frac{1}{2}$ && $1$\\
\bottomrule
\end{tabular}
\caption{Choice of parameters}
\label{table-coefficients}
\end{table*}

So far we have not considered the curvatures
$\hat R(S)$
and $\hat R(K)$  in detail.
In principle, one could find explicit expressions for them  in terms of
$\hat\phi_m{}_\a^i$ and $\hat{\frak f}_m{}_c$.
In practice, such expressions are not particularly useful since
these connections and their curvatures are always composite quantities.
Instead, it is more convenient to
follow the component technique of analyzing the component Bianchi identities,
which in our case is equivalent to performing the component projection of the corresponding
superspace curvatures.
Projecting  \eqref{new-frame_ALL-R-k} and using
\eqref{XRQ} gives
\bea
\hat{R}(S)_{ab}{}^k &=&
- \frac{\ri }{2} \hat{\slashed{\de}} \hat{R}(Q)_{ab}{}^{k}
- \frac{\ri}{3}  \g_{[a} \hat{\de}^c\hat{R}(Q)_{b]c}{}^{ k}
+\frac{\ri}{3}T^-_{cd[a} \g^{c}\hat{R}(Q)_{b]}{}^{d}{}^{ k}
\non\\
&&
+\frac{2\ri}{9}\Big(\l_2+\frac{3}{8}\Big)\g_c\tilde{\g}_{ab}\hat{\de}^{c} \chi^{ k} 
+\frac{2\ri}{9}\Big(\l_2+\frac{3}{8}\Big)T^-_{abc}\g^{c}\chi^{ k} 
~.
\label{RSdeRQ}
\eea
In a general frame, the special conformal curvature $\hat R(K)$ is still rather complicated,
which is apparent from its superspace expression \eqref{new-frame_ALL-R-l}.
In the traceless frame, it simplifies dramatically, and using \eqref{XRQ} and \eqref{YRM}
one obtains
\bea
\hat{R}(K)_{abc} &=&
 \frac{1}{4} \hat{\de}^d {\hat R}(M)_{abcd} 
-\frac{\ri}{24}\hat{R}(Q)^{de}{}^{ k}\gamma_{a b c}\hat{R}(Q)_{de}{}_{ k}
- \frac{\ri}{2}\hat{R}(Q)_{a}{}^d{}^{k}\gamma_{c}\hat{R}(Q)_{bd}{}_{ k}
\non\\
&&
 +\frac{\ri}{30}\chi^{k}\gamma_{c}\hat{R}(Q)_{ab}{}_k
~.
\eea

There are actually other component Bianchi identities that we have not analyzed so far.
These are differential conditions among the various superconformal component curvatures.
In superspace, they are given by the differential equations \eqref{deaBI}.
Their component forms can be derived by straightforward component projection.


\subsection{The supersymmetry transformations}

The supersymmetry transformations of the fundamental gauge connections of the Weyl
multiplet can be derived directly from the transformations of their corresponding
superspace one-forms, using either \eqref{transf-E_O} or \eqref{new-frame-sugra-transf}.
We are mainly interested in their form in the traceless
frame, but for comparison with \cite{BSVanP} we give the results for arbitrary
$\l_2$, $\l_3$ and $\l_5$, keeping $\l_1=2$ and $\l_4=-\tfrac{1}{2}$:
\bsubeq\label{eq:WeylSUSY2}
\bea
\delta e_m{}^a &= &
-\ri\xi_{k}\g^a\psi_m{}^{k}
~,
\label{eq:WeylSUSY2-a}
\\
\delta \psi_m{}_i
&= &
2\hat{\cD}_m \xi_i 
+\frac{1}{12}T^-{}^{abc}\tilde{\g}_{abc}\g_m\xi_i
+2\ri\tilde{\g}_m\eta_{ i}
~,
\label{eq:WeylSUSY2-b}
\\
\delta \cV_m{}^{kl} 
&=&
-4 \xi^{(k}\hat{\f}_m{}^{l)} 
-\frac{16\ri}{15}\Big(\l_2 +\frac{5}{8}\Big) \xi^{(k}\g_m\chi^{l)} 
+4 \psi_m{}^{(k}\eta^{l)}
~,
\label{eq:WeylSUSY2-c}
\\
\delta b_m &=&
\xi_i \hat{\f}_m{}^i 
+\frac{4\ri}{15}\Big(\l_2+ \frac{5}{8}\Big)\xi_i\g_m\chi^i
+\psi_m{}^i \eta_i
-2e_m{}^a  \l_a
~.
\label{eq:WeylSUSY2-d}
\eea
Here we have restricted to the  $Q$, $S$, and $K$ transformations.
In the same way, one can derive the transformations 
$\d\hat{\o}_m{}^{cd}$, $\d\hat{\f}_\a^i$ and $\d\frak{\hat{f}}_m{}_a$, which we omit.
For the covariant fields, the transformations are directly given
by \eqref{SUGRAtransformations-tensor-components}, which leads to
\bea
\d T^-_{abc}
&=&
-\frac{\ri}{8}\x^{ k}\g^{de}\g_{abc}\hat{R}(Q)_{de}{}_k
-\frac{\ri}{5}\Big(\l_2+\frac{25}{24}\Big)\x_k\g_{abc}\chi^{ k} ~,
\label{eq:WeylSUSY2-e}
\\
\d \chi^{ i}
&=&
\hf  D\x^{ i}
-\frac{3}{4} \hat{R}(J)_{ab}{}^{ij}\tilde{\g}^{ab}\x_j
+\frac{1}{4}\hat{\de}_a T^-_{bcd}\tilde{\g}^{bcd}\g^a\x^{i}
-\ri T^-_{abc}\tilde{\g}^{abc}\eta^i
~,
\label{eq:WeylSUSY2-f}
\\
\d D
&=&
-2 \ri \,\xi_j \hat{\slashed{\de}}\chi^{ j}
-4\chi^k\eta_k
~.
\label{eq:WeylSUSY2-g}
\eea
\esubeq
Upon making the appropriate choice of parameters $\l_i$, these transformations 
match those of Bergshoeff et al.\,\cite{BSVanP} up to differences in notation
and conventions given in Appendix \ref{MatchConventions}.



\section{The supersymmetric $C^3$ invariant} \label{C3action}

In \cite{BKNT16} it was shown that there were two invariants for conformal supergravity, containing 
$C^3$ and $C \Box C$ terms at the component level, where $C$ schematically represents 
the Weyl tensor. Moreover, the construction of each of these invariants employed the use of two different action 
principles formulated in terms of superforms.
Each of these action principles was built out of a constrained primary superfield, 
which, when further chosen to be composed of the super-Weyl tensor and its covariant derivatives, yields the two conformal supergravity 
invariants.

In this section, we will focus on one of these action principles, which we will call
the $A$ action principle. 
It was first constructed using superforms in \cite{ALR14}; see \cite{BKNT16} for
its construction in conformal superspace.
Here we will present its component form as
a general density formula built upon a certain composite multiplet. Afterwards, we
will give the specific choice for the multiplet that generates 
the $C^3$ invariant.

\subsection{The $A$ action principle}

The $A$ action principle is based on 
a primary dimension $9/2$ superfield $A_\a{}^{ijk}=A_\a{}^{(ijk)}$ obeying 
the reality condition $\overline{A_\a{}^{ijk}} = A_{\a\, ijk}$ and
satisfying the differential constraint\footnote{The names ``$A$ action principle'' and
``$B$ action principle'' (discussed later) have no correlation with
type A or type B anomalies.}
\be 
\nabla_{(\a}^{(i} A_{\b)}{}^{jkl)} = 0 \ . \label{Aconst}
\ee
While a superfield obeying this constraint cannot itself be directly identified as
a Lagrangian for some subspace of the full $(1,0)$ superspace, it does possess
an important geometric significance. Within the context of the 6D abelian tensor
hierarchy, the multiplet generated by $A_\a{}^{ijk}$ provides the minimal
version of a five-form gauge multiplet, whose six-form field strength obstructs
the closure of the linear multiplet's five-form field strength \cite{ALR14}.
The same observation explains why it also appears naturally as the anomaly current
multiplet for 6D gauge theories (see e.g. \cite{KNS15} for a recent discussion),
as the anomalous current is encoded in a linear multiplet.

In accordance with the superform approach to the construction of supersymmetric invariants
\cite{Ectoplasm, GGKS,Castellani} and following very similar cases in four dimensions
\cite{Girardi:1990bz, Hasler}, it was proposed in \cite{ALR14} to use a closed six-form
$J$ built out of the superfield $A_\a{}^{ijk}$ as an action principle.
The original action principle was given in the context of SU(2) superspace \cite{LT-M12}
and later extended to conformal superspace \cite{BKNT16}. 
In this latter form, it can be straightforwardly reduced to components and written
\bsubeq \label{AactionPrinc}
\be \label{compInvActPri}
I = \int \rd^6 x \, e \, \cL \ , \quad e = \det (e_m{}^a) \ ,
\ee
where $\cL$ is a scalar Lagrangian constructed as $\cL=\frac{1}{6!} \eps^{mnpqrs} J_{mnpqrs}|$,
that is as the Hodge dual of the six-form $J$ restricted to spacetime. It is explicitly given by
\begin{align}\label{A-Lagrangian}
\cL &=
	F
	- \frac{\ri}{4} (\psi_{a i} \Omega'{}^{a i})
	- \frac{\ri}{144} (\psi_{d i} \gamma^{de} \gamma^{abc} \psi_{e j})\, S_{abc}^+{}^{ij}
	\eol & \quad
	- \frac{\ri}{12} (\psi_{a i} \gamma^{abc} \psi_{b j})\, E_c{}^{i j}
	+ \frac{1}{16} (\psi_{a i} \gamma^{abc} \psi_{b j})\, (\psi_{ck} A^{ijk})
	~.
\end{align}
\esubeq
Here we have introduced the following component fields of $A_\a{}^{ijk}$:\footnote{Here we introduce 
$\Omega'_{a\a}{}^{i}$ which differs from $\Omega_{a \a}{}^{i}$ used in \cite{BKNT16}.}
\bsubeq \label{AsuperfieldDescend}
\begin{align}
S_{abc}^+{}^{ij} &:= \frac{3}{32} (\tgamma_{abc})^{\a\b} \nabla_{\a k} A_{\b}{}^{ijk}\loco \ , \qquad
E_{a}{}^{ij} := \frac{3}{16} (\tgamma_a)^{\a\b} \nabla_{\a k} A_{\b}{}^{ijk}\loco \ , 
\\
\Omega'_{a\a}{}^{i} &:=
	\frac{\ri}{32} (\tgamma_a)^{\b\g} (
	\nabla_{\b j} \nabla_{\g k} A_\a{}^{ijk}\loco
	+ \nabla_{\a j} \nabla_{\b k} A_\g{}^{ijk}\loco
	)~, \\
F &:=
\frac{\ri}{2^4 4!}\eps^{\a\b\g\d} \nabla_{\a i} \nabla_{\b j} \nabla_{\g k} A_\d{}^{ijk}\loco \ .
\label{4.15e}
\end{align}
\esubeq
When it is clear from context, we will use the same symbol $A_\a{}^{ijk}$
for both the superfield and its lowest component.

The invariance of this general density formula under special conformal transformations is obvious
once one verifies that the fields
$S_{abc}^+{}^{ij}$, $E_{a}{}^{ij}$, $\Omega'_{a \a}{}^i$ and $F$ are each annihilated by $K^a$.
Invariance under $S$-supersymmetry can be proven by using the $S$-transformation of the gravitino
\eqref{eq:WeylSUSY2-b} together with $S$-transformations of the components of $A_\a{}^{ijk}$:
\begin{subequations}
\begin{align}
\d_S A_\a{}^{ijk} &= 0 \ , \quad
\d_S S_{abc}^+{}^{ij} = 3\,\eta_{k} \tilde{\gamma}_{abc} A^{ijk} \ , \quad
\d_S E_{a}{}^{ij} = \frac{9}{2} \,\eta_{k} \tilde\gamma_{a} A^{ijk} \ , \\
\d_S \Omega'_{a\a}{}^i &= 
	- \frac{8\ri}{3} (\g_{a b}\eta_{j})_\a E^b{}^{i j}
	+  \ri (\gamma^{b c}\eta_{j})_\a S^+_{abc}{}^{ij}
, \\
\d_S F &= \eta_k \tilde\gamma^a \Omega'_a{}^k~.
\end{align}
\end{subequations}
These transformation laws follow from the definitions of the component fields \eqref{AsuperfieldDescend}.

The invariance of the $A$ action principle under $Q$-supersymmetry follows from its description as a closed superform. 
Demonstrating closure of the superform is equivalent (although generally more efficient)
than demonstrating invariance under $Q$-supersymmetry directly from the spacetime Lagrangian.
Let us briefly sketch how the latter procedure goes.
The field $A_\alpha{}^{ijk}$ transforms under supersymmetry as
\begin{align}
\delta_Q A_\a{}^{ijk}
	= -(\gamma^{a} \xi_l)_\a \, A_{a}{}^{i j k l}
	- \frac{1}{6} (\gamma^{abc} \xi^{(i})_\a \,S^+_{abc}{}^{j k)}
	+ (\gamma^{a} \xi^{(i})_\a E_{a}{}^{j k)}~.
\end{align}
The constraint \eqref{Aconst} is encoded in the absence of a term involving
$S_{abc}^+{}^{ijkl}:=(\tilde{\g}_{abc})^{\a\b}\de_\a^{(i}A_{\b}{}^{jkl)}|$
on the right-hand side. 
Now the term $A_a{}^{ijkl}$ does not appear in the action and so its $\d_Q$ transformation 
does not need to be computed directly. This can be verified by computing $\delta_Q$ of the terms involving
$A_\a{}^{ijk}$ in the action and integrating by parts any expressions 
with $\cD_m \xi^\a_i$ to give a covariant
result. The term involving $A_b{}^{ijkl}$ turns out to be multiplied by a $\psi^3 \xi$ term,
and the various contractions of Lorentz and SU(2) indices project this term out.
The advantage of the superform method is that this sequence of steps is encoded automatically.

It is clear that the constraint \eqref{Aconst} is rather weak and describes a multiplet 
that is rather long, only a few components of which have been mentioned above.
Nevertheless, the supersymmetry transformations of all components are dictated by closure of
the supersymmetry algebra, making use of the constraint \eqref{Aconst}. For example, 
\begin{align}
\delta_Q A_a{}^{ijkl}
	&= \frac{1}{2} \xi_p \gamma_a \Xi^{ijklp}
	- \xi^{(i} \Xi_a{}^{jkl)}~,
\end{align}
for some spinor $\Xi_\a{}^{ijklp}$ and some vector-spinor $\Xi_{a \alpha}{}^{ijk}$. 
While the multiplet is rather long, only a small number of the component fields appear in the 
action, so that in practice one only needs some of the supersymmetry transformations.
These are given below in the traceless frame:
\bsubeq \label{ASUSYTranshigherW}
\begin{align}
\delta_Q E_a{}^{ij}
	&=
	\frac{\ri}{2} \xi_k \gamma^b \tilde\gamma_a \hnabla_b A^{ijk}
	- \frac{\ri}{12} \xi_k \gamma_a \tilde\gamma^{bcd} A^{ijk} T^{-}_{bcd}
	- \frac{1}{4} \xi_k \gamma_{ab} \Xi^b{}^{ijk}
	\eol & \quad
	- 4 \ri \xi^{(i} \Omega'_a{}^{j)}
	- \ri \xi^{(i} \gamma_a \tilde\gamma^b \Omega'_{b}{}^{j)}
	~, 
	\\
\delta_Q S^+_{abc}{}^{i j}
	&= \frac{\ri}{4} \xi_k \gamma^d\tilde \gamma_{abc} \hnabla_d A^{ijk}
	- \frac{1}{8} \xi_k \gamma^d \tilde\gamma_{abc} \Xi_d{}^{ijk}
	+ \ri \xi^{(i} \gamma^d \tilde\gamma_{abc} \Omega_d'{}^{j)}
	~, \\
\delta_Q \Omega'_{a\a}{}^i
	&= (\gamma_a \xi^i)_\a F
	+ \frac{1}{3} (\gamma_{abc} \xi_j)_\a \hnabla^b E^c{}^{ij}
	+ \frac{1}{72} (\gamma_{a b} \gamma^{cde} \xi_j - \gamma^{cde} \tilde\gamma_{ab} \xi_j)_\alpha
		\hnabla^{b}{S^+_{c d e}{}^{i j}}
	\eol & \quad
	+ (\gamma^b \xi_j)_\a C_{b a}{}^{(ij)}
	+ (\gamma_{abc} \xi^i)_\a C^{[bc]}
	- \frac{3 \ri}{5} A_{\alpha}{}^{i j k} (\xi_j \gamma_a \c_k)
	+ \frac{3 \ri}{5} (\xi_j A^{i j k}) (\gamma_{a} \c_k)_\alpha
	\eol & \quad
	+ \frac{1}{288} S^+_{b c d}{}^{i j} T^{-}_{e f g}
		\,(\gamma_a \tilde\gamma^{efg} \gamma^{bcd}\xi_j - \g^{bcd} \tilde\g^{efg} \g_a\xi_j)_{\a}
	\eol & \quad
	+ \frac{1}{36} E_{b}{}^{i j} T^{-}_{c d e}
		\,(\gamma_a \tilde\gamma^{cde} \gamma^b\xi_j + \g^b \tilde\g^{cde} \g_a\xi_j)_{\a} 
	~, \\
\delta_Q F
	&= - \ri \, \xi_i \hat{\nabla}_a \Omega'^a{}^i 
	- \frac{\ri}{24} T^{- abc} \xi_k \g^d \tilde\g_{abc} \Omega'_d{}^k
	- \frac{\ri}{72} \xi_i \g^{ab} \g^{cde} \hat{R}(Q)_{ab}{}_j S^+_{cde}{}^{ij}
	\ .
\end{align}
\esubeq
Using the above supersymmetry transformations one can check that the invariant \eqref{AactionPrinc} is 
indeed supersymmetric. Note that the transformation of the vector-spinor $\Omega'_{a\a}{}^i$ involves
two additional fields that do not appear in the action: an antisymmetric tensor $C_{[ab]}$ and a
rank-two tensor $C_{ab}{}^{(ij)}$ symmetric in its SU(2) indices.

We underline here that these supersymmetry transformations are sufficient to prove invariance of the
action. They arise as a consequence of closure of the algebra provided the constraint \eqref{Aconst}
holds; therefore, there is no need to analyze the transformation properties of the entire multiplet.


\subsection{The supersymmetric $C^3$ invariant in components}

Having elaborated the component structure of the $A$ action principle we now provide an 
immediate application: the derivation of the supersymmetric $C^3$ invariant at the component level. 
The suitable superfield $A_\a{}^{ijk}$ was constructed in \cite{BKNT16}.
At the component level, it corresponds to\footnote{Relative to \cite{BKNT16}, 
we have renormalized the choice for $A_\a{}^{ijk}$ by a factor of $-64$.}
\begin{align}
A_{\alpha}{}^{i j k}
&= -\frac{64\ri}{15} (\gamma_{a } \c^{ (i})_\a \Big(
\frac{4}{45} \c^{j} \gamma^a \c^{ k)}
+ \hat R(Q)_{bc}{}^{j} \gamma^a \hat R(Q)^{bc\,k)}
- 3 \ri\, \hat R(J)_{b c}{}^{j k)} T^{- a b c}
\Big)
\eol & \quad
- 128 \, (\gamma_{a } \hat R(Q)_{b c}{}^{ (i})_\a \Big(
	\hat{R}(J)_{d}{}^{c\, jk)}\, T^{- a b d} 
	+ \frac{2\ri}{3} \hat R(Q)^{a d}{}^{j} \g^c \hat R(Q)_{d}{}^{b\, k)}
\Big)~.
\end{align}
One may verify that it is $S$-invariant and transforms under supersymmetry so that the
constraint $\eqref{Aconst}$ is satisfied.
The remaining composites
 $S^+_{abc}{}^{ij}$,  $E_{a}{}^{ij}$, $\Omega'_{a \a}{}^{i}$ and $F$ appearing in \eqref{A-Lagrangian} 
can be computed directly from the supersymmetry transformations \eqref{ASUSYTranshigherW} or equivalently
from the superspace definitions \eqref{AsuperfieldDescend}.
This is a straightforward task but the explicit calculation becomes technically 
quite involved, so we made use of the computer algebra program {\it Cadabra} \cite{Cadabra-1,Cadabra-2}.
In a separate supplementary file, we give the explicit expressions for each of these components.
The bosonic part of the Lagrangian is given by
\bea \label{eq:BosonicC3}
\cL
&=&
\frac{8}{3}\,{C}_{a bc d} {C}^{a b e f} {C}^{c d}{}_{e f} 
  - \frac{16}{3}\, {C}_{a b c d} {C}^{aecf} {C}^{b}{}_{ e}{}^{d}{}_{f} 
-2\, {C}_{a b c d} {\cR}^{a b}\,^{i j} {\cR}^{c d}\,_{ij}  
+ 4\, { \cR}_{a b}\,^{i j} { \cR}^{a c}\,_{i}{}^{k} { \cR}^{b}{}_{c}\,_{j k} 
  \non\\
    &&
     - \frac{32}{225} D^3
  - \frac{4}{15} \, D {C}_{a b c d} { C}^{a b c d} 
  + \frac{8}{5} D { \cR}_{a b}\,^{i j} {\cR}^{a b}\,_{ij}
    + \frac{128}{15} \, {T}^{-}_{a b c} {T}^{ - a d e} D {C}^{b}{}_{d}{}^{c}{}_{e} 
\non\\
&&
     + \frac{64}{15} \, {T}^{-}_{a b c} D {\hat{\de}}^{a}{{\hat{\de}}_{d}{{T}^{- b c d}}}
- \frac{4}{5} \, D {\hat{\cD}}^{a}{{T}^{-}_{a b c}}\,  {\hat{\cD}}_{d}{{T}^{- b c d}}
+ \frac{4}{15} \, D {\hat{\cD}}_{a}{{T}^{-}_{b c d}}\,  {\hat{\cD}}^{a}{{T}^{- b c d}}
\non\\
&&
- \frac{4}{3} \, D {\hat{\cD}}_{a}{{T}^{-}_{b c d}}\,  {\hat{\cD}}^{b}{{T}^{- a c d}}
  - \frac{16}{5} \, {T}^{-}_{a b c} {T}^{- a b d} {T}^{- c e f} {T}^-_{ d e f} D 
\non\\
&&
- \frac{32}{3}\, {T}^{-}_{a b c} { C}^{a b d e} {\hat{\cD}}^{f}{{ C}^{c}{}_{d e f}}
   + \frac{16}{3}\, { C}_{a b c d} { C}^{a b e f} {\hat{\cD}}^{c}{{T}^{- d}{}_{e f}}
\non\\
&&
    - 16\, {T}^{-}_{a b c} {\hat{\cD}}_{d}{{T}^{- a b e}}\,  {\hat{\de}}_{e}{{\hat{\de}}_{f}{{T}^{- cd f}}}   
   - 16 \, {T}^{-}_{a b c} {\hat{\cD}}_{d}{{T}^{- a d e}}\,  {\hat{\de}}_{e}{{\hat{\de}}_{f}{{T}^{- b c f}}}
   \non\\
   &&
  - 48 \, {T}^{-}_{a b c} {\hat{\cD}}_{d}{{T}^{- a d e}}\,  {\hat{\de}}^{b}{{\hat{\de}}^{f}{{T}^{- c}{}_{ e f}}}
    + 16 \, {\hat{\cD}}^{e}{{T}^{-}_{e a b}}\,  {\hat{\cD}}^{f}{{T}^{-}_{f c d}}\,  {\hat{\cD}}^{a}{{T}^{- b c d}}\,  
    \non\\
    &&
    - 40\, {T}^{-}_{a b e} {T}^{- c d e} {\hat{\cD}}_{f}{{T}^{- f a b}}\,  {\hat{\cD}}^{g}{{T}^{-}_{g c d}}
 + 16 \, {T}^{-}_{a b c} { C}^{a b d e} {\hat{\de}}^{c}{{\hat{\de}}^{f}{{T}^{-}_{d e f}} }
\non\\
&&
 - 16\, {T}^{-}_{a b c} {C}^{a b d e} {\hat{\de}}_{d}{{\hat{\de}}^{f}{{T}^{- c}{}_{e f}}}
 - 4\, {C}_{a b c d} {\hat{\cD}}_{e}{{T}^{- a b e}} {\hat{\cD}}_{f}{{T}^{- c d f}}
     \non\\
    &&
 +8\, { C}_{a b c d} {\hat{\cD}}_{e}{{T}^{- a b f}}\,  {\hat{\cD}}_{f}{{T}^{- c d e}}
- \frac{64}{3}\, {T}^{- f b}{}_{ d} \,{\hat{\cD}}^{e}{{C}_{e a b c}}\,  {\hat{\cD}}_{f}{{T}^{- a c d}}
    \non\\
    &&
+32\, {T}^{- a b}{}_{ d} \,  {\hat{\cD}}^{e}{{ C}_{e a b c}}\,  {\hat{\cD}}_{f}{{T}^{- f c d}}
    - 32\,  {T}^{-}_{f g c} {T}^{- f g d} {\hat{\cD}}^{c}{{T}^{-}_{d a b}}\, {\hat{\cD}}_{e}{{T}^{- e a b}}
   \non\\
    &&
- 8\, {\hat{\cD}}_{e}{{T}^{-}_{b a d}}\,  {\hat{\cD}}^{e}{{T}^{- c a d}}\,  {T}^{- f g b} {T}^{-}_{f g c} 
- 8\, {T}^{-}_{a b c} {T}^{- a b d} { C}^{c e f g} {\hat{\cD}}_{e}{{T}^{-}_{d f g}}
   \non\\
    &&
- \frac{8}{3}\, {T}^{-}_{a b c} { \cR}^{a b}\,^{i j} {\hat{\cD}}_{d}{{ \cR}^{c d}\,_{ij}}
+ \frac{28}{3}\, {T}^{-}_{a b c} { \cR}^{a d}\,^{i j} {\hat{\cD}}_{d}{{ \cR}^{b c}\,_{ij}}
- \frac{32}{9}\, {\cR}_{a b}\,^{i j} {\cR}_{c d}\,_{ij} {\hat{\cD}}^{a}{{T}^{- b c d}}\, 
   \non\\
    &&
+ 4 \,  {T}^{-}_{e f b} {T}^{- e f c} {T}^{-}_{g h a} {T}^{- g h}{}_{ c}\, {\hat{\cD}}_{d}{{T}^{- d a b}}
-8\,  {T}^{-}_{a b c} {T}^{- a b d} {T}^{-}_{e f g} {T}^{- e f h} { C}^{c g}{}_{ d h} 
    \non\\
    &&
+12\, {T}^{-}_{a b c} {T}^{- a d e} {\cR}^{b c}\,^{i j} {\cR}_{d e}\,_{ij} 
+ {\rm \ fermion \ terms}
~.
\eea
Note that in the previous result there 
is some hidden dependence on the special conformal connection $\hat{\mathfrak{f}}_{ab}$, which 
can be made explicit by using
\bea
\hat{\de}_a\hat{\de}_b T^{-}_{cde}
&=&
\hat{\cD}_a\hat{\cD}_b T^{-}_{cde}
-\hat{\mathfrak{f}}_{af}{[}K^f,\hat{\de}_b{]} T^{-}_{cde}
+ {\rm \ fermion \ terms}
\non\\
&=&
\hat{\cD}_a\hat{\cD}_b T^{-}_{cde}
-2\hat{\mathfrak{f}}_{ab} T^{-}_{cde}
-6\hat{\mathfrak{f}}_{a[c} T^{-}_{de]b}
+6\hat{\mathfrak{f}}_{a}{}^{f} T^{-}_{f[cd}\eta_{e]b}
+ \!\!{\rm \ fermion \ terms}
~,
\eea
and the results in section \ref{SCTC}.

We conclude this section by mentioning one useful consistency check of the previous result. 
By computing the variation of the action with respect to the $D$ field one should obtain the component 
projection of the supercurrent for the supersymmetric $C^3$ invariant. Upon doing this, one finds 
a supercurrent in the two parameter family described in \cite{BKNT16} with $c_1 = 0$.


\section{The supersymmetric $C\Box C$ invariant} \label{CboxCaction}

Another conformal supergravity invariant was
constructed in \cite{BKNT16}, corresponding to the supersymmetrization of $C_{abcd} \Box C^{abcd}$.
It makes use of a different superform based on another constrained primary superfield.
Curiously, this construction, which we denote the $B$ action principle, is not based on an
$S$-invariant superform, which leads to explicit dependence on the $S$-supersymmetry
and special conformal connections in the component action. As discussed in \cite{BKNT16},
another choice for the basic primary superfield leads to the supersymmetric $F \Box F$
invariant minimally coupled to conformal supergravity. Both results are given below.


\subsection{The $B$ action principle}

Let us begin by describing the $B$ action principle in components. Its building block is
a primary dimension 3 superfield  $B_{a}{}^{ij} = B_{a}{}^{(ij)}$, which is
pseudoreal, $\overline{B_a{}^{ij}} = B_a{}_{ij}$, and satisfies the
differential constraint
\begin{align}\label{eq:Bconstraint}
\nabla_\a^{(i} B^{\b\g jk)} = - \frac{2}{3} \d_\a^{[\b} \nabla_{\d}^{(i} B^{\g]\d jk)} \ .
\end{align}
Its corresponding superform action principle was given in \cite{BKNT16}. Its reduction
to components, following \eqref{compInvActPri}, is straightforward and leads to
\begin{align}\label{B-Lagrangian}
\cL &= 
	F'
	+ (\psi_{m i} \gamma^m \Omega'^{i})
	+ \frac{\ri}{12} (\psi_{a i} \gamma^{a b c} \psi_{b j}) E_{c}{}^{i j}
	+ \frac{1}{16} (\psi_{m i} \gamma^{mnp} \psi_{n j})
		(\psi_{p k} \rho^{i j k})
\eol & \qquad
	- 16 \,{\mathfrak{f}}^{a b} C_{a b} 
	- 8 \ri\, (\psi_{m i} \gamma^{m n} \Lambda_{a}^{i}) \, {\mathfrak f}_{n}{}^a
	+ 8\ri \,(\psi_{m i} \gamma^{mnp} \psi_{n j}) \, {\mathfrak f}_{p}{}^{a} B_{a}{}^{i j}
\eol & \qquad
	+ 2\, (\phi_{a j} \rho^{a j})
	+ \frac{1}{3} (\psi_{m j} \gamma^{m n} \gamma^{a} \tgamma^{b} \phi_{n k}) 
		C_{ab}{}^{j k}
	+ \frac{3\ri}{2}  (\psi_{m i} \gamma^{mnp} \psi_{n j})
		(\phi_{p k} \Lambda^{i j k})~.
\end{align}
The various component fields are defined by successive application of superspace
spinor derivatives,
\begin{subequations}
\begin{alignat}{2}
\L^{\a i j k} &:= \frac{\ri}{3} \nabla_{\b}^{(i} B^{\b\a j k)}\loco~, &\,
\L_{\a b}{}^i &:= \frac{2\ri}{3} \nabla_{\a j} B_b{}^{ij}\loco~, \\
C^{ijkl} &:= \frac{1}{4} \nabla_\a^{(i} \L^{\a j k l)}\loco~, &\,
C_{a b}{}^{i j} &:= 
\frac{1}{8} \Big(3\, (\gamma_{ab})_\beta{}^\alpha + \eta_{a b} \delta_\beta{}^\alpha\Big)
	\nabla_{\a k} \L^{\b i j k}\loco~, \\
C_{a b} &:= \frac{1}{8} (\tilde\gamma_{a})^{\alpha \beta} \nabla_{\a k} \L_{\b b}{}^k\loco~, \\
\rho_{\a}{}^{ijk} &:= -\frac{4\ri}{5} \nabla_{\a l} C^{i j k l}\loco~, &\,
\rho_{a}{}^{\gamma i} &:= -\frac{\ri}{6} (\tgamma_{a})^{\a\b} \nabla_{\a j} C_{\beta}{}^{\gamma i j}\loco~,\\
E_{a}{}^{i j} &:= \frac{3}{16} (\tilde\gamma_a)^{\a\b} \nabla_{\alpha k} \rho_{\beta}{}^{ijk}\loco~,\\
\Omega^{\alpha i} &:= 
	\frac{\ri}{18} \nabla_{\beta j} E^{\b\a \,i j}\loco~, &\,
F &= \frac{1}{8} \nabla_{\a j} \Omega^{\a j}\loco~,
\end{alignat}
\end{subequations}
where we abuse notation somewhat by denoting e.g. $\L^{\alpha\, ijk}$ both as a superfield
and as its lowest component. 
In the Lagrangian \eqref{B-Lagrangian}, we have written $F'$ and $\Omega'^{\alpha i}$
for the terms involving zero and one gravitini, respectively. These are given in
terms of $F$ and $\Omega^{\alpha i}$ as
\bsubeq\begin{align}
\Omega'^{\alpha i} &= 
	\Omega^{\alpha i}
	+ \frac{32}{3} B_{a}{}^{i j} \,\nabla_b R(Q)^{a b\, \alpha}{}_j
	+ \frac{3}{4} \Lambda^{\beta i j k}\, (\gamma^{a b})_\b{}^\a  \, R(J)_{a b\, j k}~, \\
F' &=
	F
	- \frac{16\ri}{3} \Lambda_{\alpha b}{}^{i} \nabla_c R(Q)^{b c\, \alpha}{}_i
	+ 4 B_{a\, i j} \nabla_b R(J)^{a b\, i j}
	+ \frac{2}{3} C_{a b\, i j} R(J)^{a b\, i j}~,
\end{align}
\esubeq
where we have written the vector derivative and gravitino curvature in the original
frame to maintain contact with \cite{BKNT16}.

As already alluded to, the connections $\mathfrak{f}_m{}^a$ and $\phi_{m}{}_\alpha{}^i$
appear explicitly within the Lagrangian \eqref{B-Lagrangian}, so invariance under both
special conformal and $S$-supersymmetry transformations holds only up to total derivatives.
It does not seem possible to eliminate this explicit dependence in the density formula,
although we will find that at least for the bosonic action, we can eliminate most of the
explicit $\mathfrak{f}_m{}^a$ connections by judiciously adding a certain total derivative.

As we are primarily working with the traceless connections when in components, it is
useful to give the density formula in that case as well. Its form is slightly modified,
\begin{align}\label{eq:B-LagrangianTraceless}
\cL &= 
	\widehat F
	+ (\psi_{a i} \widehat \Omega^{a \,i})
	+ \frac{\ri}{12} (\psi_{a i} \gamma^{a b c} \psi_{b j}) \widehat E_{c}{}^{i j}
	+ \frac{1}{16} (\psi_{m i} \gamma^{mnp} \psi_{n j})
		(\psi_{p k} \rho^{i j k})
\eol & \qquad
	- 16 \,\hat {\mathfrak{f}}^{a b} C_{a b} 
	- 8 \ri\, (\psi_{m i} \gamma^{m n} \Lambda_{a}^{i}) \, \hat {\mathfrak f}_{n}{}^a
	+ 8\ri \,(\psi_{m i} \gamma^{mnp} \psi_{n j}) \, \hat {\mathfrak f}_{p}{}^{a} B_{a}{}^{i j}
\eol & \qquad
	+ 2\, (\hat \phi_{a j} \rho^{a j})
	+ \frac{1}{3} (\psi_{m j} \gamma^{m n} \gamma^{a} \tgamma^{b} \hat \phi_{n k}) 
		C_{ab}{}^{j k}
	+ \frac{3\ri}{2}  (\psi_{m i} \gamma^{mnp} \psi_{n j})
		(\hat \phi_{p k} \Lambda^{i j k})~,
\end{align}
with the hatted fields defined by
\begin{subequations}
\begin{align}
\widehat F &= F'
	+ 2 \, C^{a b} \Big(
	\frac{2}{15} \eta_{a b} D
	-  2 \nabla^c T^-_{c a b}
	+  T^-_a{}^{c d} T^-_{b c d}
	\Big)
	+ \frac{\ri}{5} (\chi_j \gamma_a  \rho^{a j})
	~,\\
\widehat \Omega_{a}{}^i &=
	\gamma_a \Omega'^i
	+ \ri \,\gamma_{a b} \Lambda_{c}^{i} \Big(
	\frac{2}{15} \eta^{b c} D
	- 2 \nabla_d T^-{}^{d b c}
	+ T^-{}^{b d e} T^-_{de}{}^c
	\Big)
	- \frac{\ri}{30} \gamma_{a b} \gamma_c \tgamma_d \gamma^b \chi_j \,C^{c d \,i j}~, \\
\widehat E_a{}^{ij} &=
	E_a{}^{ij}
	- 12\, B^b{}^{ij} \Big(
	\frac{2}{15} \eta_{a b} D
	- 2 \nabla^c T^-_{c a b}
	+ T^-_{a}{}^{c d} T^-_{b c d}
	\Big)
	+ \frac{9\ri}{5} (\chi_k \gamma_a \L^{ijk})~.
\end{align}
\end{subequations}
The explicit supersymmetry relations between the various component fields in this
action principle are quite complicated and not very enlightening,
so we do not give them here.

For the actions that we will be considering, we will mainly be interested in
their bosonic Lagrangians, and these amount to
\begin{align}
\cL &= \widehat F - 16 \, \hat {\mathfrak f}^{a b} C_{a b}~.
\end{align}
It turns out that at least for the bosonic parts, it is possible to
eliminate the explicit appearance of the $K$-connection within the bosonic
action and render a $K$-invariant result. Adding a total derivative
\begin{align}
\cL_{\rm add} &= - \hnabla_a \Big(
	4 \hnabla_b C^{(a b)} - \hnabla^a C_c{}^c
	\Big)
	+ 16 \,\hat{\mathfrak f}_{a b} C^{(a b)}
\end{align}
has the effect of removing the symmetric part of $C_{a b}$ from the $K$-connection term, giving
\begin{align}\label{eq:FinalBLagBosonic}
\cL' &=\cL + \cL_{\rm add} =  \widehat F
	- \hnabla_a \Big(
	4 \hnabla_b C^{(a b)} - \hnabla^a C_c{}^c \Big)
	- 16 \, \hat{\mathfrak f}_{[a b]} C^{[a b]}~.
\end{align}
Henceforth, we will drop the prime on the Lagrangian.

The advantage of this modification
is that the antisymmetric (and bosonic) part of $\hat{\mathfrak f}_{ab}$
is just the curl of the dilatation gauge field $b_m$. In the circumstances we will be interested in,
$C^{[ab]}$ will be divergenceless and so this term can be dropped. Equivalently, the
$K$-transformation of the other terms in \eqref{eq:FinalBLagBosonic}
is proportional to $\hnabla^a C_{[ab]}$,
and so becomes $K$-invariant when this quantity vanishes; therefore, the explicit $K$-connection
must also vanish up to a total derivative. This is the case when
$B_a{}^{ij}$ describes a four-form field strength multiplet (with $C_{[ab]}$ the Hodge dual
of the supercovariant field strength) as studied e.g. in \cite{ALR14}. There, in addition to the
constraint \eqref{eq:Bconstraint},
the authors found that 
\begin{align}\label{eq:FourFormConstraint}
C_{a}{}^{a\, i j} = -4 \hnabla_{a} B^{a\, i j}~.
\end{align}
Both the models we construct in this paper using the $B$-action principle will actually
satisfy this extra condition.
It is possible that for this restricted class, the $B$-action principle might be significantly
simplified by eliminating a number of the $K$ and $S$ connections, but it seems impossible
to remove all of them simultaneously, so we have not attempted to massage its form any further.


\subsection{The supersymmetric $C\Box C$ invariant in components}

Using the $B$ action principle, one can directly construct the $C \Box C$ invariant.
The appropriate superfield $B_a{}^{ij}$ should be a composite built out of
the standard Weyl multiplet fields and was given in superspace (up to a change
in normalization) as \cite{BKNT16} 
\begin{align}\label{eq:SuperBforCBoxC}
B^{\a\b\, i j} &=
- 4 W^{\gamma [\alpha} Y_{\gamma}{}^{\beta] i j} 
- 32 \ri\, X_{\gamma}{}^{\alpha \delta (i} X_{\delta}{}^{\beta \gamma j)}
+ 10 \ri\, X^{\alpha (i} X^{\beta j)}~.
\end{align}
Its lowest component corresponds to
\begin{align}\label{eq:BforCBoxC}
B_a{}^{ij} &= T^-_{a b c}\, R(J)^{b c\, i j}
	+ \ri\, (\hat R(Q)_{b c}{}^i \gamma_a \hat R(Q)^{b c}{}^j)
	+ \frac{2\ri}{45} (\chi^i \gamma_a \chi^j)~.
\end{align}
The Lagrangian can then be built out of successive applications of superspace
spinor derivatives or, equivalently, supersymmetry transformations to give
the various composite objects. These are quite intricate and we made use of the
computer algebra program {\it Cadabra} to help in their construction. Their
full expressions are given in a supplementary file.
Right now, an important feature is that \eqref{eq:FourFormConstraint} holds,
and so this describes a four-form field strength multiplet. Indeed,
the bosonic part of $C_{[ab]}$ can be written
\begin{align}
C_{[ab]} &= \frac{1}{8} \veps_{a b}{}^{c d e f} \Big(
\hat R(M)_{c d}{}^{gh} \hat R(M)_{e f gh}
- \hat R(J)_{c d}{}^{i j} \hat R(J)_{e f \,i j}
\Big)
+ \hnabla^c B_{abc}~, \\
B_{abc} &=
- \frac{4}{15} D \,T^-_{a b c}
- \frac{3}{2} T^-_{[a}{}^{de} \hat R(M)_{b c] d e} 
+ 3 \,T^-_{d e [a} \hat R(M)_{b}{}^{de}{}_{c]}
\eol & \quad
+ 6 \,T^-_{d e [a} \hnabla^{d}T^-_{b c]}{}^e
- 4 \,T^-_{a d}{}^e T^-_{b e}{}^f T^-_{cf}{}^d
~,
\end{align}
and so the bosonic Lagrangian can be compactly written as \eqref{eq:FinalBLagBosonic}
without the final term.

Even with all these manipulations, the bosonic Lagrangian is fairly involved
and it turns out to be useful to perform yet another set of integrations by parts
to place it into a final form that will be useful later on. Up to a total derivative,
we find 
\begin{align}\label{eq:BosonicCBoxC}
\cL &= \tfrac{1}{3}\, C_{a b c d} \hnabla^2 C^{abcd}
- \tfrac{1}{3} C_{a b}{}^{c d} C_{c d}{}^{e f} C_{e f}{}^{a b}
- \tfrac{4}{3} C_{a b c d} C^{a e c f} C^b{}_e{}^d{}_f 
\eol & \quad
- \cR_{ab}\,^{i j} \hnabla^2 \cR^{ab}\,_{i j}
- 2\, \cR_{a}{}^{b}\,_{i}{}^{j} \cR^a{}_{c}\,_{j}{}^{k} \cR_{b}{}^{c}\,_{k}{}^{i}
+ 2 \, C^{a b c d} \,\cR_{a b}\,^{i j} \cR_{c d}\,_{ij}
\eol & \quad
+ \hat{\mathfrak f}_{a}{}^b (
	\tfrac{32}{3}\, C^{acde} C_{b c d e} 
	- 8\, \cR_{bc}\,^{i j} \cR^{ac}\,_{ij}
	) 
- 4\, \hat {\mathfrak f}_{a}{}^a (C_{bcde} C^{bcde} - \cR_{bc}\,^{i j} \cR^{bc}\,_{ij})
\eol & \quad
+ \tfrac{4}{45}\, D \hnabla^2 D
+ \tfrac{8}{225} \, D^3
+ \tfrac{2}{15} D\, C_{abcd} C^{abcd}
- \tfrac{14}{15} D\, \cR_{a b}\,^{i j} \cR^{a b}\,_{ij}
\eol & \quad
+ \tfrac{20}{3} T^-{}^{a b e} C_{a b}{}^{c d} \hnabla^f C_{f e c d}
+ 4  \,T^-{}^{abe} \hnabla^f C_{ab}{}^{cd} C_{fecd}
\eol & \quad
+ 2 \, T^-_{a b c} \hnabla_{d}{\cR^{a b}\,^{i j}}\,  \cR^{c d}\,_{ij}
+ 4 \, T^-_{a b c} \hnabla_{d}{\cR^{a d}\,^{i j}}\,  \cR^{b c}\,_{ij}
\eol & \quad
- 4 \, T^-_{a b c} \Delta^4 T^-{}^{a b c}
- \tfrac{16}{3}  C_{a b c d} T^-{}^{a b e} \hnabla_{e}{\hnabla_{f}{T^-{}^{c d f}}\, }\,  
- \tfrac{8}{3}  C_{a b c d} T^-{}^{a b e} \hnabla_{f}{\hnabla_{e}{T^-{}^{c d f}}\, }\,  
\eol & \quad
+ \tfrac{16}{3}  C_{a b}{}^{c d} T^-{}^{a e f} \hnabla^{b}{\hnabla_{c}{T^-_{d e f}}\, }\,
- 4  \, C_{a b}{}^{c d} \hnabla^{a}{T^-{}^{b e f}}\,  \hnabla_{c}{T^-_{d e f}}\, 
- 6 \, C_{a b}{}^{c d} \hnabla_{e}{T^-{}^{a b f}}\,  \hnabla_{f}{T^-{}^{c d e}}\,  
\eol & \quad
- \tfrac{16}{15} \, D\, T^-_{a b c} \hnabla^{a}{\hnabla_{d}{T^-{}^{b c d}}\, }
+ \tfrac{8}{15} D\, \hnabla^{a}{T^-_{a b c}}\,  \hnabla_{d}{T^-{}^{b c d}}\, 
\eol & \quad
- 2 \, T^-_{a b c} T^-{}^{a d e} \cR^{b c}\,_{i j} \cR_{d e}\,^{ij}
- \tfrac{4}{3} \, C_{a b e f} C^{c d e f} T^-_{a b g} T^-{}^{c d g} 
\eol & \quad
- \tfrac{1}{2}\,\hnabla^{a_1}{T^-_{a_1 ab}}\,  \hnabla^{a_2}{T^-_{a_2 cd}}\,  
\hnabla^{a_3}{T^-_{a_3 ef}}\,  \veps^{abcdef} 
- 6\, T^-_{ab}{}^g \hnabla^{a_1}{T^-_{a_1 g c}}\,  \hnabla_{d}{\hnabla^{a_2}{T^-_{a_2 e f}}\, }\,  \veps^{abcdef} 
\eol & \quad
+ 8 \,C_{a b c d} T^-{}^{e c d} T^-_{e f g} \hnabla^{a}{T^-{}^{b f g}}\,  
+ \tfrac{10}{3}\, T^-_{a b c} T^-{}^{a e d} T^-{}^{b f}{}_d \hnabla^2 T^-{}^{c}{}_{e f}
\eol & \quad
- 2\, T^-_{a b c} T^-{}^{a b d} \hnabla^{c}{T^-_{d e f}}\,  \hnabla_{g}{T^-{}^{e f g}}\,  
+ 4 \, T^-_{a b c} T^-{}^{a}{}_{d e} \hnabla^{f}{T^-{}^{b d g}}\,  \hnabla_{f}{T^-{}^{c e}{}_{g}}\,  
\eol & \quad
+ 2 \, C_{a b c d} T^-{}^{a b e} T^-{}^{c f g} T^-{}^{d}{}_{f h} T^-_{eg}{}^h 
+ \tfrac{8}{15} D \,T^-_{a b c} T^-{}^{a b d} T^-{}^{c e f} T^-_{d e f}
+ \text{fermion terms}~,
\end{align}
where we have introduced the $K$-invariant fourth order operator
\begin{align}\label{eq:Delta4T}
T^-{}^{abc} \Delta^{4} T^-_{a b c} &:= T^-{}^{abc} \Big(\hnabla_{a} \hnabla^{d} \hnabla^2 T^-_{b c d}
+ \hnabla^2 \hnabla_{a} \hnabla^{d} T^-_{b c d}
\eol & \qquad\qquad
+ \tfrac{1}{3}\, \hnabla_{a} \hnabla^2 \hnabla^{d} T^-_{b c d}
- \tfrac{4}{3}\, \hnabla_{e} \hnabla_{a} \hnabla^{d} \hnabla^{e} T^-_{b c d}\Big)~.
\end{align}

In order to extract the $K$-connections, which encode the Ricci tensor
contributions, it is useful to have the following results, which hold
up to fermionic terms:
\bsubeq
\bea
{\hat{\nabla}}_a {\hat{\nabla}}_b D &=& \hat\cD_a \hat\cD_b D - 4 \hat{f}_{ab} D~, \\
{\hat{\nabla}}_a {\hat{\nabla}}_b \cR_{cd}{}^{ij} &=& \hat\cD_a \hat\cD_b \cR_{cd}{}^{ij} 
- 4 \hat{f}_{ab} \cR_{cd}{}^{ij} + 4 \hat{f}_{a[c} \cR_{d]b}{}^{ij} 
- 4 \hat{f}_a{}^e \cR_{e [c}{}^{ij} \eta_{d] b}
\ , \\
{\hat{\nabla}}_a {\hat{\nabla}}_b C_{cdef}
&=& \hat\cD_a \hat\cD_b C_{cdef} - 4 \hat{f}_{ab} C_{cdef}
+ 4 \hat{f}_{a[c} C_{d]bef} - 4 \hat{f}_a{}^g \eta_{b[c} C_{d]gef} \non\\
&&+ 4 \hat{f}_{a[e} C_{f]bcd} - 4 \hat{f}_a{}^g \eta_{b[e} C_{f]gcd} \ , \\
{\hat{\nabla}}_a {\hat{\nabla}}_b {\hat{\nabla}}_c T^-{}^{def} 
&=& 
\hat\cD_a {\hat{\nabla}}_b {\hat{\nabla}}_c T^-{}^{def} 
- 8 \hat{f}_{a(b} \hat{\cD}_{c)} T^-{}^{def}
+ 2 \eta_{bc} \hat{f}_a{}^g \hat{\cD}_g T^-{}^{def} \non\\
&&- 12 \hat{f}_a{}^{[d} \hat{\cD}_{(b} T^-_{c)}{}^{ef]}
+ 12 \hat{f}_a{}^g \d_{(b}^{[d} \hat\cD_{c)} T^-_g{}^{ef]}
\ , \\
{\hat{\nabla}}_a {\hat{\nabla}}_b {\hat{\nabla}}^c {\hat{\nabla}}_c T^-{}^{def} 
&=& 
\hat\cD_a {\hat{\nabla}}_b {\hat{\nabla}}^c {\hat{\nabla}}_c T^-{}^{def} 
- 6 \hat{f}_{ab} \hat{\nabla}^c \hat{\nabla}_c T^-{}^{def} 
- 6 \hat{\nabla}^c \hat{\nabla}_c T^-_b{}^{[de} \hat{f}_a{}^{f]} \non\\
&&+ 6 \hat{f}_a{}^g \hat{\nabla}^c \hat{\nabla}_c T^-_g{}^{[de} \d_b^{f]}
+ 4 \hat{f}_a{}^c \hat{\nabla}_b \hat{\nabla}_c T^-{}^{def}
- 12 \hat{f}_a{}^{[d} \hat{\nabla}_b \hat{\nabla}_c T^-{}^{ef] c} \non\\
&&+ 12 \hat{f}_{ac} \hat{\nabla}_b \hat{\nabla}^{[d} T^-{}^{ef]c}
\ .
\eea
\esubeq


\subsection{The supersymmetric $F\Box F$ invariant in components}

The supersymmetrization of the $F\Box F$ action follows a similar path as $C\Box C$,
although in this case the calculations involved are significantly simpler.
We will need the description of an off-shell Yang-Mills multiplet coupled to
conformal supergravity. This was given in conformal superspace
in Appendix C of \cite{BKNT16} where we refer the reader for more details.

Let us first briefly review and elaborate on the basic ingredients that we need
for our analysis.
The non-abelian vector multiplet is described by a dimension 3/2 conformal primary
spinor field strength superfield, $\L^{\a i\, I}$, satisfying the 
constraints \cite{Siegel:1978yi,LT-M12,BKNT16}:
\bea 
S^\a_i \L^{\g k\, I}=0~,~~~
\mathbb D \L^{\g k\, I}=\frac{3}{2}\L^{\g k\, I}
~,~~~~~~
\de_\g^k \L^\g_k{}^I = 0 \ , \quad \nabla_\a^{(i}  \L^{\b j)\, I}
= \frac{1}{4} \d_\a^\b \nabla_\g^{(i} \L^{\g j)\, I}
\label{vector-Bianchies} \ .
\eea
The index $I$ is in the adjoint, and associated with $\L^{\alpha i\, I}$ is
a matrix valued in the Lie algebra of the gauge group via
$\bm \L^{\a i} = \L^{\a i\, I} {\mathbf t}_I$ with ${\mathbf t}_I$ the Hermitian
generators of the unitary gauge group, obeying
$[\mathbf t_I, \mathbf t_J] = -\ri\, f_{IJ}{}^K \mathbf t_K$.
(The generalization to non-unitary gauge groups is obvious.)
Here the covariant derivatives 
$\hat{\de}_A=(\hat{\de}_a,\de_\a^i)=E_A-\hat{\o}_A{}^{\ul b}X_b - \ri {\bm V}_A$
carry the additional Yang-Mills connections ${\bm V}:=E^A {\bm V}_A = E^A V_A{}^I \mathbf t_I$.
Their algebra is
\bea
[{\hat \nabla}_A, {\hat\nabla}_B \} 
&=&
 -\hat{T}_{AB}{}^C{  \hnabla}_C
-\hat{R}_{AB}{}^{\ul c} X_{\ul c}
- \ri   {\bm \cF}_{AB} \ ,
\eea
where the torsion and curvatures are those of conformal superspace 
and $\cF_{AB}{}^I$ is the field strength two-form.
In terms of the primary superfield $\L^{\a i\, I}$ the components of $\cF_{AB}{}^I$ are
\bea 
  \cF_\a^i{}_\b^j{}^I = 0 \ , \quad   \cF_a{}_\b^j = (\g_a)_{\a\b}   \L^{\b i\, I} \ , \quad
  \cF_{ab}{}^I = - \frac{\ri}{8} (\g_{ab})_\b{}^\a   \nabla_\a^k   \L^\b_k{}^I
  ~.
  \label{YMFs}
\eea

The various component fields are defined as follows.
The gaugino of the vector multiplet is given by the projection $\L^{\a i I}|$.
The component one-form $v_m{}^I$ and its field strength $F_{mn}{}^I$ are given by
$V_m{}^I\loco$ and $\cF_{mn}{}^I\loco$, respectively. The \emph{supercovariant}
field strength $\cF_{ab}{}^I$ is given simply by $\cF_{ab}{}^I\loco$, and as usual one finds
that $\cF_{ab}{}^I$ and $F_{mn}{}^I$ differ by gravitino terms,
\begin{align}
\cF_{ab}{}^I
= e_a{}^m e_b{}^n F_{mn}{}^I +\psi_{[a}{}_k  \g_{b]}\L^{k}{}^I
~.
\end{align}
The last physical field of the vector multiplet is a  Lorentz scalar and SU(2) triplet 
associated with the bar-projection of the following  descendant superfield
\begin{align}
  X^{ij\, I} := \frac{\ri}{4}  \nabla^{(i}_\g   \L^{\g j) I} \ .
\end{align}
In this subsection, we will denote the covariant components with exactly the same name as
the associated superfield and avoid the explicit bar-projection.

The superfields $  \L^{\a i\, I}$, $  X^{ij\, I}$, together with
$\cF_\a{}^\b{}^I = - \frac{1}{4}(\g^{ab})_\a{}^\b \cF_{ab}{}^I$,
are all annihilated by $K_a$ and
satisfy the following useful identities:
\bsubeq \label{VMIdentities}
\bea
& \nabla_\a^i   \L^{\b j}{}^I
=
- \ri \d_\a^\b   X^{ij}{}^I - 2 \ri \eps^{ij}   \cF_\a{}^\b{}^I
\ , ~~~~~~
 \nabla_\a^i   X^{jk}{}^I
=
2 \eps^{i(j}   {\hat\nabla}_{\a\b}   \L^{\b k)}{}^I
\ , \\
 & \nabla_\a^i   \cF_\b{}^\g{}^I
=
- { {\hat{\nabla}}}_{\a\b} { \L}^{\g i}{}^I
- \d^\g_\a  {\hat{\nabla}}_{\b\d}   \L^{\d i}{}^I
+ \frac{1}{2} \d_\b^\g  {\hat{\nabla}}_{\a \d}   \L^{\d i}{}^I
-\ve_{\a\b\r\t}W^{\g\r}  {\L}^{\t i}{}^I
\ ,
\\
&
S^\g_k   \cF_\a{}^\b{}^I = - 4 \ri \d^\g_\a   \L^\b_k{}^I
+ \ri \d^\b_\a   \L^\g_k{}^I \ , \qquad
S^\g_k   X^{ij}{}^I = - 4 \ri \d_k^{(i}   \L^{\g j)}{}^I
\ .
\eea
\esubeq
These imply that the $\d=\d_Q+\d_S+\d_K$ transformations of the component fields are
\bsubeq
\bea
\d   \L^{j}{}^I
&=&
- \ri \,\x_i   X^{ij}{}^I 
+\frac{\ri}{2}\tilde{\g}^{ab}\x^{j}   \cF_{ab}{}^I
~,
\\
\d   X^{jk}{}^I
&=&
-2\,\x^{(j}
{ {\hat{\slashed{\nabla}}}}
   \L^{k)}{}^I
- 4 \ri\,\eta^{(j}   \L^{k)}{}^I
~,
\\
\d    \cF_{ab}{}^I
&=&
2\,\x_i{\g}_{[a} { {\hat{\nabla}}}_{b]} { \L}^{i}{}^I
+\,T^-_{abc} \,\x_i\g^c {\L}^{ i}{}^I
+ 2\ri \,\eta^i \tilde{\g}_{ab}  \L_i{}^I
~,
\eea
\esubeq
where
\bsubeq\bea
&\hat{\de}_a\L^{i}{}^I
=
\hat{\cD}_a\L^{i}{}^I
+\frac{\ri}{2} X^{ij}{}^I \psi_a{}_j 
-\frac{\ri}{4} \tilde{\g}^{bc}\psi_a{}^{i}  \cF_{bc}{}^I
~,
\\
&\hat{\cD}_a:=
e_a{}^m \pa_m
- \frac{1}{2} \hat\omega_a{}^{cd} M_{cd}
- b_a \mathbb D
- \cV_a{}^{kl} J_{kl}
-\ri {\bm v}_a
~.
\eea
\esubeq
The transformation rule of the component connection $v_m{}^I$ can be derived from the
supergravity gauge transformation of $V^I$,
$\delta_{\cG} V{}^I = E^B \xi^{C} \cF_{C B}{}^I$, leading to
\bea
\delta v_m{}^I
=
- \xi_k \g_m\L^{k}{}^I
~.
\eea
Up to differences in conventions, these match the results of \cite{BSVanP}.

To construct the $F\Box F$ action, we again exploit the $B$-action principle.
Here the relevant composite superfield is 
\begin{align}
B_a{}^{ij} &= \frac{\ri}{4} \Tr (\bm\L^{i} \gamma_a \bm\L^{j})
	= \frac{\ri}{4} \L^{i I} \gamma_a \L^{j J}\, g_{IJ}~,
\end{align}
where $g_{IJ}$ is the Cartan-Killing metric, which we employ to raise and lower
adjoint indices. This superfield again describes a composite four-form
field strength multiplet, with the bosonic part of $C_{[ab]}$ given by
\begin{align}
C_{[ab]} = -\frac{1}{8} \veps_{abcdef} F^{cd\, I} F^{ef}{}_I~,
\end{align}
which is indeed the dual of a closed four-form. The full expressions for
the various pieces of the $B$ action principle are given in a supplementary file.

Building the bosonic Lagrangian as in \eqref{eq:FinalBLagBosonic} and
dropping a total derivative leads to
\begin{align}\label{eq:FBoxF}
\cL &=
- F_{a b}{}^{I} \hnabla^2 F^{a b}{}_I
+ X^{i j K} \hnabla^2 X_{ij K}
- \tfrac{2}{15} D\, F_{a b}\,^{I} F^{a b}\,_{I}
+ \tfrac{2}{5}\, D\,X^{i j K} X_{i j K} 
\eol & \quad
+ 4\, F_{a b}\,^{I} F^{c d}{}_I {T}^-{}^{a b e} {T}^-_{c d e}
+ C^{a b c d} \, F_{a b}\,^{I} F_{c d}\,_{I}
+ 8 T^-{}^{a b c} F_{a b}\,^{I} \hnabla^{d}{F_{c d}\,_{I}}
+ 4 T^-{}^{a c d} F_{a b}\,^{I} \hnabla^{b}{F_{c d}\,_{I}}
\eol & \quad
+ 2\, F^{a b}\,^{I} X^{i j}{}_I \hat R(J)_{a b}\,_{i j}
- 2\, F_a{}^b\,^{I} F_b{}^c\,^{J} F_c{}^a\,^{K\, } {f}_{I J K} 
+ X_i{}^j{}^{I} X_j{}^k{}^{J} X_k{}^i{}^{ K\, } f_{I J K}
\eol & \quad
+ 4 \,\hat {\mathfrak f}_{a}{}^a F_{b c}{}^{I} F^{b c}{}_I
- 8 \,\hat {\mathfrak f}_{a}{}^b F_{b c}{}^{I} F^{a c}{}_I~.
\end{align}
Some useful results, which hold up to fermions, are
\bsubeq
\bea
{\hat{\nabla}}_a {\hat{\nabla}}_b X^{ij I} &=& \hat\cD_a \hat\cD_b X^{ij I} - 4 \hat{f}_{ab} X^{ij I} \ , \\
{\hat{\nabla}}_a {\hat{\nabla}}_b F_{cd}{}^I &=& \hat\cD_a \hat\cD_b F_{cd}{}^I
- 4 \hat{f}_{ab} F_{cd}{}^I + 4 \hat{f}_{a[c} F_{d]b}{}^I - 4 \hat{f}_a{}^e F_{e [c}{}^I \eta_{d] b} 
 \ .
\eea
\esubeq


\section{The $\cN=(2,0)$ conformal supergravity invariant} \label{confSUGRAInv(2,0)}

As mentioned in the introduction, 
it is expected that there should be only 
one type B anomaly for $\cN=(2,0)$ superconformal field theories
with a specific relative coefficient between the bosonic $C \Box C$ and curvature cubed terms.
The type B anomaly should correspond to some $\cN=(2,0)$ 
conformal supergravity invariant. In this section, assuming such an 
invariant exists, we provide an 
alternative argument for its uniqueness (and, therefore, of 
the type B anomaly) and significantly extend its known bosonic terms.

The Weyl multiplet for $\cN=(2,0)$ conformal supergravity was constructed
in \cite{BSvP99}. There it was noted that many of the formulae involving the
$\cN=(2,0)$ Weyl multiplet can actually be obtained by considering their
truncations  to the $\cN=(1,0)$ case. It turns out that most of the
bosonic part of the $\cN=(2,0)$ conformal supergravity invariant can be
reverse-engineered in a similar way by considering its potential $\cN=(1,0)$
reduction. A key issue is that while there exist two $(1,0)$ conformal supergravity
invariants, only one $(2,0)$ invariant is expected. It is immediately apparent
that neither of the two $(1,0)$ invariants described in the previous sections
can \emph{alone} originate from the truncation of a $(2,0)$ invariant.
The reason for this derives from the presence of e.g. the term $D C^{abcd} C_{abcd}$ 
appearing in these actions, which cannot arise from the truncation of 
a scalar term in the $\cN=(2,0)$ case; the covariant scalar field
in the $(2,0)$ Weyl multiplet is $D^{ij}{}_{kl}$, which lies in the \rep{14} of the
USp(4) $R$-symmetry group, and so no such term can be built as a USp(4) singlet.
However, there exists a certain linear combination of the two $(1,0)$ invariants
for which all such terms cancel and it is this combination which could come from
a potential truncation. It is worth emphasizing
that since any potential $(2,0)$ conformal supergravity invariant has a $(1,0)$
truncation, this leads to a proof that there can be at most one $(2,0)$ conformal
supergravity invariant. In order to see this in more detail and uncover many of the
bosonic terms in the $(2,0)$ conformal supergravity invariant it is necessary to
first briefly review the salient details of the $(1,0)$ truncation of the
$(2,0)$ Weyl multiplet (slightly adapted to our notation and conventions).

To begin with let us recall the component structure of the Weyl multiplet of
$\cN=(2,0)$ conformal supergravity \cite{BSvP99}. 
The superconformal tensor calculus of \cite{BSvP99} is based on an off-shell gauging of the $\cN=(2,0)$ superconformal group. 
One associates the following independent fields with the local translations, $Q$-supersymmetry, USp(4) $R$-symmetry, and the dilatations:
the vielbein $e_m{}^a$, 
the gravitino $\psi_m{}^{i}$, 
the USp(4) gauge field $\cV_m{}^{ij}$, 
and the dilatation gauge field $b_m$. The remaining gauge symmetries are associated with composite connections, which include 
the spin connection $\o_m{}^{cd}$, the $S$-supersymmetry connection $\phi_m{}^i$ and the special conformal connection $\frak{f}_{m a}$. 
An off-shell representation of the conformal supersymmetry algebra is achieved by introducing three covariant matter fields: $T_{abc}{}^{ij} 
= T_{[abc]}{}^{[ij]}$, 
$\c^{i, jk} = \c^{i, [jk]}$ and $D^{ij}{}_{kl} = D^{[ij]}{}_{[kl]} = D_{kl}{}^{ij}$. 
Here $T_{abc}{}^{ij} $ is anti-self-dual with respect to its Lorentz vector indices and 
all covariant matter fields of the Weyl multiplet are traceless with respect to
the invariant antisymmetric tensor $\Omega^{ij}$ of $\rm USp(4)$. 
These covariant fields are used to build the full covariant curvatures given in \cite{BSvP99}.
In this section we have endeavored to match the conventions of \cite{BSvP99}, but the 
reader should keep in mind some minor differences explained later. 
Here we do not  provide details such as the supersymmetry transformations and expressions for the composite fields since these 
are given in \cite{BSvP99}.

We now wish to outline how to perform the truncation.
The $(2,0)$ Weyl multiplet should decompose into the following $(1,0)$ multiplets:
a Weyl multiplet on which half of the supersymmetry is manifest,
two gravitini multiplets associated with the extra spin-3/2 gauge fields,
and a Yang-Mills multiplet associated with the extra $R$-symmetry connections.
The structure of the additional gravitino multiplets and their couplings will be quite
complicated from a $(1,0)$ perspective, so we will switch them off.
However, we will elect to keep the
additional $(1,0)$ Yang-Mills multiplet, which takes its values in an SU(2) gauge group,
rather than turning it off (as in the analysis of \cite{BSvP99}).
This means that the truncation of the $\cN=(2,0)$ conformal supergravity action should also
generate the $F \Box F$ invariant in a linear combination with the
$\cN=(1,0)$ conformal supergravity invariants. This will in turn provide a useful
consistency check on our results.

The truncation follows \cite{BSvP99} very closely.
We split the USp(4) indices $i = 1, \cdots, 4$
to $(i = 1,2, i' = 1,2)$
and switch off the third and fourth gravitini $\psi_m{}^{i'} = 0$. To preserve
this last condition, we must restrict
USp(4) transformations to the block diagonal form
\begin{align} \L^{i}{}_{j} = \begin{pmatrix}
\Lambda^i{}_j & 0 \\
0 & \Lambda^{i'}{}_{j'}
\end{pmatrix} \ ,
\end{align}
where we have chosen a basis for $\Omega^{ij}$ so that\footnote{
Our conventions for $\eps_{ij}$ and $\Omega_{ij}$ differ by a sign from 
the ones of \cite{BSvP99}. However, our conventions for lowering a USp(4) 
index also differ by a sign so lowering a USp(4) index as $\l_{i} = \Omega_{ij} \l^{j}$ is 
actually equivalent to lowering the index in the conventions of \cite{BSvP99}.} 
\begin{align}
\Omega^{ij} &=
\begin{pmatrix}
\eps^{i j} & 0 \\
0 & \eps^{i' j'}
\end{pmatrix} \ , \quad
\Omega_{ij} =
\begin{pmatrix}
\eps_{i j} & 0 \\
0 & \eps_{i' j'}
\end{pmatrix} \ .
\end{align}
The above conditions ensure that $\Lambda^{i}{}_j$ and $\Lambda^{i'}{}_{j'}$
parametrize $\rm SU(2) \times SU(2)$ local gauge transformations. Considering the
supersymmetry transformation of the gravitini one must also impose
$V_m{}^{ij'} = 0$.  We keep separately the $\rm SU(2)$ gauge fields
$V_m{}^{ij} = \hat{\cV}_m{}^{ij}$ and $V_m{}^{i'j'}$ in what follows. 
Since we have turned off the extra gravitini, it is necessary to constrain some of the 
covariant fields so that the $Q$- and $S$-supersymmetry transformations are 
consistent.\footnote{We take as in \cite{BSvP99} the truncated supersymmetry parameters 
$\e^{i} \rightarrow (\e^{i} , 0)$ and $\eta^{i} \rightarrow (\eta^i, 0 )$.} 
The non-vanishing covariant fields are
\bsubeq
\begin{align}
T_{abc}{}^{i j} &= \eps^{i j} T^-_{abc}~, \qquad T_{abc}{}^{i'j'} = -\eps^{i'j'} T^-_{abc} \ , \\
\c_i{}^{jk} &= \eps^{jk} \c_i \ , \quad \c_i{}^{j'k'} = - \eps^{j'k'} \c_i \ , \quad
\c_{i'}{}^{j'k} = - \frac{15}{4} (\Omega^k)_{i'}{}^{j'} + \hf \d_{i'}^{j'} \c^k \ , \\
D^{ij}{}_{kl} &= - \eps^{ij} \eps_{kl} D \ , \quad
D^{ij}{}_{k'l'} = \eps^{ij} \eps_{k'l'} D \ , \quad
D^{i'j'}{}_{k'l'} = - \eps^{i'j'} \eps_{k'l'} D \ , \\
D^{ij'}{}_{kl'} &= - \hf \d^i_k \d^{j'}_{l'} D + \frac{15}{2} (Y^i{}_k)^{j'}{}_{l'} \ ,
\end{align}
\esubeq
where $(\Omega^k)^{i'j'}$, $(Y^{ij})^{i' j'}$ are the covariant component fields of the 
additional SU(2) Yang-Mills multiplet.

It is important to keep in mind that the conventional constraints chosen
for the  $(2,0)$ Weyl multiplet in \cite{BSvP99} are not ``traceless'', because there
is a contribution of the form $T_{acd}{}^{ij} T^{bcd}{}_{ij}$ to the  $(2,0)$ special conformal
gauge field $\mathfrak f_a{}^b$. Thus upon truncating, one would have to switch
to the ``traceless'' frame by extracting the term quadratic in $T_{abc}^-$ in order
to match formulae. We will deal with this by writing all  $(2,0)$ formulae below in
terms of a ``traceless''  $(2,0)$ gauge field
$\hat{\frak f}_a{}^b := \frak{f}_a{}^b - \frac{1}{32} T_{acd ij} T^{bcd ij}$.

Now by considering the truncation of various terms built out of the $\cN=(2,0)$ Weyl
multiplet, one can identify the linear combination of the supersymmetric $C^3$, $C \Box C$
and $F \Box F$ invariants that permits an uplift to $(2,0)$. This combination is
\be \label{eq:(2,0)Uplift}
I_{C \Box C} + \hf I_{C^3} + I_{F \Box F} \ ,
\ee
where the additional SU(2) generators are taken in the fundamental so that 
$\ri \,V_m{}^I (t_I)^{i'}{}_{j'} = V_m{}^{i'}{}_{j'}$.
As already described, this linear combination of
$C\Box C$ and $C^3$ is the only choice that eliminates the $D C^{abcd} C_{abcd}$ term
appearing in both the  $(1,0)$ invariants.
The additional contribution $I_{F \Box F}$ can be determined by uplifting the first two
invariants to  $(2,0)$ and then reducing back to  $(1,0)$; this actually provides
an independent check of the entire $I_{F \Box F}$ invariant.
The result of the uplift gives most of the bosonic terms in the corresponding
$\cN=(2,0)$ conformal supergravity invariant:
\begin{align} \label{(2,0)Lagrangian}
\cL &=
\frac{1}{3}\, C_{a b c d} {\hat \nabla}^2 C^{abcd}
+ C_{a b}{}^{c d} C^{a b}{}^{e f} C_{c d e f} 
- 4\, C_{a b c d} C^{a e c f} C^b{}_e{}^d{}_f \non\\
&- R(V)_{ab}\,^{i j} {\hat \nabla}^2 R(V)^{ab}\,_{i j}
- 2 \, R(V)_{a}{}^{b}\,_{i}{}^{j} R(V)^a{}_{c}\,_{j}{}^{k} R(V)_{b}{}^{c}\,_{k}{}^{i}
+ C^{a b c d} \,R(V)_{a b}\,^{i j} R(V)_{c d}\,_{ij} \non\\
&+ {\hat{\frak f}}_{a}{}^b \Big(
	\frac{32}{3}\, C^{acde} C_{b c d e} 
	- 8\,R(V)_{bc}\,^{i j} R(V)^{ac}\,_{ij}
	\Big)
- 4\, {\hat{\frak f}}_{a}{}^a (C_{bcde} C^{bcde} - R(V)_{bc}\,^{i j} R(V)^{bc}\,_{ij}) \non\\
&+ \frac{1}{225} D^{ij}{}_{kl} \hnabla^2 D^{kl}{}_{ij}
- \frac{2}{3375} D^{ij}{}_{kl} D^{kl}{}_{pq} D^{pq}{}_{ij}
- \frac{2}{15} D^{i j}{}_{kl} R(V)_{a b}{}^{k}{}_i R(V)^{a b}{}^{l}{}_j \non\\
&+ 4 T_{abc}{}^{ij} \hnabla_{d} R(V)^{ab}{}_{jk} R(V)^{cd}{}^k{}_i
+ 8 T_{abc}{}^{ij} \hnabla_{d} R(V)^{ad}{}_{jk} R(V)^{bc}{}^k{}_i
\non\\
&- \, T_{a b c}{}^{ij} \Delta^4 T^{a b c}{}_{ij}
+ \frac{8}{3}\, C_{a b c d} T^{a b e}{}_{ij} {\hat \nabla}_{e}{{\hat \nabla}_{f}{T^{c d f\, ij}}\, }\,  
+ \frac{4}{3}\, C_{a b c d} T^{a b e}{}_{ij} {\hat \nabla}_{f}{{\hat \nabla}_{e}{T^{c d f\, ij}}\, }\,  \non\\
&- \frac{8}{3}\, C^{a b c d} T_{a e f\, ij} {\hat \nabla}_{b}{{\hat \nabla}_{c}{T_{d}{}^{ef\,ij}}\, }\,
+ 2\, C_{a b}{}^{c d} \hnabla^{a}{T^{b e f}{}_{ij}}\,  \hnabla_{c}{T_{d e f}{}^{ij}}\, 
+ 3\, C_{a b}{}^{c d} \hnabla_{e}{T^{a b f}{}_{ij}}\,  \hnabla_{f}{T^{c d e}{}^{ij}}\,   \non\\
&- \frac{4}{3}\, C_{a b e f} C^{c d e f} T_{a b g}{}^{ij} T^{c d g}{}_{ij}
+ 4 \a \, T_{abc}{}^{ij} T^{ade\, kl} R(V)^{bc}{}_{ik} R(V)_{de}{}_{jl} \non\\
&+ 2 (1 - \a) T_{abc}{}^{ij} T^{ade}{}_{ij} R(V)^{bc}{}_{kl} R(V)_{de}{}^{kl}
\non\\ &
+ \frac{2}{15} D^{i j}{}_{k l} \Big(
	T_{a b c}{}^{kl} {\hat \nabla}^a {\hat \nabla}_d T^{b c d}{}_{i j}
	- \frac{1}{2} \hnabla^a T_{a b c}{}^{kl} \hnabla_d T^{b c d}{}_{i j}
	\Big)
\non\\
&- \frac{1}{60} D^{i j}{}_{kl} T_{a b c}{}^{kl} T^{a b d}{}_{ij}
	T^{c e f\,}{}_{pq} T_{d e f}{}^{pq}
+ \cO(T^4) \ ,
\end{align}
where we have introduced
$\hat{\nabla}_a := \cD_a - \hat{\frak{f}}_a{}^b K_b$,
and the $K$-invariant operator $\D^4$ is defined formally the same as
\eqref{eq:Delta4T}.
This action is exactly determined up to linear order in the covariant 
field $T_{abc}{}^{ij}$, while one combination of two terms quadratic in $T_{abc}{}^{ij}$ 
is left undetermined by the uplift and parametrized by an unknown real constant $\a$. 
No terms cubic in $T_{abc}{}^{ij}$ may appear for group-theoretic reasons, and indeed
one finds that in the corresponding $(1,0)$ invariant \eqref{eq:(2,0)Uplift}, all terms cubic in $T_{abc}^-$ 
cancel. There is only one possible term involving four $T_{abc}{}^{ij}$
and a single $D^{ij}{}_{kl}$, which we have given here explicitly since it
can be determined by its $(1,0)$ descendant. All other terms involving
four $T_{abc}{}^{ij}$ are denoted here $\cO(T^4)$: these involve
terms with four $T_{abc}{}^{ij}$ and two derivatives (or a
Weyl tensor or a USp(4) curvature tensor) and cannot be uniquely determined
by matching to  $(1,0)$.

Note that our $(2,0)$ conventions differ from \cite{BSvP99} in the same way as
our $(1,0)$ conventions differ from \cite{BSVanP}, see Appendix \ref{MatchConventions}.
In particular, one should keep in mind a sign difference in the Lorentz curvature
tensor from the one in \cite{BSvP99}, as well as the redefinition of the 
special conformal gauge field. Finally, the USp(4) gauge field
and curvature are scaled by a factor of 
$2$ so that
\be R(V)_{ab}{}^{kl} = 2 \,e_a{}^m e_b{}^n (\partial_{[m} V_{n]}{}^{kl} + V_{[m}{}^{j(k} V_{n]}{}^{l)}{}_{j}) \ .
\ee


\section{Discussion}
\label{conclusion}

In this paper we have described the component actions for the two conformal supergravity
invariants constructed  in \cite{BKNT16}. As shown by a supercurrent analysis \cite{BKNT16},
all conformal supergravity invariants must be given by the linear combination
$I = a I_{C^3} + b I_{C \Box C}$, where $a$ and $b$ are some constants, 
and $I_{C^3}$ and $I_{C \Box C}$ are the supersymmetric $C^3$ and $C \Box C$ invariants.
The relative coefficients of the three purely gravitational
pieces \eqref{eq:ConfInvs} then obey the linear relation already known in the literature,
see e.g. \cite{Beccaria:2015uta} and references therein.
As discussed in the previous section, amongst the invariants there exists a special choice
which corresponds to the $\cN=(1,0)$ truncation of the $\cN=(2,0)$ conformal supergravity
invariant. Identifying the choice of coefficients as the
one-parameter family that permits an uplift to $\cN=(2,0)$ supergravity allowed us to
prove uniqueness of the $(2,0)$ conformal supergravity invariant and construct many of its
bosonic terms. 
It would be interesting to develop an alternative method to construct 
this invariant
and recover all terms.
This would be interesting not only in the context of Weyl anomalies, but also
in the context of higher-derivative gravity
theories \cite{Lu:2011ks,Pang:2012rd,Lu:2013hx}, where the combination
\eqref{eq:Intro(2,0)} of Weyl invariants particular to  $(2,0)$ theories was observed to
correspond to 6D critical gravity.
It would be interesting to understand how the covariant matter fields of the
standard Weyl multiplet affect the dynamics of these models when extended to
supergravity.
We expect that the development of $(2,0)$ conformal superspace in six dimensions together with the ideas advocated in 
\cite{BKNT16} for the $(1,0)$ case will provide a viable means to complete the construction
of the $(2,0)$ invariant. We hope the results in this paper will be useful
for these applications.

While the supersymmetric $C^3$ and $C \Box C$ invariants have been constructed 
in the standard Weyl multiplet,
it is interesting to note that one could construct 
these invariants with a variant Weyl multiplet, known as the dilaton-Weyl multiplet (or type II formulation) \cite{BSVanP}.\footnote{
See e.g. \cite{Coomans:2011ih,Bergshoeff:2012ax} for a recent discussion.}  
The dilaton-Weyl multiplet is obtained by coupling the standard Weyl multiplet to a tensor
multiplet \cite{Howe:1983fr,Koller:1982cs} and exchanging the covariant matter fields of the standard
Weyl multiplet with those of the tensor multiplet, which include a scalar field 
$\sigma$, a gauge 2-form $B_{ab}$, and a negative chirality spinor $\chi_\a^i$. This
procedure can directly be performed for the supersymmetric $C^3$ and $C \Box C$ invariants and
would lead to new higher-derivative invariants in the dilaton-Weyl multiplet.

Because the dilaton-Weyl multiplet possesses a built-in Weyl compensator in the form of
the scalar field $\sigma$, it is evident that there are many other curvature invariants one
can construct that have no analogues in the standard Weyl multiplet. For example,
one can construct a supersymmetric $\cR_{abcd} \Box \cR^{abcd}$ action in addition to
$C_{abcd} \Box C^{abcd}$. To see this, it is important to realize that the supersymmetry
transformations of certain fields built out of the dilaton-Weyl multiplet can be formally
mapped to those of a Yang-Mills multiplet taking values in the 6D Lorentz algebra \cite{BR}.
This property leads to a natural correspondence between an action based on 
a Yang-Mills multiplet and an action in terms of the dilaton-Weyl multiplet.
Using this correspondence (given in component form in \cite{BR}),
the supersymmetric $F \Box F$ action can be converted into an invariant containing
$\cR_{abcd} \Box \cR^{abcd}$.

It is worth mentioning that the YM correspondence can also be exhibited
in superspace.
First, one must adapt superspace to the dilaton-Weyl multiplet by introducing
a compensating tensor multiplet $\Phi$ (with lowest component $\sigma$), which satisfies $\nabla_\a^{(i} \nabla_\b^{j)} \Phi = 0$,
and then make use of the modified derivatives $\mathscr D_\a^i$ and the associated torsion
components that were introduced in section 3.4 of \cite{BKNT16} (with $X = \Phi$).\footnote{The
resulting superspace geometry is equivalent to using the SU(2) 
superspace formulation of conformal supergravity in \cite{LT-M12} with the torsion
component $C_{a}{}^{ij}$ switched off. This is also equivalent to the
superspace of \cite{BR}.} 
One can then construct a primary superfield $\bm \L^{\a i}{}_\b{}^\g$ satisfying the following
constraints (which are formally the same constraints as those of a vector multiplet 
valued in the Lorentz algebra):\footnote{The superspace results given here are similar to the 5D $\cN = 1$ description of 
the Riemann curvature squared 
invariant in the dilaton-Weyl multiplet given in \cite{BKNT-M5D}.}
\bea
 \bm \L^{\a i}{}_{\b}{}^\b=0~,~~~
{\mathscr D}_\a^{(i} {\bm \L}^{\b j)}{}_\g{}^\d - \frac{1}{4} \d_\a^\b {\mathscr D}_\r^{(i} {\bm \L}^{\r j)}{}_\g{}^\d = 0 
~,~~~
{\mathscr D}_{\a i} \bm \L^{\a i}{}_{\b}{}^\g = 0 \ .
\eea
The appropriate primary superfield $\bm \L^{\a i}{}_\b{}^\g$ is given by
\be \label{YMtrick-0}
\bm \L^{\a i} := \bm \L^{\a i}{}_{\b}{}^\g M_\g{}^\b
= \Phi^{3/4} \Big( {\mathscr D}_\b^i \mathscr W^{\a\g} 
- \frac{2}{3} \eps^{\a\g\d\r} \mathscr D_\d^i \mathscr N_{\r \b} 
- \frac{1}{3} \d^\a_\b {\mathscr D}_\d^i \mathscr W^{\g\d} \Big) M_\g{}^\b
\ .
\ee
We can now describe a 
supersymmetric $\cR_{abcd} \Box \cR^{abcd}$ invariant in the dilaton-Weyl multiplet in 
a completely analogous way as the 
supersymmetric $F \Box F$ action by using the $B$ action principle with 
$B^{\a \b ij} = \ri \Tr \big( \bm \L^{\a (i} \bm \L^{\b j)} \big)$. 
It would be interesting to carry out a detailed 
analysis of this supersymmetric invariant elsewhere.
As a side note, the primary superfield \eqref{YMtrick-0} may be used to construct other 
invariants. For instance, one can describe a topological invariant
containing the 6D Euler term using the $A$ action principle 
with $A_\a{}^{ijk} = \eps_{\a\b\g\d} \, \Tr \big( \bm \L^{\b (i} \bm \L^{\g j} \bm \L^{\d k)}\big)$.

A natural question to ask is whether other curvature invariants with fewer than
six derivatives may be built when compensating superfields, such as the tensor superfield
of the dilaton-Weyl multiplet, are present. It turns out that it is possible to construct
all of the curvature-squared invariants using either tensor or linear multiplet compensators,
just as in five dimensions \cite{OP13}.
In six dimensions, there is a topological action, corresponding to the supersymmetrization
of $B_2 \wedge H_4$, that couples a tensor multiplet (with two-form potential $B_2$)
to a four-form field strength multiplet (containing a closed four-form $H_4$). It can be built
by starting with the $A$ action principle with the specific choice\footnote{We refer
the reader to \cite{BKNT16} for details about the gauge invariance of this action.}
\be A_\a{}^{ijk} = \eps_{\a\b\g\d} V^{\b (i} B^{\g\d jk)} \ , \label{BHaction}
\ee
where $V^{\a i}$ is the prepotential for the tensor multiplet
(see \cite{Sokatchev:PhiPrepot, LT-M12} for details)
with the superfield $B^{\a\b}{}^{ij}$ satisfying the constraints
\eqref{eq:Bconstraint} and \eqref{eq:FourFormConstraint}
corresponding to a four-form field strength
multiplet \cite{ALR14}.
A special form of this action was already constructed
in \cite{BSVanP}, for the case where $H_4 = \Tr (\bm{F} \wedge \bm{F})$ is built out of
a non-abelian vector multiplet. Its component action contains the terms
$\sigma \Tr (\bm{F}_{ab} \bm{F}^{ab})$ and
$\sigma \Tr (\bm{X}_{ij} \bm{X}^{ij})$.
Employing the YM correspondence, the first of these terms gives a
curvature-squared invariant in the gauge $\sigma=1$ via
$\Tr (\bm{F}_{ab} \bm{F}^{ab}) \rightarrow \cR_{abcd} \cR^{abcd}$,
as shown in \cite{BR} (see also \cite{BSS1, BSS2}).
This construction corresponds to choosing the superfield
$B^{\a \b ij} = \ri \Tr \big( \bm \L^{\a (i} \bm \L^{\b j)} \big)$ 
within \eqref{BHaction} and imposing the gauge $\Phi = 1$.

A second curvature squared invariant can be built by choosing a composite
abelian vector multiplet built out of a linear multiplet $G^{ij}$ \cite{OzkanThesis}.
In superspace, this corresponds to taking $B^{\a\b ij} = \ri \mathbb W^{\a (i} \mathbb W^{\b j)}$,  
with ${\mathbb W}^{\a i}$ built from the superfield $G^{ij}$ as
\bea
{\mathbb W}^{\a i} 
&=& \frac{1}{G} \nabla^{\a\b} \U_\b^i 
+ \frac{4}{G} \big( W^{\a\b} \U_\b^i +10 \ri X^\a_j G^{ji} \big) 
- \frac{1}{2 G^3} G_{jk}  (\nabla^{\a\b} G^{ij}) \U_\b^k
+ \frac{1}{2 G^3} G^{ij} F^{\a\b} \U_{\b j} \non\\
&&+ \frac{\ri}{16 G^5} \eps^{\a\b\g\d} \U_{\b j} \U_{\g k} \U_{\d l} G^{ij} G^{kl} \ ,
\eea
where $\U_\a^i := \frac{2}{3} \nabla_{\a j} G^{ij}$ and $F_{\a\b} := \frac{\ri}{4} \nabla_{[\a}^k \U_{\b] k}$.
Provided $G^{ij}$ satisfies the linear multiplet constraint $\nabla_\a^{(i} G^{jk)} = 0$,
$\mathbb W^{\a i}$ describes a composite abelian vector multiplet.\footnote{The component
form of the vector multiplet $\mathbb W^{\a i}$ appeared originally in \cite{BSVanP}.
The corresponding result in Minkowski superspace was given in \cite{GL84}.}
The isotriplet $\mathbb X^{ij}$ turns out to include a term
$\cR G^{ij} / G$, and the term $\mathbb X_{ij} \mathbb X^{ij}$ within the
component action generates $\cR^2$.

A third curvature-squared invariant is possible but unlike the previous two examples,
it actually requires the supersymmetric version
of $B_2 \wedge H_4$ where the four-form field strength is \emph{not} a product of
Yang-Mills curvature two-forms. One must take the four-form multiplet built from the
superfield $B^{\a\b\, ij}$ given in \eqref{eq:SuperBforCBoxC}.
It is not hard to see that the component action
must contain a term $C_{abcd} C^{abcd}$, which then completes the set of curvature-squared
invariants. 
Because the use of the four-form field strength multiplet has to our knowledge not been considered, 
the last curvature-squared invariant remained
undiscovered until now.
The three curvature squared invariants described here, 
corresponding to supersymmetric extensions of $\cR_{abcd} \cR^{abcd}$, 
$\cR^2$, and $C_{abcd} C^{abcd}$, extend the analogous 5D examples constructed
in components \cite{OP13} and superspace \cite{BKNT-M5D},
and span the supersymmetric extensions of all possible curvature 
squared terms. Another
curvature-squared invariant was partially constructed in
\cite{BSS1, BSS2} and may correspond to a linear combination of these.
It would also be interesting to find a connection with the curvature-squared
invariants of \cite{Nishino:1986da} involving supergravity coupled to matter.
We leave a further discussion and analysis of
these invariants, including the component expression for $C_{abcd} C^{abcd}$,
for future work.


\acknowledgments
\noindent
We thank Sergei M.~Kuzenko and Stefan Theisen for discussions
and collaboration at an early stage of this work.
The work of DB was supported in part by the ERC Advanced Grant no. 246974,
{\it ``Supersymmetry: a window to non-perturbative physics''}
and by the European Commission Marie Curie International Incoming Fellowship 
grant no. PIIF-GA-2012-627976. 
JN acknowledges support 
from GIF -- the German-Israeli Foundation for Scientific Research and Development.
The work of GT-M is supported by the Interuniversity Attraction Poles Programme
initiated by the Belgian Science Policy (P7/37), in part by COST Action MP1210 
``The String Theory Universe,'' and was also in part supported by the
Australian Research Council (ARC) Discovery Project DP140103925.
DB also thanks Kasper Peeters for frequent correspondence regarding
the use of {\it Cadabra}, and Ergin Sezgin and the Mitchell Institute at Texas A\&M
for kind hospitality and support at the end stage of the project.
GT-M is grateful to the School of Physics at the University of Western Australia for kind hospitality
and support during the final part of this project.
We thank the Galileo Galilei Institute for Theoretical Physics for hospitality and the INFN for partial support during the 
completion of this work in September 2016.


\appendix


\section{6D $\cN=(1,0)$ conformal superspace}
\label{conf-superspace}

In this appendix we review and expound on 
 the 6D $\cN=(1,0)$ conformal superspace of \cite{BKNT16}
focusing on the ingredients relevant to our presentation in this paper.


\subsection{The superconformal algebra}

The 6D $\cN=(1,0)$ superconformal algebra contains the generators of
translation ($P_{a}$), Lorentz  ($M_{ab}$), 
special conformal ($K^{a}$), dilatation ($\mathbb{D}$), $\rm SU(2)$ ($J_{ij}$),
$Q$-supersymmetry ($Q_\a^i$) and $S$-supersymmetry ($S^\a_i$) 
transformations.\footnote{
For our spinor conventions and notation we refer the reader to Appendix A of \cite{BKNT16}.} 
Their (anti)commutation algebra is
\bsubeq\label{SCA}
\bea
&[M_{a b} , M_{c d}] = 2 \eta_{c [a} M_{b] d} - 2 \eta_{d [ a} M_{b ] c} 
\ ,\quad
[J^{ij} , J^{kl}] = \eps^{k(i} J^{j) l} + \eps^{l (i} J^{j) k} \ ,
~~~~~~
 \\
&[M_{a b}, P_c] = 2 \eta_{c [a} P_{b]} 
\ , \quad 
[M_{a b} , K_c] = 2 \eta_{c [a} K_{b]} 
\ , \quad
[\mathbb{D} , P_a] = P_a 
\ , \quad
 [\mathbb{D} , K_a] = - K_{a} 
 \ ,~~~~~~ 
 \\
  &{[}M_{ab} , Q_\g^k {]} = - \hf (\g_{ab})_\g{}^\d Q_\d^k 
\,, \quad 
[{\mathbb D} , Q_\a^i ] = \hf Q_\a^i 
\ ,\quad
[J^{ij} , Q_\a^k ] = \eps^{k (i} Q_\a^{j)} \ ,~~~~~~~~
 \\
&{[}M_{ab} , S^\g_k {]} =- \hf (\tilde{\g}_{ab})^\g{}_\d S^\d_k  
\ , \quad 
[{\mathbb D} , S^\a_i ] = - \hf S^\a_i
\ , \quad [J^{ij} , S^\a_k ] = \d_k^{(i} S^{\a j)} 
\ ,~~~~~~~~
\\
&\{ Q_\a^i , Q_\b^j \} 
=
- 2 \ri \eps^{ij} (\g^{c})_{\a\b} P_{c} 
\ , \quad
\{ S^\a_i , S^\b_j \} = - 2 \ri \eps_{ij} (\tilde{\g}^{c})^{\a\b} K_{c} 
~,~~~~~~
\\
&\{ S^\a_i , Q_\b^j \} = 2 \d^\a_\b \d^j_i \mathbb D - 4 \d^j_i M_\b{}^\a + 8 \d^\a_\b J_i{}^j 
\ ,
\quad
[K_a , P_b] = 2 \eta_{a b} {\mathbb D} + 2 M_{ab} 
\ , ~~~~~~
\\
& [ K_a , Q_\a^i ] = - \ri (\g_a)_{\a\b} S^{\b i} \ , \quad [S^\a_i , P_a ] = - \ri (\tilde{\g}_a)^{\a\b} Q_{\b i} 
 \ , ~~~~~~
\eea
\esubeq
with all other (anti)commutators vanishing.
Note that the generator $M_\a{}^\b = - \frac{1}{4} (\g^{ab})_\a{}^\b M_{ab}$
acts on $Q_\g^k$ and $S^\g_k$ as follows
\bea
&[ M_\a{}^\b , Q_\g^k ] = - \d_\g^\b Q_\a^k + \frac{1}{4} \d^\b_\a Q_\g^k \ , \quad 
[ M_\a{}^\b , S^\g_k ] = \d_\a^\g S^\b_k - \frac{1}{4} \d^\b_\a S^\g_k \ .
\eea

We can group together the translation and $Q$-supersymmetry as 
generators of supertranslations, $P_A = (Q_\a^i,P_a )$.
Similarly we group together the special conformal and $S$-supersymmetry transformations
by denoting $K^A = (S^\a_i,K^a )$ and the closed subset of generators that do not
contain $P_{{A}}$ by $X_{\underline{a}} = (M_{ab} , J_{ij}, \mathbb D , K^A)$. The 
superconformal algebra takes the form 
\bea
[X_{\underline{a}} , X_{\underline{b}} \} = -f_{\underline{a} \underline{b}}{}^{\underline{c}} X_{\underline{c}} \ , \quad
[X_{\underline{a}} , P_{{B}} \} &=
- f_{\underline{a} {{B}}}{}^{{C}} P_{{C}}
 -f_{\underline{a} {{B}}}{}^{\underline{c}} X_{\underline{c}}
	\ , \quad
[P_{{A}} , P_{{B}} \} = -f_{{{A}} {{B}}}{}^{{{C}}} P_{{C}}
	\ .
	~~~~~~
	\label{SCA_compact}
\eea
The group associated with the subalgebra generated by $X_{\ul a}$ is denoted by $\cH$.


\subsection{Gauging the superconformal algebra in conformal superspace} 
\label{Gauging}

The 6D $\cN=(1,0)$ conformal superspace is parametrized
by local bosonic $(x)$ and fermionic coordinates $(\theta_i)$, $z^M = (x^{m},  \q^\mu_i)$, 
where $m = 0, 1, 2, 3, 4, 5$, $\mu = 1,2,3,4$ and $i = 1,2$. 
In gauging the superconformal group we associate to each generator 
a connection one-form. 
In particular, we associate with the supertranslation $P_A$ a vielbein one-form 
$E^A = (E^\a_i , E^a)=\rd z^ME_M{}^A$, while
to the generators $X_{\underline{a}} = (M_{ab} , J_{ij}, \mathbb D , S^\a_i , K^a)$ 
we associate the connection one-forms 
$\omega^{\underline{a}} = (\Omega^{ab} , \Phi^{ij}, B , \frak{F}_\a^i , \frak{F}_a)
= \rd z^{M} \omega_M{}^{\underline{a}}=E^A\o_{A}{}^{\ul a}$. 
These are used to construct the covariant derivatives, which 
have the form
\be
\nabla_A  
:=
 E_A - \o_A{}^{\ul c} X_{\ul c}
= E_A - \hf \Omega_A{}^{cd} M_{cd} - \Phi_A{}^{kl} J_{kl} - B_A \mathbb D
	- \mathfrak{F}_A{}_B K^B 
	\ ,
\ee
with $E_A = E_A{}^M \partial_M$ the inverse vielbein. 

The superconformal algebra is gauged in superspace\footnote{Note that in the conventional superspace approach to 
6D supergravity the locally superconformal structure is encoded in super-Weyl transformations 
\cite{Gates6D,LT-M12} analogously to the 4D case \cite{HoweTucker}.}
by the following local transformations of the vielbein and the connections
\bsubeq\label{transf-E_O}
\bea
\d_\cK E_M{}^{A}
&=&
\pa_M\x^A
+E_M{}^C\x^B {T}_{BC}{}^A
+ {\o}_M{}^{\ul{c}}\x^B{f}_{B\ul{c}}{}^A
+E_M{}^C {\L}^{\ul b}{f}_{\ul bC}{}^A
~,
\\
\d_\cK  {\o}_M{}^{\ul{a}}
&=&
\pa_M {\L}^{\ul{a}}
+E_M{}^{C}\x^B {R}_{BC}{}^{\ul{a}}
+ {\o}_M{}^{\ul{c}}\x^{B} {f}_{B\ul{c}}{}^{\ul{a}}
+E_M{}^{C} {\L}^{\ul{b}} {f}_{\ul{b}C}{}^{\ul{a}}
+ {\o}_M{}^{\ul{c}} {\L}^{\ul{b}}{f}_{\ul{b}\ul{c}}{}^{\ul{a}}
~.~~~~~~
\eea
\esubeq
Here $\x^A=\x^A(z)$ parametrizes the covariant general coordinate transformations 
and $\L^{\ul a}=\L^{\ul a}(z)=(\L^{ab},\L^{ij},\s,\L_\a^i,\L^a)$ 
are the gauge parameters associated with the structure group $\cH$.
It is important to observe that in \eqref{transf-E_O} 
${f}_{\ul{b}\ul{c}}{}^{\ul{a}},\,{f}_{\ul{b}C}{}^{\ul{a}}$, and ${f}_{\ul bC}{}^A$
are components of the structure constants of the superconformal algebra
\eqref{SCA} and \eqref{SCA_compact}.
The superfields 
${T}_{BC}{}^A$ and ${R}_{BC}{}^{\ul{a}}$ are respectively the torsion and curvature tensors 
that appear as components of the two-forms
\begin{subequations} \label{TRexpComp}
\begin{align}
T^C &:= \hf E^B \wedge E^A T_{AB}{}^C 
= \rd E^C - E^B \wedge \omega^{\underline{a}} \,f_{\underline{a} B}{}^C 
\ , \\
R^{\underline{c}} 
&:= \hf E^B \wedge E^A R_{AB}{}^{\underline{c}} 
= \rd \omega^{\underline{c}}
- E^B \wedge \omega^{\underline{a}} \, f_{\underline{a} B}{}^{\underline{c}}
- \hf \omega^{\underline{b}} \wedge \omega^{\underline{a}} \,f_{\underline{a} \underline{b}}{}^{\underline{c}}
 \ .
\end{align}
\end{subequations}
From the explicit structure constants of the superconformal algebra, these tensors become\begin{subequations} 
\label{torCurExp}
\bea
T^a &=& \rd E^a + E^b \wedge \Omega_b{}^a + E^a \wedge B 
\ , 
\label{torCurExp_a}
\\
T{}^\a_i &=& \rd E^\a_i 
+ E^\b_i \wedge \Omega_\b{}^\a 
+ \hf E^\a_i \wedge B - E^{\a j} \wedge \Phi_{ji} 
- \ri \, E^c \wedge \mathfrak{F}_{\b i} (\tilde{\g}_c)^{\a\b} 
\ ,~~~~~~~~~~~ 
\label{torCurExp_b}
\\
\RD &=& \rd B + 2 E^a \wedge \mathfrak{F}_a + 2 E^\a_i \wedge \mathfrak{F}_\a^i 
\ , 
\label{torCurExp_c}
\\
\RM^{ab} &=& \rd \Omega^{ab} + \Omega^{ac} \wedge \Omega_c{}^b 
- 4 E^{[a} \wedge \mathfrak{F}^{b]} 
+ 2 E^\a_j \wedge \mathfrak{F}_\b^j (\g^{ab})_\a{}^\b 
\ ,
\label{torCurExp_d}
 \\
\RJ^{ij} &=& \rd \Phi^{ij} - \Phi^{k (i} \wedge \Phi^{j)}{}_k - 8 E^{\a (i} \wedge \mathfrak{F}_{\a}^{j)} 
\ ,
\label{torCurExp_e}
 \\
\RK^a &=& \rd \mathfrak{F}^a 
+ \mathfrak{F}^b \wedge \Omega_b{}^a 
- \mathfrak{F}^a \wedge B 
- \ri \mathfrak{F}_\a^k \wedge \mathfrak{F}_{\b k} (\tilde{\g}^a)^{\a\b} 
\ , 
\label{torCurExp_f}
\\
\RS_\a^i &=& \rd \mathfrak{F}_\a^i 
- \mathfrak{F}_\b^i \wedge \Omega_\a{}^\b
- \hf \mathfrak{F}_\a^i \wedge B 
- \mathfrak{F}_\a^j \wedge \Phi_j{}^i 
- \ri E^{\b i} \wedge \mathfrak{F}^c (\g_c)_{\a\b}    \ .
\label{torCurExp_g}
\eea
\end{subequations}
This gauging leads to a consistent
 modification of the superconformal algebra \eqref{SCA_compact}
described by the (anti)commutation relations
\bsubeq\label{gauging-SCA-1}
\bea
{[}X_{\ul a},X_{\ul b}\}
&=&
-f_{\ul a\ul b}{}^{\ul c}X_{\ul c}
~,
\label{gauging-SCA-1_a}
\\
{[}X_{\ul a},{\de}_B\}
&=&
-f_{\ul aB}{}^{C}{\de}_{C}
-{f}_{\ul aB}{}^{\ul c}X_{\ul c}
~,
\label{gauging-SCA-1_b}
\\
{[}{\de}_A,{\de}_B\}
&=&
-{T}_{AB}{}^{C}{\de}_{C}
-{R}_{AB}{}^{\ul c}X_{\ul c}
~,
\label{gauging-SCA-1_c}
\eea
\esubeq
where the generators of the supertranslations, $P_A$, are replaced by the covariant derivatives $\de_A$.
The $\cK$ transformations, \eqref{transf-E_O},
 can then be described by the following variation of the covariant derivatives
\be
\delta_\cK \nabla_A = [\cK,\nabla_A] 
\ ,\quad 
\cK :=\xi^C \nabla_C+\L^{\ul c}X_{\ul c}
= \xi^C \nabla_C + \hf \L^{cd} M_{cd} + \L^{kl} J_{kl} + \s \mathbb D + \L_A K^A
~,
\label{SUGRAtransmations}
\ee
provided that one interprets the action of the covariant derivatives on the parameters as
\be \nabla_A \xi^B := E_A \xi^B + \omega_A{}^{\underline{c}} \xi^D f_{D \underline{c}}{}^{B} \ , \quad 
\nabla_A \L^{\underline{b}} := E_A \L^{\underline{b}}
+ \omega_A{}^{\underline{c}} \xi^D f_{D \underline{c}}{}^{\underline{b}}
+ \omega_A{}^{\underline{c}} \L^{\underline{d}} f_{\underline{d}\underline{c}}{}^{\underline{b}} 
\ .
\ee

A {\it covariant} superfield $\Phi$
is such that under $\cK$ transformations it varies with
no derivatives on the parameters and can be represented as
\be 
\d_{\cK} \Phi =\cK \Phi 
\ .
\label{SUGRAtransformations-tensor}
\ee
Due to \eqref{SUGRAtransmations}, covariant derivatives of $\F$ transform covariantly 
$\d_\cK(\de_A\F)=\cK\de_A\F$.
The torsion and curvatures superfields, $T_{AB}{}^C$ and $R_{AB}{}^{\ul c}$,
are necessarily covariant.
A superfield $\Phi$ is said to be \emph{primary} if it is annihilated by the special
conformal generators, $K^A \Phi = 0$.
Due to \eqref{SCA}, $S^\a_i\F=0$ is a sufficient condition for $\F$ to be primary.

In \cite{BKNT16} it was proven that a consistent description of the 
6D $\cN=(1,0)$ Weyl multiplet in conformal superspace can be achieved by:
\begin{itemize}
\item[i)] choosing the gauging in conformal superspace described before that leads to the algebraic structures in 
\eqref{gauging-SCA-1};
\item[ii)] requiring the covariant derivative algebra to resemble the one of 6D $\cN=(1,0)$
super Yang-Mills theory \cite{Siegel:1978yi,Nilsson,Wess81,HST}, which takes the form
\bsubeq
\bea \label{eq:A.12a}
\{ \nabla_\a^i , \nabla_\b^j \} &=& - 2 \ri \eps^{ij} (\g^a)_{\a\b} \nabla_a 
\ , \\ 
\left[ \nabla_a , \nabla_\a^i \right] &=& (\g_a)_{\a\b} \cW^{\b i} 
\ ,\\
\left[\nabla_a , \nabla_b\right] &=& - \frac{\ri}{8} (\g_{ab})_\a{}^\b \{ \nabla_\b^k , \cW^\a_k \}
~,
\eea
\esubeq
where $\cW^{\a i}$ is a primary dimension $3/2$ operator 
such that
\be
{[}K^A,\cW^{\a i}\}=0
\ ,\quad
\{ \nabla_\a^{(i} , \cW^{\b j)} \} = \frac{1}{4} \d^\b_\a \{ \nabla_\g^{(i} , \cW^{\g j)} \} 
\ , \quad \{ \nabla_\g^k , \cW^\g_k \} = 0
\ ; 
\ee
\item[iii)]
constraining the operator $\cW^{\a i}$ to be of the form
\be 
\cW^{\a i} = 
W^{\a\b} \nabla_\b^i
+ \hf \cW(M)^{\a i ab} M_{ab}
+ \cW(J)^{\a i jk} J_{jk}
+ \cW(\mathbb D)^{\a i} \mathbb D
+ \cW(K)^{\a i}{}_B K^B
 \ ,
\ee
where $W^{\a\b}=\tfrac{1}{3!}(\tilde{\g}^{abc})^{\a\b}W_{abc}$ 
is the super-Weyl tensor \cite{GatesSiegel-1979,AwadaTownsendSierra,LT-M12} such that
\be
S^\g_kW^{\a\b}=0 
\ , \quad
\mathbb D W^{\a\b}=W^{\a\b}
\ .\\
\ee
\end{itemize}

It turns out that, under the previous assumptions,
 the super-Jacobi identities for the 
algebra \eqref{SCA_compact} uniquely fix all the superfields
$\cW(M)^{\a i ab},\ \cW(J)^{\a i jk},\ \cW(\mathbb D)^{\a i},\ \cW(K)^{\a i}{}_B$
together with  the torsion and curvatures
in terms of $W^{\a\b}$ and its covariant derivatives.
Moreover, $W^{\a\b}$ satisfies the Bianchi identities 
\bsubeq\label{WBI}
\bea 
&\nabla_\a^{(i} \nabla_{\b}^{j)} W^{\g\d} = - \d^{(\g}_{[\a} \nabla_{\b]}^{(i} \nabla_{\r}^{j)} W^{\d) \r} 
\ , \\
&
\nabla_\a^k \nabla_{\g k} W^{\b\g} - \frac{1}{4} \d^\b_\a \nabla_\g^k \nabla_{\d k} W^{\g\d}
 = 8 \ri \nabla_{\a \g} W^{\g \b} 
 \ .
\eea
\esubeq
We refer the reader to \cite{BKNT16} for more details.


\subsection{Different superspace frames}

It is worth underlining that the action of the generators $X_{\ul a}$ on $\de_A$,
eq. \eqref{gauging-SCA-1_b}, was chosen in \cite{BKNT16} to be identical to the action on
$P_A$, eq. \eqref{SCA_compact}. 
This condition leads for example to a simple choice \eqref{eq:A.12a} for the form of the
supersymmetry algebra, but gives more complicated constraints on the vector
curvatures \eqref{compConstraintsTR}, which contribute a number of covariant fields
into the composite connections.
As described in the main body of the paper, for components applications, different choices
of conventional constraints can be more convenient.
In conformal superspace this results in a framework where 
the structure constants ${f}_{\ul aB}{}^{\ul c}$ are replaced by structure functions
$\hat{f}_{\ul aB}{}^{\ul c}$.\footnote{The reader can find a pedagogical discussion of 
structure functions in the textbook \cite{FVP}.}

For applications in this paper we  use a change of frame where only the vector covariant derivatives
are modified. In particular, we 
introduce the $\hat{\de}_A$ derivatives as\footnote{Note that, due to $W_{a[b}{}^{e}W_{cd]e}=0$ the 
useful relation $\de_aW_{bcd}=\hat{\de}_aW_{bcd}$ holds.}
\bsubeq\label{new-frame}
\bea
\hat{\de}_\a^i&:=&\de_\a^i
~,
\\
\hat{\de}_a
&:=& \nabla_a
- \hf\l_1 W_{a}{}^{b c} M_{bc}
-\ri \l_2 (\gamma_a)_{\alpha \beta} X^{\alpha j} S^{\beta}_j
\non\\
&&
-\left(
\l_3 Y \eta_{ac}
+\l_4 \de^b W_{a b c}
+\l_5 W_{a}{}^{ e f} W_{e f c}
\right)
 K^c
~,
\eea
\esubeq
where $\l_1,\l_2,\l_3,\l_4,\l_5$ are arbitrary real constant parameters
and the dimension 3/2 covariant superfield $X^\a_i$ is defined as
\bea
X^{\a i} &:=& -\frac{\ri}{10} \nabla_{\b}^i W^{\a\b} \ .
\eea

The new 
$\hat{\de}_A$ derivatives
have the same vielbein of ${\de}_A$, $E_M{}^A$, but have modified connections
\be
\hat{\de}_A:=E_A{}^M\left(\pa_M-\hat{\omega}_M{}^{\ul a}X_{\ul a}\right)
~,~~~~~~
\hat{\omega}_M{}^{\ul a}
=
\omega_M{}^{\ul a}
+E_M{}^B\cM_B{}^{\ul a}
~.
\label{change_frame-1}
\ee
For the change of frame \eqref{new-frame}, $\cM_A{}^{\ul c}$  is given by 
\bsubeq
\bea
\cM_a{}^{ cd}&=&\l_1W_a{}^{cd}
~,
\\
\cM_a{}_\g^k&=&-\ri\l_2 (\g_a)_{\g\d} X^{\d k} 
~,
\\
\cM_a{}_c&=&
\l_3 Y \eta_{ac}
+ \l_4 \de^b W_{a b c}
+\l_5 W_{a}{}^{ e f} W_{e f c}
~,
\eea
\esubeq
with all the other components identically zero.

Given the algebra \eqref{gauging-SCA-1} 
of $(X_{\ul a},\de_A)$ and the relations \eqref{new-frame}, 
it is straightforward 
to show that
the (anti)commutation relations satisfied by
$X_{\ul a}$ and $\hat{\de}_A$  have the following form\bsubeq\label{gauging-SCA-2}
\bea
{[}X_{\ul a},X_{\ul b}\}
&=&
-f_{\ul a\ul b}{}^{\ul c}X_{\ul c}
~,
\label{gauging-SCA-2_a}
\\
{[}X_{\ul a},\hat{\de}_B\}
&=&
-f_{\ul aB}{}^{C}\hat{\de}_{C}
-\hat{f}_{\ul aB}{}^{\ul c}X_{\ul c}
~,
\label{gauging-SCA-2_b}
\\
{[}\hat{\de}_A,\hat{\de}_B\}
&=&
-\hat{T}_{AB}{}^{C}\hat{\de}_{C}
-\hat{R}_{AB}{}^{\ul c}X_{\ul c}
~.
\label{gauging-SCA-2_c}
\eea
\esubeq
Here
$f_{\ul a\ul b}{}^{\ul c},\ f_{\ul aB}{}^{C}$ match the structure constants of
the superconformal group, but
${f}_{\un a}{}_A{}^{\un c}$
becomes a nontrivial structure function
$\hat{f}_{\un a}{}_A{}^{\un c}$ that has dependence on $W_{abc}$ and its descendant superfields.
In the new frame, the torsion and curvatures,
 $\hat{T}_{AB}{}^{C}$ and $\hat{R}_{AB}{}^{\ul c}$, 
have the same form as their unhatted partners in \eqref{TRexpComp}, 
with $\o_{\underline{a}}\to\hat{\o}_{\underline{a}}$ and 
$f_{\underline{a}B}{}^{\underline{c}}\to \hat{f}_{\underline{a}B}{}^{\underline{c}}$.

It turns out that by properly tuning the parameters $\l_1$ and $\l_4$ as 
\bea
\l_1=2~,~~~~~~
\l_4=-\hf~,
\label{simpl-1}
\eea 
we can set to zero the torsion $\hat{T}_{ab}{}^c$ and the dilatation curvature
\bea
\hat{T}_{ab}{}^c=0
~,~~~~~~
\hat{R}(\mathbb D)_{ab}
=0~.
\eea
Moreover, the choice of parameters \eqref{simpl-1}
removes terms of the form $\hat{\de}^c W_{a bc}$  from $\hat{R}_{ab}{}^{cd}(M)$
and ensures that $\hat{f}^a{}_{b}{}^{\ul c}={f}^a{}_{b}{}^{\ul c}$,
so that
\bea
[K_a, \hat{\de}_b]
=
2\eta_{ab}\mathbb D
+2M_{ab}
~.
\eea
The choice \eqref{simpl-1} simplifies the component analysis and we will assume it
from now on.
We  leave $\lambda_2$, $\lambda_3$ and $\lambda_5$ unfixed for the moment 
although two sets of choices, highlighted 
in table \ref{table-coefficients},
are particularly interesting.
The first, that we denoted as ``Traceless,'' gives rise to a superspace geometry whose
projection to components, as described in section \ref{new-frames}, 
leads to convenient constraints on the component curvatures.
The second choice leads to a superspace whose component constraints
are identical to the ones
originally used in \cite{BSVanP}.

With the choice \eqref{simpl-1}, the structure constants $\hat{f}_{\un a}{}_A{}^{\un c}$
turn out to have the following nontrivial components
\bsubeq\label{structure-functions}
\bea
\hat{f}^\a_i{}_b{}_\g^k
&=&
\left(
\hf
+\frac{8}{5} \lambda_2
\right)
W_{bcd}\,\d_i^k (\tilde{\g}^{cd})^\a{}_\g 
~,
\\
\hat{f}^\a_i{}_b{}_c&=&
-2(\l_2+2\l_3)  X^{\a}_{ i}\eta_{bc}
-\left(1+ 2\lambda_2\right) (\tilde{\g}_{bc})^{\a}{}_{\b} X^{\b}_{ i}
-\frac{1}{2}({\g}_{bc})_\b{}^{\g}X_{\g i}{}^{\b\a} 
~,
\eea
\esubeq
while all the other components of $\hat{f}_{\ul a b}{}^{\ul c}$ are identical to the ones of the superconformal 
algebra.
Here the dimension 3/2 covariant superfield $X_{\g}^k{}^{\b\a}$ is 
\cite{BKNT16}
\bea
X_\g^k{}^{\a\b} 
&=& 
-\frac{\ri}{4}\nabla_\g^k W^{\a\b} - \d^{(\a}_{\g} X^{\b) k} 
~,~~~~~~
X_\g^k{}^{\a\b}=X_\g^k{}^{\b\a}
~,~~~
X_\g^k{}^{\a\g}=0
~.
\eea
Note that for the choice of frame we are considering,  
the commutator $[S^\a_i , \hat{\nabla}_b]$ is
\bea
[S^\a_i , \hat{\nabla}_b]
&=&
 -\ri (\tilde{\g}_b)^{\a\b} \hat{\de}_{\b i} 
 - \left(
\hf
+\frac{8}{5} \lambda_2
\right)
W_{bcd}(\tilde{\g}^{cd})^\a{}_\g 
S^\g_i
+2 (\l_2+2\l_3)  X^{\a}_{ i}K_b
\non\\
&&
+\Big{[}
\left(1+ 2\lambda_2\right) (\tilde{\g}_{bc})^{\a}{}_{\b} X^{\b}_{ i}
+\frac{1}{2}({\g}_{bc})_\b{}^{\g}X_{\g i}{}^{\b\a} \Big{]}K^c
~.
\eea

The torsion $\hat{T}_{AB}{}^C$
and curvatures $\hat{R}_{AB}{}^{\ul c}$ 
in the new frame can be computed by using the
(anti)commutation relations of the $\de_A$ derivatives derived in \cite{BKNT16}
together with \eqref{new-frame}.
The anticommutator of two spinor derivatives,
$\{\hat{\de}_\a^i,\hat{\de}_\b^j\}$, has the following torsion and curvatures\bsubeq\label{new-frame_spinor-spinor}
\bea
\hat{T}_{\a}^i{}_\b^j{}^c
&=&
2\ri\ve^{ij}(\g^c)_{\a\b}
~,
\\
\hat{R}(M)_{\a}^i{}_\b^j{}^{cd}
&=&
4\ri\ve^{ij}(\g_a)_{\a\b} W^{acd}
~,
\\
\hat{R}(S)_{\a}^i{}_\b^j{}_\g^k
&=&
4\l_2\ve^{ij}\ve_{\a\b\g\d} X^{\d k} 
~,
\\
\hat{R}(K)_{\a}^i{}_\b^j{}_c
&=&
2\ri\ve^{ij}(\g^a)_{\a\b}\left(
\lambda_3 \eta_{ac}Y
+\lambda_4 \hat{\de}^b W_{a b c}
+\lambda_5 W_{a}{}^{ef} W_{cef}
\right)
~,
\eea
\esubeq
where the omitted components vanish.
The non-zero torsion and curvatures in the commutator ${[}\hat{\de}_a,\hat{\de}_\b^{j}]$ are:
\bsubeq\label{new-frame_vector-spinor}
\bea
\hat{T}_{a}{}_\b^j{}^\g_k
&=&
-\frac{1}{2} (\g_{a})_{\b \d} W^{\d \g}\d^j_k
~,
\\
\hat{R}(\mathbb D)_a{}_\b^j
&=&
- 2\ri \left(\lambda_2 + \frac{5}{8}\right) (\gamma_{a})_{\b\g} X^{\g j}
~,
\\
\hat{R}(M)_a{}_\b^j{}^{cd}
&=&
- 2\ri\left(\lambda_2 + \frac{3}{8}\right) (\gamma_{a}{}^{ c d})_{\b\g} X^{\g j} 
- 4\ri \left(\lambda_2 + \frac{1}{8} \right) \delta_{a}^{[c} (\g^{d]})_{\b\g} X^{\g j} 
\non\\
&&
-\ri(\gamma_{a}{}^{ c d})_{\g \d} X_{\b}^j{}^{\g \d}
+2\ri(\gamma_{a})_{\b\g} (\gamma^{c d})_{\d}{}^{\r}X_{\r}^j{}^{\gamma \delta }
~,
\\
\hat{R}(J)_a{}_\b^j{}^{kl}
&=&
8\ri \left(\lambda_2 + \frac{5}{8}\right) (\gamma_{a})_{\b\g} X^{\g (k}\ve^{l)j} 
~,
\\
\hat{R}(S)_a{}_\b^j{}_\g^k
&=&
\ri\left(\frac{5}{16} + \frac{1}{2}\lambda_2 - \lambda_3\right)(\gamma_{a})_{\b \g} \veps^{jk} Y
- \frac{\ri}{4} (\gamma_{a})_{ \b \d} \, Y_{\g}{}^{\d}{}^{ jk} 
- \frac{2\ri}{5} \lambda_2  (\gamma_{a})_{\g\d} Y_{\b}{}^{\d}{}^{ jk} 
\non\\
&&
- \frac{\ri}{8} (\gamma_{a})_{\b \d} \hat{\de}_{\g \r}{W^{\d \r}}  \veps^{jk}
-\ri\left( \frac{1}{8}+\frac{2}{5} \lambda_{2}\right)
(\gamma_{a})_{\g\d} \hat{\de}_{\b\r}{W^{\d \r}}\veps^{jk}
\non\\
&&
- \frac{\ri}{4} \lambda_5 (\gamma_{a})_{\d \epsilon}\, \veps_{\b \r \t \g}\,W^{\d \r} W^{\epsilon \t} \veps^{jk} 
~, \\
\hat{R}(K)_a{}_\b^j{}_c
&=&
\frac{\ri}{4} (\gamma_{c})_{\b \g} \hat{\de}_{a}{X^{\g j}} 
+2 \ri \left(\lambda_3 - \frac{1}{8}\right)\eta_{ac} \hat{\de}_{\b\g}{X^{\g j}} 
- \frac{\ri}{4} (\gamma_{a c d})_{\g \d} \hat{\de}^{d}{X_{\b}^j{}^{\g \d }} 
\non\\
&&
+ \frac{\ri}{3} (\gamma_{a})_{\b \d} (\gamma_{c d})_{\r}{}^{\g}\hat{\de}^{d}{X_{\g}^j{}^{\d \r }} 
+\ri\left(\lambda_2 + \frac{1}{2}\lambda_5+\frac{1}{8}\right) (\gamma_{a})_{\d \r}(\gamma_{c})_{\b \g}W^{\g \d}X^{\r j} 
\non\\
&&
+\ri \left(\lambda_{2} + \frac{1}{2} \lambda_5\right)(\gamma_{a})_{\b \g} (\gamma_{c})_{\d \r}W^{\g \d} X^{\r j} 
+ \frac{5\ri}{12} (\gamma_{a})_{\b\r}(\gamma_{c})_{\g \epsilon} W^{\g \d} X_{\d}^j{}^{\r \epsilon } 
\non\\
&& 
+ \frac{\ri}{4} ( \gamma_{a})_{\g \r} (\gamma_{c})_{\b \epsilon} 	W^{\g \d} X_{\d}^j{}^{\r \epsilon }
- \ri\lambda_{5} (\gamma_{a})_{\g \r} (\gamma_{c})_{ \d \epsilon}W^{\g\d}X_{\b}^j{}^{\r\epsilon } 
~.
\eea
\esubeq
 In the new frame,
 the commutator of two vector derivatives,
$[\hat{\de}_a,\hat{\de}_b]$,
has the following non-vanishing torsion and curvatures:
\bsubeq\label{new-frame_vector-vector}
\bea
\hat{T}_{ab}{}^\gamma_k 
&=&
(\g_{ab})_\b{}^\a \Big{[}
X_{\a k}{}^{\b\g} 
-2\left(\l_2+ \frac{3}{8}\right) \d^\g_\a X^\b_k 
\Big{]}
~,
\label{new-frame_ALL-T-c}
\\
\hat{R}(M)_{a b}{}^{c d} &=&
Y_{a b}{}^{cd}
+ 8\left(\lambda_3-\frac{1}{8}\right) Y \d_{a}^{[c}\d_b^{d]}
+ 8\left(\lambda_5 - \frac{1}{2}\right) W_{a b f} W^{fcd} 
~,
\label{new-frame_ALL-R-i}
\\
\hat{R}(J)_{ab}{}^{kl} 
&=& 
\frac{1}{2}(\g_{ab})_\d{}^\g Y_\g{}^\d{}^{kl} = Y_{a b}{}^{kl}
~,
\\
\hat{R}(S)_{ab}{}_\gamma^k 
&=&
- \frac{\ri}{3} (\gamma_{a b})_{\delta}{}^{\alpha} \hat{\de}_{\gamma \beta} X_{\alpha}^k{}^{\beta \delta }
- \frac{\ri}{6} (\gamma_{a b c})_{\alpha\beta} \hat{\de}^c X_{\gamma}^k{}^{\alpha \beta }
- \frac{\ri}{6} \veps_{\gamma \beta \epsilon \rho} (\gamma_{a b})_{\delta}{}^\rho
W^{\alpha \beta} X_{\alpha}^k{}^{\delta \epsilon}
\non\\
&&
+\ri \left(2\lambda_2 + \frac{3}{4}\right) \hat{\de}_{[a} X^{\alpha k} (\gamma_{b]})_{\alpha\gamma}
- \frac{\ri}{2}\left(\lambda_2 + \frac{3}{8}\right)
\veps_{\gamma \beta \delta \epsilon} (\gamma_{a b})_{\alpha}{}^{\epsilon}
W^{\alpha\beta} X^{\delta k}
~,~~~~~~~~~~~~
\label{new-frame_ALL-R-k}
\\
\hat{R}(K)_{ab}{}_{c}
&=& 
\frac{1}{4} \hat{\de}^d Y_{a b c d}
+\frac{\ri}{3} X_{\alpha}^k{}^{\beta \gamma } X_{\beta k}{}^{\alpha \delta}(\gamma_{a b c})_{\gamma\delta}
+\ri (\gamma_{a b})_{\epsilon}{}^\alpha
(\gamma_{c})_{\gamma\delta} X_{\alpha}^k{}^{\beta \gamma } X_{\beta k}{}^{\delta \epsilon}
\non\\
&&
+ \frac{5\ri}{3}\left(\lambda_2 + \frac{3}{8}\right)
X^{\gamma k} X_{\gamma k}{}^{\alpha\beta} (\gamma_{a b c})_{\alpha\beta}
-4 \ri \left(\lambda_2 + \frac{5}{16}\right) X^{\alpha k} X_{\beta k}{}^{\gamma \delta}
(\gamma_{a b})_{\gamma}{}^{\beta} (\gamma_{c})_{\alpha\delta}
\non\\
&&
+2 \ri \left(\lambda_2 + \frac{3}{8} \right) \left(\lambda_2 + \frac{5}{8}\right)
X^{\alpha k} X^{\beta}_k (\gamma_{a b c})_{\alpha\beta}	
+ 2 \left(\lambda_3 - \frac{1}{8}\right) \hat{\de}_{[a} Y \, \eta_{b] c} 
\non\\
&&
+ \frac{1}{2} \left(\lambda_5 - \frac{1}{2}\right)
W^{\alpha\beta} \hat{\de}_{[a} W^{\gamma \delta} (\gamma_{b]})_{\alpha\gamma} (\gamma_{c})_{\beta \delta}
~.
\label{new-frame_ALL-R-l}
\eea
\esubeq
Note that we have introduced the following higher dimension descendant superfields constructed 
from spinor derivatives of $W^{\a \b}$:
\bsubeq
\bea
Y_\a{}^\b{}^{ij} &:=& - \frac{5}{2} \left( \nabla_\a^{(i} X^{\b j)} - \frac{1}{4} \d^\b_\a \nabla_\g^{(i} X^{\g j)} \right)
 \ , \\
Y &:=& \frac{1}{4} \nabla_\g^k X^\g_k \ , \\
Y_{\a\b}{}^{\g\d} &:=& 
\nabla_{(\a}^k X_{\b) k}{}^{\g\d}
- \frac{1}{3}  \nabla_\r^k X_{(\a k}{}^{\r(\g}\d_{\b)}^{\d)}
\ .
\eea
\esubeq
By using \eqref{WBI} and the previous definitions, 
one can derive the following relations for the descendant superfields: 
\bsubeq \label{eq:Wdervs}
\bea
\nabla_\a^i X^{\b j} 
&=& 
- \frac{2}{5} Y_\a{}^\b{}^{ij} 
- \frac{2 }{5} \eps^{ij} \hat{\nabla}_{\a \g} W^{\g\b}
- \hf \eps^{ij} \d_\a^\b Y 
\ , \\
\nabla_\a^i X_\b^j{}^{\g\d}
&=& 
\hf \d^{(\g}_\a Y_\b{}^{\d)}{}^{ij} - \frac{1}{10} \d_\b^{(\g} Y_\a{}^{\d)}{}^{ij}
- \hf \eps^{ij} Y_{\a\b}{}^{\g\d}
- \frac{1}{4}\eps^{ij} \hat{\nabla}_{\a\b} W^{\g\d} 
\non\\
&&
+ \frac{3}{20} \eps^{ij} \d_\b^{(\g} \hat{\nabla}_{\a \r} W^{\d) \r}
-\frac{1}{4}\eps^{ij} \d^{(\g}_\a \hat{\nabla}_{\b \r} W^{\d) \r} 
\ , \\
\nabla_\a^i Y 
&=& 
- 2 \ri \hat{\nabla}_{\a \b} X^{\b i} 
\ , \\
\nabla_\g^k Y_{\a}{}^\b{}^{ij} 
&=& 
\frac{2}{3} \eps^{k (i} \Big( 
- 8 \ri\,\hnabla_{\g\d}{X^{j)}_{\alpha}{}^{\beta \delta}} 
- 4 \ri\, \hnabla_{\a\d}{X^{j)}_{\gamma}{}^{\beta \delta}}
+ 3 \ri \,\hnabla_{\g\a}{X^{\beta j)}} 
+ 3 \ri\,\delta^{\beta}_{\gamma}\,\hnabla_{\a\d}{X^{\delta j)}} 
\non\\ &&~~~~~~~~~
- \frac{3\ri}{2} \delta_{\alpha}^{\beta} \,\hnabla_{\g\d}{X^{\delta j)}} 
- 3 \ri \,\veps_{\alpha \gamma \delta \epsilon} W^{\beta \delta} X^{\epsilon j)} 
+ 4 \ri\, \veps_{\alpha \gamma \epsilon \rho} \,W^{\delta \epsilon} X^{j)}_{\delta}{}^{\beta \rho} 
\Big)\ , \\
\nabla_\e^l Y_{\a\b}{}^{\g\d} &=& 
- 4 \ri \,\hnabla_{\e(\a}{X_{\beta)}^l{}^{\gamma\delta}} 
+ \frac{4\ri}{3} \delta_{(\alpha}^{(\gamma} \hnabla_{\b)\rho}{X_{\epsilon}^l{}^{\delta) \rho}} 
+ \frac{8\ri}{3} \delta_{(\alpha}^{(\gamma} \hnabla_{|\epsilon\rho|}{X_{\beta)}^l{}^{\delta) \rho}} 
+ 8 \ri\, \delta_{\epsilon}^{(\gamma} \hnabla_{\rho(\alpha}{X_{\beta)}^l{}^{\delta) \rho}} 
\non\\ &&
- \frac{4\ri}{3} \, W^{\rho \sigma} \,
	\delta_{(\alpha}^{(\gamma} \veps_{\beta) \epsilon \sigma \tau} X_{\rho}^l{}^{\delta) \tau} 
- 8 \ri\, \veps_{\epsilon \rho \sigma (\alpha} W^{\rho (\gamma} X_{\beta)}^l{}^{\delta) \sigma}~.
\eea
\esubeq
These relations define the $Q$-supersymmetry transformations of the descendant superfields of the super-Weyl
tensor.
Their $S$-supersymmetry transformations are instead given by the following relations
\cite{BKNT16}:
\bsubeq\label{S-on-X_Y}
\bea
S^\a_i X^{\b j} &=& \frac{8\ri}{5} \d_i^j W^{\a\b} ~, 
~~~~~~
S^\a_i X_\b^j{}{}^{\g\d} 
=
-\ri\d^j_i \d^\a_\b W^{\g\d} 
+\frac{2\ri}{5} \d^j_i \d^{(\g}_{\b} W^{\d) \a} 
~,
\label{S-on-X_Y-a}
\\
S^\g_k Y_\a{}^{\b}{}^{ij} 
&=& 
-\d^{(i}_k \left( 
16 X_\a^{j)}{}^{\g\b} 
- 2 \d_\a^\b X^{\g j)}
+ 8 \d^\g_\a X^{\b j)} \right) 
~, 
\label{S-on-X_Y-b}
\\
S^\r_j Y_{\a\b}{}^{\g\d} 
&=& 
24 \left( \d^\r_{(\a} X_{\b) j}{}^{\g\d}
- \frac{1}{3} \d^{(\g}_{(\a} X_{\b) j}{}^{\d) \r} \right) 
~ ,~~~~~~
S^\a_i Y = - 4 X^\a_i 
~ .
\label{S-on-X_Y-c}
\eea
\esubeq

Note that the Bianchi identities for the $\hat{\de}_A$ derivatives
are identically satisfied due to \eqref{WBI},
\eqref{eq:Wdervs} and the following useful relations
\bsubeq\label{deaBI}
\bea
\hat{\de}_{\gamma (\alpha} Y_{\beta)}{}^{\gamma i j}
&=&
0~, \qquad
\hat{\de}^{\gamma (\alpha} Y_{\gamma}{}^{\beta) i j} 
=
4\ri\left(5 + 8 \lambda_2\right) X^{\gamma (i} X_{\gamma}^{ j)}{}^{\alpha \beta}
~,
\\
\hat{\de}^{\delta (\alpha} X_{\delta}^i{}^{\beta \gamma)}
&=&
W^{\delta (\alpha} X_{\delta}^i{}^{\beta \gamma)}
+ 6 \left(\lambda_2 + \tfrac{3}{8}\right) W^{(\alpha \beta} X^{\gamma) i}
~,
\\
\hat{\de}_{\delta (\alpha} Y_{\beta \gamma)}{}^{\delta \epsilon} 
&=& 
0~, 
\\
\hat{\de}^{\delta (\alpha} Y_{\delta \epsilon}{}^{\beta \gamma)} &= &
24\ri X_{\epsilon}^k{}^{\delta (\alpha } X_{\delta k}{}^{\beta \gamma)}
- 8\ri X_{\rho}^k{}^{\delta (\alpha} \delta_{\epsilon}^{\beta} X_{\delta k}{}^{\gamma) \rho}
+ 96\ri \left(\lambda_2 + \tfrac{3}{8}\right) X^{(\alpha k} X_{\epsilon k}{}^{\beta \gamma)}
\non\\
&&
- 16 \ri\left(\lambda_2 + \tfrac{3}{8} \right)X^{\delta k} X_{\delta k}{}^{(\beta \gamma} \delta_\epsilon^{\alpha)}
~.
\eea
\esubeq

We conclude this appendix by mentioning that
in the new frame
the supergravity gauge transformations of the vielbein, the connections
and of a covariant tensor superfield $\F$, respectively, are
 (compare with \eqref{transf-E_O} and \eqref{SUGRAtransformations-tensor})
\bsubeq\label{new-frame-sugra-transf}
\bea
\d_\cK E_M{}^{A}
&=&
\pa_M\x^A
+E_M{}^C\x^B\hat{T}_{BC}{}^A
+\hat{\o}_M{}^{\ul{c}}\x^B{f}_{B\ul{c}}{}^A
+E_M{}^C\hat{\L}^{\ul b}{f}_{\ul bC}{}^A
~,
\\
\d_\cK \hat{\o}_M{}^{\ul{a}}
&=&
\pa_M\hat{\L}^{\ul{a}}
+E_M{}^{C}\x^B\hat{R}_{BC}{}^{\ul{a}}
+\hat{\o}_M{}^{\ul{c}}\x^{B}\hat{f}_{B\ul{c}}{}^{\ul{a}}
+E_M{}^{C}\hat{\L}^{\ul{b}}\hat{f}_{\ul{b}C}{}^{\ul{a}}
+\hat{\o}_M{}^{\ul{c}}\hat{\L}^{\ul{b}}{f}_{\ul{b}\ul{c}}{}^{\ul{a}}
~,~~~~~~
\\
\d_\cK\F
&=&
\cK\F~,
\eea
\esubeq
where the operator $\cK$ is
$\cK=(\x^A\hat{\de}_A+\hat{\L}^{\ul a}X_{\ul a})$ and
the gauge parameters $\hat{\L}^{\ul a}$ and $\L^{\ul a}$ are related to each other by
\bea\label{new-frame_gauge-parameters}
\hat{\L}^{\ul a}
=
\L^{\ul a}
+\x^A\cM_A{}^{\ul a}
~.
\eea


\section{Relating notation and conventions}
\label{MatchConventions}

As underlined in Section \ref{SCTC} and Appendix \ref{conf-superspace},
the ``hat'' frame described in section \ref{new-frames} is equivalent to the one employed originally
in \cite{BSVanP} by choosing the parameters as follows
\bea
\lambda_1=2
~,~~~
\lambda_2=-\frac{5}{16}
~,~~~
\lambda_3=\frac{5}{32}
~,~~~
\lambda_4=-\frac{1}{2}
~,~~~
\lambda_5=1
~.
\label{B1}
\eea
This is true up to a choice of notation and conventions. In this appendix we describe the 
relevant notational differences 
and show how to obtain the results of \cite{BSVanP} from the ones in section \ref{new-frames}.

\begin{table*}[!t]
\centering
\renewcommand{\arraystretch}{1.4}
\resizebox{14cm}{!}{
\begin{tabular}{@{}cc@{}} \toprule
Our notation&  Bergshoeff et al. \cite{BSVanP}  \\ \midrule
$\eta^{ab}$, 
$\veps^{abcdef}$, $\G^a$, $\G^{a_1\cdots a_n}$, $\cdots$
&~~~~~~
$\eta^{ab}$,
$\veps^{abcdef}$, $-\ri \gamma^a$, $(-\ri)^n\g^{a_1\cdots a_n}$, $\cdots$
\\
\midrule
$P_a$, $K^a$, $Q^i$,  $S_i$ 
&~~~~~~
$P_a$, $K^a$, $-2Q^i$,  $2S_i$ 
 \\
$M_{ab}$,
$J_{ij}$,
$\mathbb D$
&~~~~~~
$-2M_{ab}$,
$2U_{ij}$,
$D$
  \\
\midrule
$e_m{}^a$,
$\hat{\frak{f}}_m{}^\ha$,
$\psi_m{}^i$,
$\hat\phi_m{}^i$ 
& ~~~~~~
 $e_\mu{}^a$,
$f_\mu{}^a$,
 $\psi_\mu{}^i$,
 $\phi_\mu{}^i $  
 \\
$\hat\omega_m{}^{ab}$,
$\cV_m{}^{ij}$, 
$b_m$
&~~~~~~
 $-\omega_\mu{}^{ab}$,
$\hf V_\mu{}^{ij}$,
$b_\mu$
 \\
\midrule
$\hat R(P)_{ab}{}^{c}$,
$\hat R(K)_{ab}{}_c$,
$\hat R(Q)_{ab}{}^i$,
$\hat R(S)_{ab}{}_i$ 
&~~~~~~
$\hat R(P)_{ab}{}^{c}$,
 $\hat R(K)_{ab}{}_c$,
$\frac{1}{2} \hat R(Q)_{ab}{}^i$, 
$\frac{1}{2} \hat R(S)_{ab}{}_i$ 
\\
$\hat R(M)_{ab}{}^{cd}$,
$\hat R(J)_{ab}{}^{ij}$,
$\hat R(\mathbb D)_{ab}$,
&~~~~~~
$-\hat R(M)_{ab}{}^{cd}$,
$\hf\hat R(U)_{ab}{}^{ij}$,
$\hat R(\mathbb D)_{ab}$
\\
\midrule
$\x^a$,
$\l^a$,
$\x_i$,
$\eta^i$
&~~~~~~
$\x^a$,
$\L_K^a$,
$\hf\ve_i$,
$\hf\eta^i$
\\
$\l^{ab}$,
$\l^{ij}$,
$\s$
&~~~~~~
$-\ve^{ab}$,
$\hf\L^{ij}$,
$\L_D$
\\
\bottomrule
\end{tabular}}
\caption{Translation of notation and conventions}
\label{tab:ConvComp}
\end{table*}

First of all, note that throughout our paper we have used chiral four-component spinor notation
while in \cite{BSVanP} eight-component spinor notation is used. 
To match the results, one should first reinterpret  our formulae using eight component  spinors.
This is straightforward by using Appendix A of \cite{BKNT16}
where we refer the reader for more details.
Our $8\times 8$ Dirac spinors $\Psi$ and matrices $\G^a$ are
\begin{align}
\Psi =
\begin{pmatrix}
\psi^\alpha \\
\chi_\alpha
\end{pmatrix}
~, \qquad
\Gamma^a =
\begin{pmatrix}
0 & (\tilde\gamma^a)^{\alpha\beta} \\
(\gamma^a)_{\alpha\beta} & 0 
\end{pmatrix}
~,\qquad
\Gamma_* =
\begin{pmatrix}
\delta^\a_\b & 0 \\
0 & -\delta_\a^\b
\end{pmatrix}
~,
\end{align}
where $\G_*$ obeys
$\Gamma_{[a} \Gamma_b \Gamma_c \Gamma_d \Gamma_e \Gamma_{f]} = \veps_{abcdef} \Gamma_*$.
Similarly, there is a direct relation between ${\g}^{a_1\cdots a_n}$,
$\tilde{\g}^{a_1\cdots a_n}$ and $\G^{a_1\cdots a_n}:=\G^{[a_1}\G^{a_2}\cdots\G^{a_n]}$
since a product of chiral $\g$s are straightforwardly lifted to a product of Dirac $\G$s.
The eight component spinor generators of the 6D $\cN=(1,0)$ superconformal algebra 
are 
\bea
Q^i=\left(
\begin{array}{c}
0\\ 
Q_{\a}^i
\end{array}
\right)
~,~~~~~~
S_{i}= \left(
\begin{array}{c}
S^\a_{ i}
\\
0
\end{array}
\right)
~.
\eea
Similarly, all the (anti)chiral spinor fields are straightforwardly lifted to eight components, such that, e.g.
$\psi_m{}^{\a}_{i}Q_{\a}^{i}\to\bar{\psi}_m{}_{i}Q^{i}$,
${\hat{\f}}_m{}_\a^{i}S^\a_{i}\to\bar{\hat{\f}}_m{}^{i}S_{i}$.
The results of \cite{BSVanP} for the 6D $\cN=(1,0)$ Weyl multiplet 
may be obtained from the results of section \ref{SCTC} by lifting to eight-component spinors, 
fixing the parameters as in \eqref{B1} and renaming the fields in accordance with
Table \ref{tab:ConvComp}.
We have normalized the covariant fields $T^-_{abc}$, $\chi^i$ and $D$ to match those of
\cite{BSVanP}.

\begin{footnotesize}

\end{footnotesize}

\end{document}